\documentclass[twocolumn,showpacs,preprintnumbers,amsmath,amssymb]{revtex4-2}

\usepackage{CJK}

\usepackage{graphicx}

\usepackage{dcolumn}

\usepackage{bm}
\usepackage[normalem]{ulem}
\usepackage{color}

\usepackage{epstopdf}

\numberwithin{equation}{section}

\def\rf#1{(\ref{#1})}
\newcommand{\beps}{\bm{\epsilon}}
\newcommand{\bew}{\begin{widetext}}
\newcommand{\ew}{\end{widetext}}

\newcommand{\eps}{\epsilon}
\newcommand{\dw}{\delta\omega}
\newcommand{\nn}{\nonumber}
\newcommand{\re}{{\rm Ei}}
\newcommand{\bJ}{\bf{J}}

\newcommand{\lak}{\lambda_K}

\newcommand{\hn}{\hat{\bf n}}
\newcommand{\dr}{\delta\rho}
\newcommand{\ii}{{\rm i}}
\newcommand{\lnl}{L_{_{NL}}}

\newcommand{\lv}{L_v}
\newcommand{\ls}{\ell^*}
\newcommand{\lk}{\lambda_k}

\newcommand{\bp}{\mathbf{p}}
\newcommand{\bq}{\mathbf{q}}
\newcommand{\bv}{\mathbf{v}}

\newcommand{\br}{\mathbf{r}}
\newcommand{\bR}{\mathbf{R}}
\newcommand{\bff}{\mathbf{f}}

\newcommand{\hx}{\bf{\hat{x}}}

\newcommand{\tri}{\triangle}

\newcommand{\sep}{ \ \ \ , \ \ \ }

\newcommand{\beq}{\begin{equation}}
\newcommand{\eeq}{\end{equation}}
\newcommand{\beqn}{\begin{eqnarray}}
\newcommand{\eeqn}{\end{eqnarray}}
\newcommand{\pp}{\partial}
\newcommand{\dd}{{\rm d}}
\newcommand{\ee}{{\rm e}}

\newcommand{\cO}{{\cal O}}

\newcommand{\cG}{{\cal G}}

\newcommand{\la}{\langle}

\newcommand{\ra}{\rangle}

\newcommand{\vnab}{{\bf \nabla}}

\begin{document}

\begin{CJK*}{GBK}{}
%\preprint{APS/123-QED}

%

\title{Hydrodynamic Theory of Two-dimensional Chiral Malthusian Flocks}
\author{Leiming Chen}
	\email{leiming@cumt.edu.cn}
	\affiliation{School of Materials and Physics, China University of Mining and Technology, Xuzhou Jiangsu, 221116, P. R. China}
	\author{Chiu Fan Lee}
	\email{c.lee@imperial.ac.uk}
	\affiliation{Department of Bioengineering, Imperial College London, South Kensington Campus, London SW7 2AZ, U.K.}
	\author{John Toner}
	\email{jjt@uoregon.edu}
	\affiliation{Department of Physics and  Institute for Fundamental
 Science, University of Oregon, Eugene, OR $97403$}
	
\date{\today}
\begin{abstract}
We study the hydrodynamic behavior of two-dimensional chiral dry Malthusian flocks; that is, chiral polar-ordered active matter with neither number nor momentum conservation.
We show that, in the absence of fluctuations, such systems generically   form a ``time cholesteric", in which the velocity of the entire system rotates uniformly at a fixed frequency $b$. Fluctuations about this state belong to the universality class of ($2+1$)-Kardar-Parisi-Zhang (KPZ) equation, which implies short-ranged orientational order in the hydrodynamic limit.  We then  show that, in the limit of weak chirality, the hydrodynamics of a system with reasonable size  is expected to governed by  the linear regime of the KPZ equation, exhibiting quasi-long-ranged orientational order. Our predictions for the velocity and number density correlations are testable in both simulations and experiments.
\end{abstract}
%\pacs{05.65.+b, 64.60.Ht, 87.18Gh}
\maketitle
\end{CJK*}

\section{Introduction}{\label{Intro}}

Active matter \cite{book, Active1, Active2, Vicsek, TT1, TT3, birdrev, Active3, Active4, tissue, PS, Nematics, Chate1, Chate2} differs from equilibrium systems due to a variety of intrinsically non-equilibrium effects, including self-propulsion \cite{book, Vicsek, TT1, TT3, birdrev} and birth and death \cite{Toner_prl12, Chen_prl20, Chen_pre20}. These non-equilibrium effects, which are most commonly found in biological systems, lead to many surprising phenomena, including long-ranged order in two-dimensional (2D) systems with spontaneously broken continuous symmetries \cite{book, Vicsek, TT1, TT3, birdrev} (i.e., 2D flocks), and hydrodynamic instabilities in very viscous systems \cite{sppprl, activerheo}.

While it is not an intrinsically non-equilibrium phenomenon, chirality \cite{glove} is also a ubiquitous feature of biological systems. It has recently been shown \cite{Chate_chiral,Liebchen_epl2022,Liebchen_prl2017}  that chirality in a collection of self-propelled particles (hereafter ``flockers") moving in two dimensions - i.e., a 2D ``flock"- can lead to a very unusual ``steady-state", in which the flockers move coherently in circles. That is, at any instant of time, all of the flockers are moving coherently in the same direction, but that coherent direction rotates at a constant angular velocity.

As a result, each flocker moves in a circle around a center unique to itself, in synchrony with all of the other flockers.

Mathematically, the coarse-grained velocity field $\bv(\br,t)$ of a chiral flock in this state is spatially uniform (that is, independent of two-dimensional position $\br$), and periodic in time:
\beq
\bv=v_0[\cos(bt+\phi)\hat{\bf x}-\sin(bt+\phi)\hat{\bf y}] \,,%,,~~~ \rho=\rho_0,
\label{SSI}
\eeq
where $b$ is an intrinsic frequency characteristic of the system. An achiral system must have $b=0$, since, without chirality, it has no way to decide whether to rotate right or left.
 Hence, we can use $b$ as a measure of the chirality of the system  \cite{glove1}.

Note that any ``snapshot" of such a flock (i.e., a flock with its velocity field given by \rf{SSI}) has perfect infinite ranged orientational order; that is, at every instant of time, all of the particles are moving in precisely the same direction.

One very simple and natural way to simulate a chiral flock is to modify the ``Vicsek" algorithm by altering the direction selecting step of the algorithm to make the  flockers select, not the average direction of its neighbors, but a direction that is, at all times and for all flockers, some fixed angle $\delta$ to the right of that mean direction (obviously, by choosing $\delta$ negative, one can make the flockers turn consistently to the left). Indeed, such simulations have been done \cite{Chate_chiral}, and do, indeed, find a state like \rf{SSI}.

In this paper, we
 investigate whether this long-ranged order is stable in the presence of noise, and furthermore determine the nature, size, and scaling of the fluctuations induced by the noise. In particular, we focus on 2D chiral ``dry Malthusian`` flocks - that is, flocks in which neither momentum or flocker number are conserved. The absence of conservation of momentum is due to the presence of a frictional substrate over which the self-propelled flockers move, while the absence of number conservation is a consequence of birth and death of flockers while in motion.
 
  Some of our results are also given in \cite{ASP}.

\begin{figure}
		\begin{center}
	\includegraphics[scale=.3]{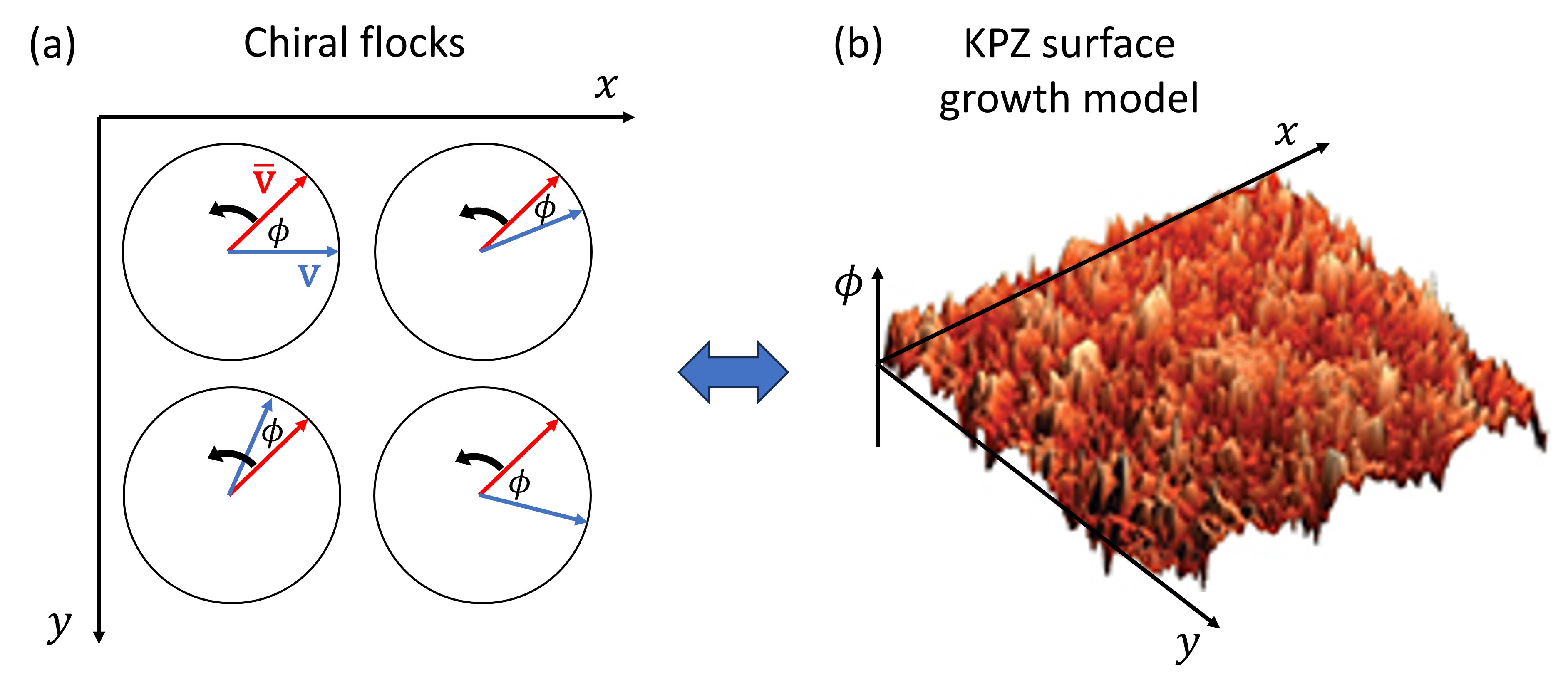}
		\end{center}
		\caption{(a) In a generic 2D chiral Malthusian flock, the mean velocity of the whole flock is denoted by $\bar{\bv}$ (red) , where $\bar{\bv}=v_0[\cos(bt)\hat{\bf x}-\sin(bt)\hat{\bf y}]$, while the spatio-temporally fluctuating velocity field is denoted by $\bv$ (blue) where $\bv=v_0[\cos(bt+\phi)\hat{\bf x}-\sin(bt+\phi)\hat{\bf y}]$. Ignoring variations in $|\bv|$, all fluctuations can be captured by the angular field $\phi(x,y,t)$. (b) Here, we show that the hydrodynamic behavior of $\phi$ can generically be mapped on the Kardar-Parisi-Zhang surface growth model \cite{KPZ} in (2+1) dimensions. The image in (b) is reproduced with permission from EPL {\bf 109}, 46003 (2015) \cite{cartoon}. 
		}
		\label{rect}
	\end{figure}

To study the effects of noise, we allow  the phase  $\phi$ in \rf{SSI} to vary slowly in space and time; i.e.,  we take the velocity to be
\beq
\bv(\br,t)=v_0[\cos(bt+\phi(\br,t))\hat{\bf x}-\sin(bt+\phi(\br,t))\hat{\bf y}]  \,.
\label{v_fluctuateI}
\eeq
We find that $\phi(\br,t)$ is the Goldstone mode of  a 2D chiral flock. Its equation of motion (EOM) proves to be the (1+2)-dimensional-KPZ equation\cite{KPZ}:
\beqn
\pp_t\phi=\nu\nabla^2\phi+{\lambda_{_K}\over 2}(\nabla\phi)^2+f_\phi\,,
\label{EOM:phi}
\eeqn
 where  $f_\phi$ is a Gaussian, zero-mean white noise with the following statistics:
\beq
\la f_\phi (\br,t) f_\phi(\br',t') \ra =2D_\phi \delta^2(\br-\br')\delta(t-t')
\,.
\label{fphicorr}
\eeq
This mapping onto the KPZ equation implies universal scaling relations between length scales,  time scales, and phase fluctuations. More specifically, characteristic  time scales $t(L)$ on length scales $L$ grow like $L^z$, while phase fluctuations scale like $L^\chi$. The current estimate of the values of
$\chi$ and $z$ based on simulations are \cite{kpzexp1, kpzexp2,kpzexp3, kpzexp4,kpzexp5, kpzexp6} $\chi=0.388\pm.002$,  $z=1.622\pm.002$.
 The analogy between a 2D chiral flock and the KPZ surface growth model is illustrated in Fig. \ref{rect}.

We note that the KPZ universality class (UC) encompasses an amazing array of diverse systems--recent additions to this include incompressible polar active fluids \cite{chen_natcomm16}, driven Bose-Einstein condensates \cite{diessel_prl22}, and non-reciprocal systems \cite{pisegna_pnas24,daviet_a24}. 
It is nonetheless  surprising, in our opinion, that all generic 2D chiral Malthusian flocks   necessarily fall into the KPZ UC as well.

The parameters $\nu$ and $\lambda_{_K}$ respectively diverge and vanish in the limit of small chirality $b$; specifically, we find, for the ``racemic" case  \cite{glove1}
\beq
\nu(b)\propto b^{-\eta_\nu} \sep \lambda_{_K}(b)\propto b \,,
\label{mubI}
\eeq
where the universal exponent $\eta_\nu$ is related to the dynamical exponent\cite{Toner_prl12,Chen_prl20} $z_a$ of ``achiral`` Malthusian flocks via the exact relation
\beq
\eta_\nu={2\over z_a}-1\approx{3\over 5} \,.
\label{etanu}
\eeq
Here $z_a$ is defined much as $z$ for the KPZ equation; specifically, characteristic  time scales $t(L)$ on length scales $y$ in the direction orthogonal to the mean flock motion direction (which remains fixed in the achiral case) grow like $|y|^{z_a}$ \cite{Toner_prl12,Chen_prl20,Chen_pre20}.
The numerical value given here is based on numerical simulations of the noisy hydrodynamic equation for achiral Malthusian flocks by \cite{Chate_prl24}. 
The other parameter in the KPZ equation, which is the noise strength $D_\phi$, is independent of chirality $b$ in the limit of small chirality.

As is well known \cite{KPZ}, the fluctuations of a field $\phi$ described by the (2+1)-dimensional-KPZ equation diverge in the limit of large distances. More quantitatively, the mean squared difference between the phases at widely separated  spatio-temporal positions $(\br, t)$ and $(\br^\prime, t^\prime)$ diverge algebraically as $|\br-\br^\prime|$ or $|t-t^\prime|\to\infty$: 
\beq
\langle[\phi(\br,t)-\phi(\br',t^\prime)]^2\rangle=A_\phi |\br-\br'|^{2\chi}\cG_\phi\left(B_\phi|t-t^\prime|\over|\br-\br'|^z\right)\,,
\label{phi_correlI_v0} %\label{KPZphiflucs}
\eeq
where
%$\chi$ is the roughness exponent for (2+1)-KPZ equation}
$A_\phi$ and $B_\phi$ are non-universal positive constants, and the universal scaling function $\cG_\phi\left(X\right)$ has the following scaling behavior:
\begin{eqnarray}
\cG_\phi\left(X\right)=\left\{
\begin{array}{ll}
1\,,&X\ll 1\,,\\
X^{2\chi\over z}\,,&X\gg1\,.
\end{array}
\right.
\end{eqnarray}
This form of the scaling function implies 
\begin{eqnarray}
\langle[\phi(\br,t)-\phi(\br',t)]^2\rangle\propto\left\{
\begin{array}{ll}
|\br-\br'|^{2\chi}\,,&{B_\phi|t-t^\prime|\over|\br-\br'|^z}\ll 1\,,\\
%|\br-\br'|
|t-t^\prime|^{2\chi\over z}\,,&{B_\phi|t-t^\prime|\over|\br-\br'|^z}\gg1\,.
\end{array}
\right.~~~
\end{eqnarray}

The current estimate of the values of
%most accurate numerical result for}
$\chi$ and $z$ based on simulations are \cite{kpzexp1, kpzexp2,kpzexp3, kpzexp4,kpzexp5, kpzexp6} $\chi=0.388\pm.002$,  $z=1.622\pm.002$.

This divergence \rf{phi_correlI_v0} implies that equal-time velocity correlations will be short-ranged. That is, chirality destroys the long-ranged orientational order present \cite{Toner_prl12,Chen_prl20,Chen_pre20} in an achiral  Malthusian flock.

In fact, 2D chiral Malthusian flocks are probably even more disordered  than this would suggest. This is because, in contrast to the usual KPZ equation, the achiral Malthusian flock  maps onto the {\it compact} $(2+1)$-dimensional KPZ equation.

What we mean by ``compact``  is that, unlike, say, a growing interface, in which the KPZ variable $\phi$ is the height of the interface, and every value of the height represents a physically distinct local state, our KPZ variable is a {\it phase}. That is, a state with a given local value $\phi(\br,t)$ is locally physically indistinguishable from state $\phi(\br,t)+2\pi n$, for any integer $n$, as can be seen directly from equation \rf{v_fluctuateI}.

This is the same symmetry that is present in the equilibrium 2D XY model. Here, as there, it allows   topological defects -i.e., vortices \cite{KT}.  These become important on length scales larger than the mean inter-vortex distance  $L_v$. These destroy the order even more thoroughly than the ``spin-wave" (i.e., vortex free) fluctuations that lead to the growth of $\phi$ fluctuations embodied in equation \rf{phi_correlI_v0}; indeed, they make it impossible \cite{KT} to even {\it define} a single-valued phase field $\phi(\br,t)$ over all space $\br$.

The exact behavior of vortices in the KPZ equation remains an open question \cite{Ehud}. 
Indeed, it is an open question for many active systems. For example, it has recently been shown that active smectics \cite{apolar Malthusian}, which, at the spin wave level, look exactly like an equilibrium XY model, exhibit completely different  behavior of topological defects \cite{JPJTFJ}. Our results here for the vortex length $L_v$ and the behavior of a compact KPZ equation on length scales longer than $L_v$ are based upon our speculation that this does {\it not} happen in KPZ like systems. We believe this is reasonable, since the surprising behavior of topological defects in active smectics is intimately connected with the existence of {\it two} spontaneously broken symmetries in those systems (translation {\it and} rotation invariance), while in our system only one continuous symmetry (time translation invariance) is broken. But our results for the behavior in the presence of vortices are somewhat speculative.

\begin{figure}
		\begin{center}
			\includegraphics[scale=.25]{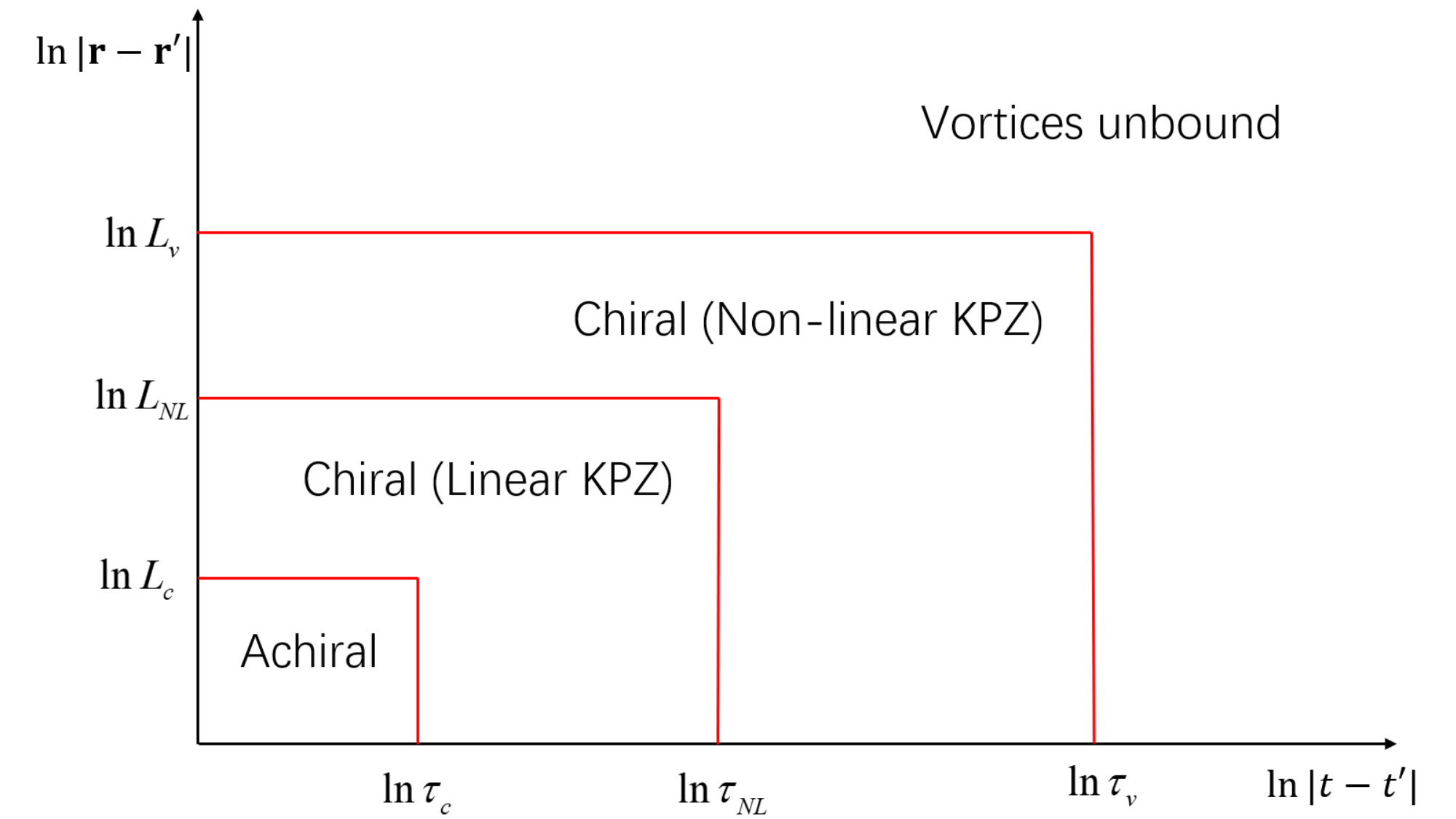}
		\end{center}
		\caption{ Regimes of different behavior in the limit of weak chirality. For small chirality (even only {\it moderately } small chirality), the longest length scales $L_{_{NL}}$ and $L_v$ and time scales $\tau_{_{NL}}$ and $\tau_v$ will be long enough to allow several decades of length and time scale to lie in each region, making our scaling predictions for each region experimentally accessible. See the text for more details.}
		\label{ltsc}
	\end{figure}

Despite our fundamental conclusion that chirality destroys long-ranged order in 2D chiral flocks, for {\it weak} chirality, order persists out to quite large distances. In fact, we find that for weak chirality, there is a hierarchy of length and time scales, illustrated in Fig. \ref{ltsc},  separating regimes of quite different scaling behavior of the velocity correlations $\langle\bv(\br, t)\cdot\bv(\br^\prime, t^\prime)\rangle$. Some of these regions will be effectively ordered.

All of these length and time scales diverge in the limit of small chirality $b$, according to the following scaling laws: 

\begin{widetext}
\begin{table*}
    \begin{tabular}{ | c | c | c | }
    \hline
           Exponent & definition and/or defining equation &  expression for and/or numerical value \\ 
           \hline
  achiral dynamical exponent $z_a$ & $t\propto |y|^{z_a}$ &$\approx5/4$ \\
     \hline
    achiral anisotropy exponent $\zeta_a$ & $|x|\propto |y|^{\zeta_a}$ & $\approx3/4$ \\
 \hline
       $\eta_\nu$ & $\nu(b)\propto  b^{-\eta_\nu}$ \,, \rf{mubI} &$\eta_\nu=2/z_a-1\approx3/5$ \\
     \hline
   $\chi$ & (2+1)-dimenional-KPZ  ``roughness" exponent  & $\approx 0.388\pm.002$\\
        \hline
   $z$ & (2+1)-dimenional-KPZ  ``dynamical" exponent  & $\approx 1.622\pm.002$ \\
           \hline
   $z_{_L}$ & Linearized  KPZ  (Edwards-Wilkinson) ``dynamical`` exponent  & $2$ \\
   \hline
       $\eta_g$ & $g_0(b)\propto  b^{\eta_g}$ \,, \rf{g0} &$\eta_g=6/z_a-1\approx19/5$ \\
     \hline
   $\eta_y$ & $\lv\propto\exp\left[-{\rm constant} \, b^{-\eta_y}\right]$  \,, 
\rf{lv}
 & $\eta_y=\eta_g+\eta_\nu\approx22/5$ \\
  \hline
      $\alpha$ & \rf{v_corre2I} \,,  \rf{v_corre3I}&$\alpha={D_\phi\over 2\pi\nu}$ \\
        \hline
          \end{tabular}
        \caption[Table caption text]{
  Exponents that appear in this paper, expressions for them, and estimates of their numerical values. The achiral exponents are those that give the relations between length and time scales in an {\it achiral} flock; their numerical values  were obtained   by \cite{Chate_prl24} from their numerical  solution of  the noisy hydrodynamic equation for achiral flocks. The numerical values of the KPZ exponents $z$ and $\chi$ are taken from the numerical work of references \cite{kpzexp1, kpzexp2,kpzexp3, kpzexp4,kpzexp5, kpzexp6}}
        \label{tab1}
\end{table*}
\end{widetext}

\beq
L_c(b)\propto b^{-1/z_a} \sep 1/z_a\approx4/5 \,,
\label{lchirI}
\eeq

\beq
L_{_{NL}}(b)\propto \exp\left(C_{_{NL}}b^{-\eta_g}\right)  \sep \eta_g\approx19/5 \,,
\label{lnlI}
\eeq

\beq
L_v(b)\propto \exp\left(C_{_{vL}}b^{-\eta_y}\right)  \sep \eta_y\approx22/5 \,,
\label{lvI}
\eeq

\beq
\tau_c(b)\propto b^{-1} \,,
\label{taucI}
\eeq

\beq
\tau_{_{NL}}(b)\propto \exp\left(2C_{_{NL}}b^{-\eta_g}\right) \,,
\label{taunlI}
\eeq

\beq
\tau_v(b)\propto \exp\left( C_{v\tau}b^{-\eta_y}\right) \,,
\label{tauvI}
\eeq
where $C_{_{NL}}$, $C_{_{vL}}$,  and $C_{v\tau}$ are all non-universal constants.

The universal exponents $\eta_\nu$, $\eta_g$ and $\eta_y$ are related to the equally universal exponent $z_a$ by the relations given in Table \rf{tab1}. Here $z_a$ is the ``dynamical exponent'' for 2D {\it achiral} Malthusian flocks, which is defined by the scaling of characteristic times $t$ with length scale $y$ in the direction perpendicular to the mean flock motion; specifically $t\propto y^{z_a}$.

The  velocity correlations 
$\langle\bv(\br,t)\cdot\bv(\br^\prime, t^\prime)\rangle$ behave as follows in each of these regimes of length and time scales:

\vspace{.2in}

\noindent1) Achiral regime: $|\br-\br^\prime|\ll L_c$  and $|t-t^\prime|\ll\tau_c$.
\vspace{.2in}

In this regime, the velocity correlations are simply those of an achiral Malthusian flock, as given in %\rf{achiral refs}.
%{n \cite{chen_natcomm16,chen_pre24}}
\cite{Toner_prl12,Chen_prl20,Chen_pre20}, exhibiting long-ranged order.

\vspace{.2in}

\noindent2) Linear KPZ/ Edwards-Wilkinson (EW) regime:  $L_c\ll  |\br-\br^\prime|\ll L_{_{NL}}$ and $|t-t^\prime|\ll\tau_{_{NL}}$, or $|\br-\br^\prime|\ll L_{_{NL}}$ and $\tau_c\ll |t-t^\prime|\ll\tau_{_{NL}}$.
\vspace{.2in}

In this regime, the effects of the $\lambda_{_K}$ non-linearity in the KPZ equation are negligible. As a result, that term can be dropped, leaving the dynamics of the $\phi$ field to be described by the linearized version of the KPZ equation, which is also known \cite{EA} as the Edwards-Wilkinson equation:
\beqn
\pp_t\phi=\nu\nabla^2\phi+f_\phi\,.
\label{EA}
\eeqn

Being linear, this EOM is easily solved to obtain, {\it inter-alia}, the mean-squared phase fluctuations: 
\bew
\beqn
\langle[\phi(\br,t)-\phi(\br',t')]^2\rangle%&=&2\int\,{\dd\omega\over 2\pi}\int\,{\dd^2q\over(2\pi)^2}\langle |\phi(\omega,\bq)|^2\rangle\bigg(1-\cos[\omega(t-t')-\bq\cdot(\br-\br^\prime)]\bigg) e^{-(qL_c)^2}\nonumber\\
&=&{D_\phi\over 2\pi\nu}\left[ -{\rm Ei}\left(-{|\br-\br'|^2\over4(\nu |t-t^\prime|+L_c^{2})}\right) + 2\ln \left( \frac{|\br-\br^\prime|}{2L_c}\right) +\gamma\right]
\nonumber\\
&\approx&\left\{
\begin{array}{ll}
\frac{D_\phi}{\pi \nu} \left[ \ln \left( \frac{|\br-\br'|}{L_c}\right) + \gamma/2\right]\,,
&\frac{|t-t'|}{\tau_c}\ll\left( \frac{|\br-\br'|}{L_c}\right)^2\,,\\
{D_\phi\over 2\pi\nu}\left[ \ln \left( \frac{|t-t'|}{\tau_c}\right) \right]\,,
&\frac{|t-t'|}{\tau_c}\gg\left( \frac{|\br-\br'|}{L_c}\right)^2\,,
\end{array}
\right.
\label{phi_corre2I}
\eeqn
\ew
\noindent where $\gamma=.57721566...$ is the Euler-Mascheroni constant\cite{EM}, and ${\rm Ei}(x)\equiv -\int_{-x}^\infty\left({e^{-u}\over u}\right)\dd u$ is the exponential integral function, whose well-known \cite{GR} asymptotic limits imply the limiting behaviors on the last two lines.

It follows from this, \rf{v_fluctuateI}, and the linearity of the EA equation,  that the velocity correlations oscillate rapidly in time and decay algebraically in space and time: 
\bew
\beqn
\langle\bv(\br,t)\cdot\bv(\br',t')\rangle
&=&v_0^2\cos\left[b(t-t')\right]\big\langle\exp\big\{\ii\left[\phi(\br,t)-\phi(\br',t')\right]\big\}\big\rangle\nonumber\\
&=&
v_0^2\cos\left[b(t-t')\right]\exp\bigg\{-{1\over2}\left\langle\left[\phi(\br,t)-\phi(\br^\prime
, t')\right]^2\right\rangle\bigg\}\nonumber\\
&=&v_0^2 \ee^{-\alpha \gamma/2} 
%{ \sqrt{\ee^{\alpha\gamma}}}
\cos\left[b(t-t')\right]\left(|\br-\br'|\over  2 L_c\right)^{-\alpha} \exp\bigg[{\alpha\over2}{\rm Ei}\left(-{|\br-\br'|^2\over4(\nu |t-t^\prime|+L_c^{2})}\right)\bigg]
\nonumber\\
 &\approx&  v_0^2 
\cos\left[b(t-t')\right]\times\left\{
\begin{array}{ll}
%\sqrt{\ee^{\alpha\gamma}}
\ee^{-\alpha \gamma/2} \left(|\br-\br'|\over 2L_c\right)^{-\alpha}\,,
& \frac{|t-t'|}{\tau_c}\ll\left( \frac{|\br-\br'|}{L_c}\right)^2\,,\\
\left(|t-t'|\over \tau_c\right)^{-{\alpha\over 2}}\,,
&\frac{|t-t'|}{\tau_c}\gg\left( \frac{|\br-\br'|}{L_c}\right)^2\,,
\end{array}
\right.
\label{v_corre2I}
\eeqn
\ew
where the exponent $\alpha$ is non-universal and given by
\beqn
\alpha={D_\phi\over 2\pi\nu}=C_\alpha b^{\eta_\nu} \sep \eta_\nu\approx{3\over 5}\,,
\label{aI}
\eeqn
where $C_\alpha$ is a non-universal constant.

 To investigate the order of the system, we take a snapshot of the system at any time and focus on the velocity correlations on that snapshot. These  velocity correlations are equal-time correlations, which can be easily gotten by setting $t=t'$ in (\ref{v_corre2I}). Doing this we find
  \beqn
\langle\bv(\br,t)\cdot\bv(\br',t)\rangle
=v_0^2\left(|\br-\br'|\over L_c\right)^{-\alpha}\times O(1)\,.\label{v_corre3I}
\eeqn

Note that for small chirality $b$, the exponent $\alpha$ will be small. Hence, the algebraic decay of velocity correlations \rf{v_corre3I} will be extremely slow; this is what we meant by our earlier cryptic comment that ``some regimes are effectively ordered". Strictly speaking, the order in this regime is ``quasi-long-ranged".

\vspace{.2in}

The behavior of the purely temporal Fourier transform of the velocity-velocity correlation function, that is,
\beq
I(\bR, \omega)\equiv\int_{-\infty}^\infty  \langle\bv(\br,t)\cdot\bv(\br+\bR,t+T)\rangle \, e^{\ii\omega T} \,\dd T
\label{idef}
\eeq
is also interesting  in this regime (and experimentally measurable). 
Unlike the equal-time case, the  non-equal-time velocity correlations carry a frequency $\omega=b$ (see the cosine factor in (\ref{v_corre2I}) and (\ref{gaussianI1})). This is reminiscent of the density correlations in smectics which oscillate in real space with period $a$ (the layer spacing) along the layer normal $\hat{n}$. The Fourier component of this density correlation function at the vicinity of the ``Bragg peaks" $\bq=(2m\pi/ a)\hat{n}$ ($m=1,2,...$) is proportional to the X-ray scattering intensity. Here we make a similar prediction on the Fourier component of the  velocity correlation function near the ``Bragg peaks`` at 
%k?
{ $\omega=\pm b$ in frequency. We find that, in the weak chirality limit,  this Fourier component 
%\br',t)\rangle
has a scaling form over a substantial range of frequencies:  
\beq
I(\bR, \omega)={v_0^2\over4\pi}\left({e^{\gamma/2} \over2L_c}\right)^{-\alpha}\left({R^{2-\alpha}\over4\nu}\right)F_h\left({\delta\omega R^2\over 4\nu}\right) \,,
\label{Iscale}
\eeq
where $\delta\omega\equiv\omega-b$ near the peak at $\omega=b$, and 
$\delta\omega\equiv\omega+b$ near the peak at $\omega=-b$.
Here the scaling function $F_h(S)$ is given by
\beq
F_h(S)=\int_{-\infty}^\infty \exp\left[\ii Su+\left({\alpha\over2}\right)\re\left(-{1\over|u|}\right)\right] \,\dd u \,.
\label{gpeak}
\eeq
Its limiting behaviors for large and small arguments are:
\bew
\beqn
F_h(S)\approx\left\{
\begin{array}{ll}
2e^{\alpha\gamma\over2}\Gamma\left(1-{\alpha\over 2}\right)\sin\left({\alpha\pi\over4}\right)|S|^{-1+\alpha/2} \,, &|S|\ll1\\
\sqrt{\pi}\alpha|S|^{-5/4}e^{-\sqrt{2|S|}}\sin\left(\sqrt{2|S|}+\pi\over8\right)\,, &|S|\gg1
\end{array}
\right.
\,.
%\nn\\
\label{golim}
\eeqn
\ew
}

% Inserting \rf{golim} into \rf{Iscale} gives
%\bew
%\beqn
%I(\bR,\omega)\approx\left\{\begin{array}{ll}\left(v_0^2\over 2\pi\right)\left(L_c^2\over\nu  \right)^{\alpha\over 2}
%\Gamma\left(1-{\alpha\over2}\right)\sin\left({\alpha\pi\over4}\right)
%|\dw|^{\alpha/2-1}\,,&\tau_{_{NL}}^{-1}\ll|\dw|\ll{\nu/R^2}\\{v_0^2\over 4\sqrt{\pi}}\left(2L_c\over R\ee^{\gamma/2}\right)^\alpha
%\left(4\nu\over R^2 \right)^{1\over 4}
%\sin\left(\sqrt{R^2|\delta\omega|\over 2\nu}+{\pi\over8}\right)
%\exp\left(-\sqrt{R^2|\delta\omega|\over 2\nu}\right)|\delta\omega|^{-{5\over 4}}\,,&|\dw|\gg{\nu/R^2}\end{array}\right.\,,\nn\\\label{Intensity}\eeqn\ew
%}

This scaling behavior is cut off for $|b-\omega|\ll\tau_{NL}^{-1}$; in this range of frequencies, the peak is rounded and featureless.

%\beqn
%I(\omega)\propto\left\{
%\begin{array}{ll}
%v_0^2\tau_c^{\alpha\over 2}  |\omega-b|^{-(1-\alpha/2)},&\tau_{NL}^{-1}\ll|b-\omega|\ll \tau_c^{-1}\\
%v_0^2\tau_{NL}\left(\tau_{NL}\over \tau_c\right)^{-{\alpha\over 2}},&|\omega-b|\ll\tau_{NL}^{-1}
%\end{array}
%\right.
%\,.\label{Fourier1}
%\eeqn

The power-law decaying ``Bragg peak" for $\tau_{_{NL}}^{-1}\ll|b-\omega|\ll  {\nu\over R^2}$ implies quasi-long-ranged order at time scales  less than $\tau_{_{NL}}$, while the broad converging ``Bragg peak" for $|b-\omega|\ll\tau_{_{NL}}^{-1}$ implies short-ranged order at time scales beyond $\tau_{_{NL}}$.

\begin{figure}
		\begin{center}
	\includegraphics[scale=.25]{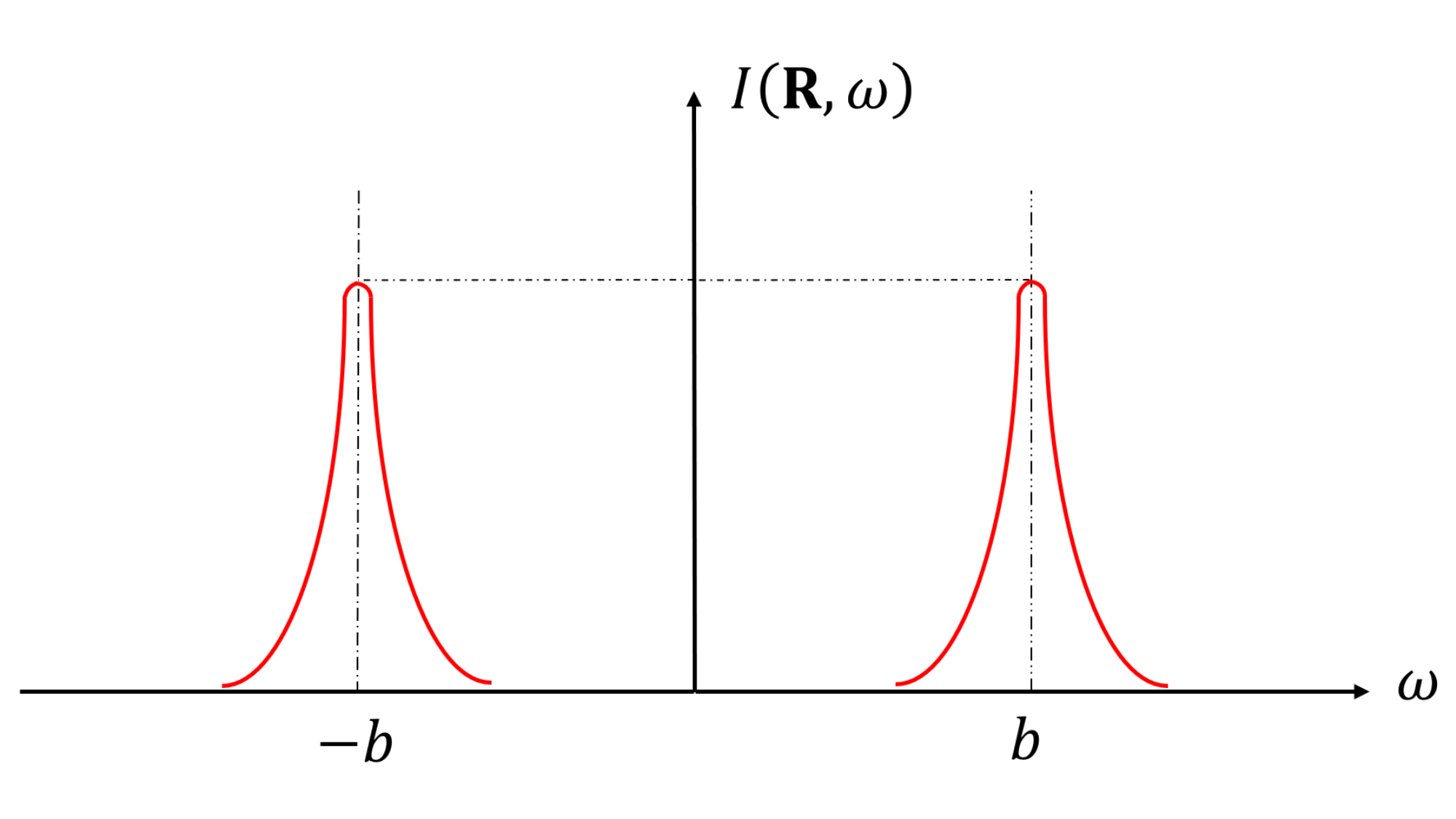}
		\end{center}
		\caption{Plot of the temporally Fourier transformed correlation function $I(\bR, \omega)$ versus $\omega$ for a weakly chiral system. Each peak has a power law divergence $|\dw|^{\alpha/2-1}$, where $\dw$ is the difference between the frequency and the peak centers, which lie at $\pm b$. This divergence is cut off sufficiently close to the peak (specifically, once $|\dw|\lesssim\tau_{_{NL}}^{-1}$, leading to a finite peak height at $\dw=0$. The oscillating tails for $|\dw|\gg{\nu/R^2}$ are invisible on the scale of this figure.
}
		\label{iw}
	\end{figure}

The behavior of the purely temporal Fourier transform of the velocity-velocity correlation function is illustrated in Fig. \ref{iw}. 

%\bew
%\beqn
%I(\bR,\omega)\approx\left\{\begin{array}{ll}\left(v_0^2\over 2\pi\right)\left(L_c^2\over\nu  \right)^{\alpha\over 2}
%\Gamma\left(1-{\alpha\over2}\right)\sin\left({\alpha\pi\over4}\right)
%|\dw|^{\alpha/2-1}\,,&\tau_{_{NL}}\ll|\dw|\ll{\nu/R^2}\\{v_0^2\over 4\sqrt{\pi}}\left(2L_c\over R\ee^{\gamma/2}\right)^\alpha
%\left(4\nu\over R^2 \right)^{1\over 4}
%\sin\left(\sqrt{R^2|\delta\omega|\over 2\nu}+{\pi\over8}\right)
%\exp\left(-\sqrt{R^2|\delta\omega|\over 2\nu}\right)|\delta\omega|^{-{5\over 4}}\,,&|\dw|\gg{\nu/R^2}\end{array}\right.\,,\nn\\\label{}\eeqn\ew

\vspace{.2in}

\noindent3) Nonlinear KPZ regime: $L_{_{NL}}\ll |\br-\br'| \ll L_v$ and $|t-t'| \ll\tau_{v}$, or $|\br-\br'| \ll L_v$ and $\tau_{_{NL}}\ll |t-t'| \ll\tau_{v}$.
\vspace{.2in}

In this regime, the effects of the $\lambda_{_K}$ non-linearity in the KPZ equation dominate. The mean squared difference between the phases at widely separated spatial and temporal positions $(t, \br)$ and $(\br^\prime, t^\prime)$ are given by (\ref{phi_correlI_v0}). 
%Simply quoting long-established results} \cite{KPZ, kpzexp1, kpzexp2,kpzexp3, kpzexp4,kpzexp5, kpzexp6}  for the $(2+1)$ dimensional KPZ equation, we have}
%\beqn
%\langle[\phi(\br,t)-\phi(\br',t)]^2\rangle{=A_\phi} |\br-\br'|^{2\chi}\,,\label{phi_correlI}
%\eeqn
%where $A_\phi$ is a non-universal positive constant. The current estimate of the value of  $\chi$ based on simulations is
%$\chi=0.388\pm.002$} \cite{kpzexp1, kpzexp2,kpzexp3, kpzexp4,kpzexp5, kpzexp6}.

%[LC: I removed an equation here since we had it before]}
%The equal-time correlation of the velocity is
%\beqn
%C(\br, \br^\prime)\equiv\langle\bv(\br,t)\cdot\bv(\bm{\br^\prime},t)\rangle &=& v_0^2\Big\{\langle\cos[bt+\phi(\br,t)\cos[bt+\phi(\br',t)]\rangle
%+\langle\cos[bt+\phi(\br,t)\cos[bt+\phi(\br',t)]\rangle\Big\}
%\nn\\
%&=&2v_0^2\big\langle\exp\big\{\ii\left[\phi(\br,t)-\phi(\br',t)\right]\big\}\big\rangle\,.\label{v_correlI}
%\eeqn

 If the fluctuations of the field $\phi$ were Gaussian, then we could say that the second equality in (\ref{v_corre2I}) still applies. Inserting (\ref{phi_correlI_v0}) into this equality we get
 \bew
 \beqn
 \langle\bv(\br,t)\cdot\bv(\br',t')\rangle=v_0^2\cos\left[b(t-t')\right]\exp\left[-\left(A_\phi\over 2\right) |\br-\br'|^{2\chi}\cG_\phi\left(|t-t^\prime|\over|\br-\br'|^z\right)\right]\,.
 \label{gaussianI1}
 \eeqn
 \ew
 The parameter $A_\phi$ can be calculated by arguing that the phase correlations (\ref{phi_correlI_v0}) and (\ref{phi_corre2I}) must connect at the crossovers, for example, at $|\br-\br'|=L_{_{NL}}$, $|t-t'|=0$. This
 argument gives 
 \beqn
 \label{eq:Aphi}
 A_\phi=2\alpha L_{_{NL}}^{-2\chi}\ln\left(L_{_{NL}}\over L_c\right)
 \propto b^{-{16\over 5}}\exp\left[-\left(b^{-3.8}\right)\times{\rm O(1)}\right]\,.
 \nn\\
 \eeqn

 Again focusing on the equal-time correlation, we'd now have:
 \beqn
 \langle\bv(\br,t)\cdot\bv(\br',t)\rangle=v_0^2\exp\left[-\left(A_\phi\over 2\right) |\br-\br'|^{2\chi}\right]\,,~~~~~
 \label{gaussianI2}
 \eeqn
%\beqn
%\big\langle\exp\big\{\ii\left[\phi(\br,t)-\phi(\br^\prime,t)\right]\big\}\big\rangle%\nonumber\\
%=\exp\bigg\{-{1\over2}\bigg\langle\left[\bigg(\phi(\br,t)-\phi(\br^\prime%\br?
%, t)\bigg)^2\right]\bigg\rangle\bigg\}=\exp\left(-\left({A_\phi\over2}\right)|\br-%\br'
%\br^\prime|^{2\chi}\right)\,,
%\label{gaussianI}
%\eeqn
which would imply (stretched) exponentially decaying velocity correlations.
However, since the $\phi$ fluctuations are non-Gaussian, due to the relevant nonlinearity $\lambda_{_K}$ in the KPZ equation, we can say very little, beyond noting that we expect velocity correlations to decay rapidly with increasing  separation $|\br-\br^\prime|$ once that separation is large enough. 
%that %$\langle[\phi(\br,t)-\phi(\br',t)]^2\rangle\gtrsim O(1)$} $A_\phi |\br-\br'|^{2\chi}\gtrsim O(1)$. Using \rf{phi_correlI}, w}
Naively,  this  would seem to imply that correlations will decay rapidly for $|\br-\br^\prime|>\xi_v$
where
\beq
\label{eq:xi_v}
\xi_v\equiv A_\phi^{-{1\over2\chi}}\approx A_\phi^{-1.29}
\ .
\eeq

While this may be true for very chiral systems, we find that, for small chirality, 
$\xi_v\ll L_{_{NL}}$. 
To see this, simply  plug the first equality of \rf{eq:Aphi}
into \rf{eq:xi_v}; this gives 
\beq
\xi_v\propto b^{8\over5\chi}L_{_{NL}}\ll L_{_{NL}}\,.
\eeq
This implies that we won't be able to see this characteristic length in  the nonlinear KPZ regime. The reason for this is simply that the velocity correlations will already have decayed to essentially zero before we even reach the non-linear regime.

 For very chiral (i.e., large $b$) systems, however,  $\xi_v$ will be the ``velocity correlation length'', although it must be kept in mind that the decay of velocity correlations is probably not exponential, nor even a simple stretched exponential like equation \rf{gaussianI2}.

\vspace{.2in}

\noindent4) Vortices unbound regime: $L_{v}\ll |\br-\br^\prime|$ or $\tau_{v}\ll |t-t^\prime|$.

\vspace{.2in}

As we discussed earlier,
the fact that this regime even exists is a consequence of the fact that, strictly speaking,  the 2D chiral Malthusian flock maps onto the {\it compact} $(2+1)$-dimensional KPZ equation, which allows vortices in the system. We believe, based on the previously discussed analogy between 2D chiral Malthusian flocks and the 2D equilibrium XY model, that the typical spacing between vortices is $L_v$, and that for length scales longer than $L_v$ 
 it is therefore  impossible \cite{KT} to even {\it define} a single-valued phase field $\phi(\br,t)$ over all space $\br$.

Our discussion so far has focussed on fluctuations of the velocity field.
%., or, equivalently, the phase $\phi$ in equation  \rf{v_fluctuateI}.}} 
The local density $\rho(\br, t)$ of flockers ``inherits" some interesting behavior from the velocity fluctuations as well. In the achiral regime, fluctuations of $\rho(\br, t)$ are ``enslaved" to $\bv(\br,t)$ \cite{Toner_prl12,Chen_prl20,Chen_pre20}:
\beqn
\delta\rho\propto\nabla\cdot\bv
\eeqn
where $\delta\rho=\rho(\br,t)-\rho_0$, where $\rho_0$ is the value of the steady state density. In the chiral regimes, we find that fluctuations of $\rho$ are `` enslaved" to those of $\phi$:
\beqn
\delta\rho=e_3\nabla^2\phi+e_4|\nabla\phi|^2\,,\label{Eq:rho_2I}
\eeqn
where
$e_3$ and $e_4$ are non-universal (i.e., system specific) constants.

As a result, the  density correlations exhibit the following behaviors in the various regimes:

\vspace{.2in}

\noindent1) Achiral regime: $|\br-\br'|\ll L_c$ and $|t-t'|\ll \tau_c$. The density correlations are just those given in \cite{Chen_prl20,Chen_pre20}.
\vspace{.2in}

%\beqn
%\langle\dr(\br,t)\dr(\br^\prime,t')\rangle\propto{1\over |r-r^\prime|^4} \,.
%\label{}
%\eeqn
%}

\vspace{.2in}

\noindent2) Linear KPZ/ Edwards-Wilkinson (EW) regime: $L_c\ll  |\br-\br^\prime|\ll L_{_{NL}}$ and $|t-t^\prime|\ll\tau_{_{NL}}$, or $|\br-\br^\prime|\ll L_{_{NL}}$ and $\tau_c\ll |t-t^\prime|\ll\tau_{_{NL}}$ .  Density correlations grow algebraically with separation:
%$L_c\ll L\ll \lnl$ or $\tau_c\ll t\ll\tau_{NL}$}.
\vspace{.2in}
\beq
\langle\dr(\br,t)\dr(\br^\prime,t^\prime)\rangle\propto\left\{
\begin{array}{ll}
{1\over |\br-\br^\prime|^4}\sep&\left({|\br-\br^\prime|\over L_c}\right)^2\gg {|t-t^\prime|
\over \tau_c}\,,\\
{1\over |t-t^\prime|^2}\sep &\left({|\br-\br^\prime|\over L_c}\right)^2\ll {|t-t^\prime|\over\tau_c}
\end{array}
\right.
\label{rho_corre1I}
\eeq

\vspace{.2in}

%\vspace{.2in}

\noindent3) Nonlinear KPZ regime: $L_{_{NL}}\ll  |\br-\br^\prime|\ll L_v$ and $|t-t^\prime|\ll\tau_v$, or $|\br-\br^\prime|\ll L_v$ and $\tau_{_{NL}}\ll |t-t^\prime|\ll\tau_v$. Once again, density correlations grow algebraically, but now with a non-trivial exponent related to the ``roughness" exponent $\chi$ of the $(2+1)$-dimensional KPZ equation: 
%$L_{NL}\ll L\ll L_v$ or $\tau_{NL}\ll t\ll\tau_{v}$}.
\vspace{.2in}
\bew
\beqn
\langle\dr(\br,t)\dr(\br^\prime,t)\rangle\propto\left\{
\begin{array}{ll}
|\br-\br'|^{4\chi-4}\approx |\br-\br'|^{-2.448} \,, &\left({|\br-\br^\prime|\over L_{_{NL}}}\right)^z\gg {|t-t^\prime|\over\tau_{_{NL}}}\\
%|\br-\br^\prime|^z\gg |t-t'|\,,\\
|t-
t^\prime|^{4\chi-4\over z}\approx |t-
t^\prime|^{-1.509} \,, &\left({|\br-\br^\prime|\over L_{_{NL}}}\right)^z\ll {|t-t^\prime|\over\tau_{_{NL}}}
%|t-t'|\,.
\end{array}
\right.
\label{rho_corre2I}
\eeqn
\ew

Unfortunately, in neither of these regimes is the decay of density correlations slow enough to
lead to ``Giant number fluctuations". That is, in Malthusian flocks, the mean squared number fluctuations $\langle
(N-\langle N \rangle)^2\ra$, where $N$ is the number of particles in some large volume,  obey
the usual ``law of large numbers" scaling $\langle (N-\langle N \rangle)^2\ra \propto\langle
N\rangle$, in contrast to the anomalous ``Giant number fluctuations" scaling law
$\langle (N-\langle N \rangle)^2>\propto\langle N\rangle^\beta$ with $\beta>1$ which occurs in the absence of momentum conservation for both achiral polar ordered flocks 
%with out} {n [[shoud it be ``without"?]]}  momentum conservation \cite{GNF}} 
and active nematics\cite{GNF, AN}.

\vspace{.2in}
\noindent4) Vortices unbound regime: $L_{v}\ll |\br-\br^\prime|$ or $\tau_{v}\ll  |t-t^\prime|$.
In this regime, the flock is completely disordered, and so the density correlations are short-ranged in space and time, with correlation length $L_v$ and correlation time $\tau_v$.
\vspace{.2in}

The remainder of this paper is organized as follows: in section \rf{model}, we formulate a simplified hydrodynamic model for 2D chiral Malthusian flocks. In section \rf{bs}, we identify the broken
symmetry state of the model. In section \rf{kpz}, we show that the dynamics of this model maps onto the KPZ equation for the phase $\phi$ introduced above. In section \rf{rho}, we derive the relation \rf{Eq:rho_2I} above between the density $\rho$ and the phase $\phi$.
In section \rf{kpzc}, we derive the correlations of  the phase $\phi$, the velocity $\bv$, and the density $\rho$ in the Non-linear KPZ regime.
In section \rf{sc}, we consider the small chirality limit, and derive the scaling laws \rf{lchirI}, \rf{lnlI}, \rf{taucI}, and \rf{taunlI}  above relating the crossover lengths $L_c$, and $L_{_{NL}}$,  and times
$\tau_c$ and $\tau_{_{NL}}$, to the chirality $b$.
We also derive the scaling laws \rf{phi_corre2I} and \rf{v_corre2I} that hold in the linear regime in section \rf{pc}.  In section \rf{F},  we calculate the  purely temporal Fourier transform of the velocity correlation function. In section \rf{DC}, we calculate the density correlations in the linear regime.
In section \rf{v}, we consider the effects of vortices in our compact KPZ equation, and derive the scaling \rf{lvI} and \rf{tauvI}  of the vortex length $L_v$ and time $\tau_v$ with $b$.
In section \rf{s}, we summarize our results.
In Appendix \rf{genmod}, we show that generalizing our simplified model to include all relevant terms allowed by symmetry changes none of our results. We discuss ``non-racemic" mixtures in Appendix \rf{nonrac}. In Appendix \rf{cphi}, we give the mathematical details of our calculation of the phase correlation function in the linear regime. Finally, appendix \rf{Steep_Descent_I} presents
the details of our calculation of the crossover
function  $F_h(S)$ for the purely temporally Fourier transformed velocity-velocity correlation function.

%\vspace{.2in}

\section{Model}{\label{model}} 

Our approach is to begin with the generic hydrodynamic EOM for achiral flocks in which neither momentum nor particle number are conserved, and then supplement this with the additional terms allowed when chiral symmetry is broken. The achiral hydrodynamic EOM is:
%with non-conservation of momentum and numbers}
\cite{Toner_prl12,Chen_prl20,Chen_pre20}:
\bew
\beqn
\pp_t \bv +\lambda_1 (\bv \cdot {\bf \nabla})\bv + \lambda_2(\nabla\cdot\bv)\bv%+\lambda_3\nabla(v^2)
&=& \mu_1 \nabla^2 \bv+\mu_2\nabla(\nabla\cdot\bv)+\mu_3(\bv\cdot\nabla)^2\bv+L(|\bv|)\bv +\bff \,,
\label{eq:original_1}
\eeqn
\ew
where $\bv(\br,t)$ is the local velocity field, $L(|\bv|)$ is a Lagrange multiplier that enforces a ``fixed length`` or ``constant speed`` constraint on the velocity that $|\bv|=v_0={\rm constant}\ne0$ everywhere\cite{fixed}.
%The parameters $\lambda_{1,2,3}$,
%The pressure $p$ }
%, and the diffusion coefficients $\mu_{1,2,3}$
%is a function of $\rho$ and $|\bv|$ in general. The function $U(|\bv|)$ determines the magnitude of $\bv$ at the steady state. Here we assume both  { that} $U(|\bv|)=0$ and { that} $U(|\bv|)\over\dd|\bv|$ { is} significantly negative at some nonzero speed
%$|\bv|=v_0$, so that the steady state { of the system in the absence of noise is firmly lodged} deeply sits} at $|\bv|=v_0$. }
The random force $\bff$ is a zero-mean ``white noise" with spatio-temporal statistics:
\beq
\la f_m (\br,t) f_n(\br',t') \ra =2D \delta_{mn} \delta^2(\br-\br')\delta(t-t')
\ .
\eeq
%The $\rho$-dependent function $g(\rho)$ is the birth rate of the particles and $g(\rho_0)=0$ where $\rho_0$ is the value of $\rho$  at the steady state.}

Next we incorporate the chirality. Our approach to doing so is the same as that used in constructing the EOMs (\ref{eq:original_1}) for the achiral case, which can be summarized by saying that we include in the velocity EOM (\ref{eq:original_1}) every vector that can be made out of the velocity field and its spatial derivatives. In the presence of chirality, however, we can also include in the velocity EOM any vector obtained by rotating one of the vectors that was allowed in the achiral case to the right by 90 degrees.  Such rotations can be done in two dimensions
%by introducing}
using the anti-symmetric rank-2 tensor:
%into the achiral model equations (\ref{eq:original_1},\ref{eq:original_2})in two dimensions (2D):}
%Assuming the particles have right-hand chirality,
%This tensor is described by the following matrix }
\beq
\bm{\epsilon} =
\left(\begin{array}{rr}
	0 & 1
	\\
	-1 & 0
	\end{array}
\right)\ .
\eeq
If we act with this tensor on an arbitrary vector $\mathbf{w}$, the product $\bm{\epsilon}\cdot\bm{w}$ is just the vector obtained by rotating $\bm{w}$ clockwise by 90 degrees, which is clearly a chiral operation.

In principle the hydrodynamic EOM for the chiral case should include all the terms in the achiral EOM plus those breaking the chiral symmetry. There are infinitely many such terms. For the sake of simplicity, we will begin by keeping only enough terms to allow us to obtain the most general ultimate final hydrodynamic model, which proves to be the (1+2)-dimensional-KPZ equation.
Later we will show that those (infinitely many!)  neglected terms do not alter the conclusion that the ultimate model is the KPZ equation; rather, they only change the values of  two of the parameters of that equation.
%will not change the final results.}

Our truncated EOM for 2D chiral flocks with non-conservation of momentum and numbers is:
\bew
\beqn
\pp_t \bv +\lambda_1 (\bv \cdot {\bf \nabla})\bv +\lambda'_1(\bv \cdot \bm{\epsilon} \cdot {\bf \nabla})\bv+ \lambda_2(\nabla\cdot\bv)\bv+\lambda_3\nabla(v^2)
%+\lambda_3 \bm{\epsilon} \cdot \left[(\bv \cdot {\bf \nabla})\bv\right]
&=&\mu_1 \nabla^2 \bv+\mu_2\nabla(\nabla\cdot\bv)+\mu_3(\bv\cdot\nabla)^2\bv+\mu' \bm{\epsilon}\cdot \nabla^2 \bv\nonumber\\
&&+\left[L(|v|)+b\bm{\epsilon}\cdot\right]\bv +\bff
\ ,\label{EOM:chiral_1}
\eeqn
\ew
where $b>0$ if the particles in our system tend to turn right, while $b<0$
if the particles in our system tend to turn left.  Note that the random force here can also be chiral in principle, but the relevant part of its two-point correlations have the same form \cite{Toner_prl12}  as those of its  achiral counterpart, i.e.,
\beq
\la f_m (\br,t) f_n(\br',t') \ra =2D \delta_{mn}\delta^2(\br-\br')\delta(t-t')
\ .
\eeq
As noted in the introduction, one very simple and natural way to simulate a chiral flock is to modify the ``Vicsek'' algorithm by altering the direction selecting step of the algorithm to make the particles select, not the average direction of their neighbors, but a direction $\delta$ clockwise of that direction, with $\delta$ an adjustable parameter of the model. For small $\delta$, we expect $b\propto\delta$. Hence, all of the $b$ dependence that we calculate below can be directly mapped onto identical dependence on $\delta$.

\section{The broken symmetry state}{\label{bs}}
In the absence of noise, the EOM has  an infinite family of spatially uniform, temporally oscillating solutions: %The ``steady-state`` solution of this model is readily solved by mean-field theory, neglecting all the terms involving spatial derivatives and the noise. It is given by}
\beq
\bv=v_0[\cos(bt+\phi)\hat{x}-\sin(bt+\phi)\hat{y}] \,.%,,~~~ \rho=\rho_0,
\label{SS}
\eeq
%where we have assumed $\bv$ is along $\hat{x}$ at $t=0$.}
Note that these solutions differ trivially by a constant phase $\phi$.

Any ``snapshot" of such a flock has perfect infinite ranged orientational order; that is, at every instant of time, all of the particles are moving in precisely the same direction.

We now wish to investigate whether this long-ranged order is stable in the presence of noise, and to furthermore determine the nature, size, and scaling of the fluctuations induced by the noise.

Since equation (\ref{SS}) is a solution of the noiseless equations of motion  for any {\it constant} value of the phase $\phi$, we expect that a velocity configuration in which $\phi$ varies slowly in space; i.e.,
\beqn
\bv(\br,t)=v_0[\cos(bt+\phi(\br,t))\hat{x}-\sin(bt+\phi(\br,t))\hat{y}] \,,
\nn\\
 \label{v_fluctuate}
\eeqn

%\vspace{.2in}

\noindent with $\phi(\br,t)$ varying slowly in {\it space} $\br$, will exhibit a slow variation of $\phi(\br,t)$ with {\it time} $t$ as well. This is, of course, just the characteristic of a Goldstone mode. We will now show that this reasoning is correct, and that $\phi(\br,t)$ is, indeed, the Goldstone mode for our problem.

\section{The Hydrodynamics of the phase field}{\label{kpz}}

Inserting (\ref{v_fluctuate}) into EOM (\ref{EOM:chiral_1}),
%and adding a Lagrangian multiplier $L(|\bv|)\bv$ to (\ref{EOM:chiral_1}) to enforce the constant-magnitude restriction on $\bv$,}
we get, in tensor form,
\beq
\pp_tv_i=(b+\pp_t\phi)\epsilon_{ik}v_k \sep \pp_jv_i=\epsilon_{ik}v_k\pp_j\phi
\label{vderivs}
\eeq

Using these in \rf{EOM:chiral_1}, we obtain the EOM for the phase $\phi$:
\bew
\beqn
\epsilon_{ij}v_j\pp_t \phi
%+\lambda_3 \bm{\epsilon} \cdot \left[(\bv \cdot {\bf \nabla})\bv\right]
&=&-\lambda_1 \epsilon_{ik}v_kv_j\pp_j\phi-\lambda'_1\left[v_i(v_k\pp_k\phi)-v_0^2(\pp_i\phi)\right] -\lambda_2\epsilon_{jk}v_iv_k(\pp_j\phi) +\mu_1\left[\epsilon_{ik}v_k\pp_j\pp_j\phi-v_i(\pp_j\phi)(\pp_j\phi)\right]\nonumber\\
&&+\mu_2\left[\epsilon_{jk}v_k\pp_i\pp_j\phi-v_j(\pp_i\phi)(\pp_j\phi)\right]
+\mu_3\left[\epsilon_{i\ell}v_\ell v_jv_k\pp_j\pp_k\phi-v_iv_kv_j(\pp_k\phi)(\pp_j\phi)\right]
\nonumber\\&&-\mu'\left[v_i\pp_k\pp_k\phi+\epsilon_{im}v_m(\pp_k\phi)(\pp_k\phi)\right]+L(|v|)v_i +f_i
\ .\label{EOM:chiral_3}
%\pp_t (\delta\rho) &=&-\rho_0\pp_iv_i
%-v_i(\pp_i\delta\rho) -\pp_i\left(c_1\epsilon_{ij}v_j+c_2v_i\pp_jv_j+c_3\epsilon_{ij}v_j\pp_kv_k\right)
%-\vnab \cdot \left[c_2\bv(\nabla\cdot\bv)+c_3\bm{\epsilon}\cdot\bv(\nabla\cdot\bv)\right] -g'(\rho_0)\delta\rho\,,\label{EOM:chiral_4}
\eeqn
\ew

%where we have absorbed $U(|\bv|)$ into $L(|\bv|)$. In deriving (\ref{EOM:chiral_3},\ref{EOM:chiral_4}) we have expanded $p(\rho,|\bv|=v_0)$ and $g(\rho)$ around $\rho=\rho_0$ to linear order in $\delta\rho$. We have also neglected the terms involving spatial derivatives of $\delta\rho$ in (\ref{EOM:chiral_4}) since they are always irrelevant compared to the term $g'(\rho_0)\delta\rho$.}

Recall that we expect $\phi$ to vary slowly in time. This means the Fourier transform of $\phi(\br,t)$ in time will be dominated by frequencies much less than the characteristic oscillation frequency $b$ in \rf{v_fluctuate}. Therefore, the left hand side of \rf{EOM:chiral_3} is dominated by frequencies near $b$, since the velocity factor  $v_j$ on that side oscillates at that frequency, while the $\pp_t\phi$ factor oscillates much more slowly.

To explicitly display, and calculate, the slow evolution of $\phi$,
%To get the hydrodynamic EOM for $\phi$,}
we need to extract the Fourier components of the right hand side of (\ref{EOM:chiral_3}) that also are near $b$.  To do this, we begin by multiplying both sides of (\ref{EOM:chiral_3}) by $\epsilon_{in}v_n$  with the usual implied summation convention on the repeated index $n$. This amounts in vector notation to taking the dot product of both sides of
(\ref{EOM:chiral_3}) with $\beps\cdot\bv$, which is orthogonal to $\bv$ itself. This eliminates all terms along $\bv$ in (\ref{EOM:chiral_3}) (that is, in tensor notation, all terms proportional to $v_i$, which includes the Lagrange multiplier term $L(|v|)v_i $, along with several others). One can also see the vanishing of these terms directly in tensor notation by using the asymmetry of $\epsilon_{ij}$ under interchange of $i$ and $j$. We are left with
\bew
\beqn
v_0^2\pp_t \phi
%+\lambda_3 \bm{\epsilon} \cdot \left[(\bv \cdot {\bf \nabla})\bv\right]
&=&-\lambda_1 v_0^2v_j\pp_j\phi+\lambda'_1v_0^2\epsilon_{in}v_n\pp_i\phi +\mu_1v_0^2\nabla^2\phi
%\nonumber\\
%&&
+\mu_2\left[v_0^2\nabla^2\phi-v_iv_ j\pp_i\pp_j\phi-\epsilon_{in}v_nv_j(\pp_i\phi)(\pp_j\phi)\right]
+\mu_3v_0^2 v_jv_k\pp_j\pp_k\phi
\nonumber\\&&-\mu'v_0^2|\nabla\phi|^2 +\epsilon_{in}v_nf_i
\ ,\label{EOM:chiral_4}
%\pp_t (\delta\rho) &=&-\rho_0\pp_iv_i
%-v_i(\pp_i\delta\rho) -\pp_i\left(c_1\epsilon_{ij}v_j+c_2v_i\pp_jv_j+c_3\epsilon_{ij}v_j\pp_kv_k\right)
%-\vnab \cdot \left[c_2\bv(\nabla\cdot\bv)+c_3\bm{\epsilon}\cdot\bv(\nabla\cdot\bv)\right] -g'(\rho_0)\delta\rho\,,\label{EOM:chiral_4}
\eeqn
\ew
where we have made liberal use of the following identities obeyed by the anti-symmetric tensor $\beps$:
\beq
\epsilon_{ij}\epsilon_{kl}=\delta_{ik}\delta_{jl}-\delta_{il}\delta_{jk}\,,~~~ \epsilon_{ik}\epsilon_{jk}=\delta_{ij} \,,
\label{epsid}
\eeq
the second of which clearly follows by taking a trace of the first when we remember that
we're
working in spatial dimension $d=2$.

Next, we time-average equation (\ref{EOM:chiral_4})  over  one  oscillation cycle of $\bv$. For the purposes of such time averages,  $\phi(\br,t)$ and all of its spatiotemporal derivatives can be treated as constants, since $\phi(\br,t)$ varies slowly on the time scale $b^{-1}$ of the oscillating velocity.
They can therefore be removed from the averages over one cycle, which we'll denote by $\langle\rangle_c$. It is then fairly straightforward to show that time averages of products of two and four components of $\bv$ are given by:
\beqn
\begin{aligned}
&\langle v_iv_j\rangle_c={v_0^2\delta_{ij}\over2} \sep \\
&\langle v_iv_jv_kv_l\rangle_c={v_0^4\over8}\bigg(\delta_{ij}\delta_{kl}+\delta_{ik}\delta_{jl}+\delta_{il}\delta_{jk}\bigg) \,.
\end{aligned}
\label{angaves}
\eeqn
Averages over a full cycle of products of any odd number of components of $\bv$ clearly vanish. This immediately eliminates all of the $\lambda$-terms, since they have an odd number of components of $\bv$. This is one of the radical changes introduced by the chirality, since those $\lambda$  terms are known \cite{Toner_prl12, Chen_prl20,Chen_pre20}  to dominate the physics of the achiral problem.

Using \rf{angaves} on the average of (\ref{EOM:chiral_4})
gives
\beqn
\pp_t\phi=\nu\nabla^2\phi+{\lambda_{_K}\over 2}(\nabla\phi)^2+f_\phi\,,\label{EOM:phi}
\eeqn
where
\beqn
\nu&=&\mu_1+\left({\mu_2+\mu_3v_0^2\over 2}\right)\,,%+{e_2p'(\rho_0)\over 2}
%=\mu_1+{\mu_3v_0^2\over 2}+{1\over 2}\left[\mu_2+{\rho_0p'(\rho_0)\over g'(\rho_0)}\right]+{b\left[c_1g'(\rho_0)-\rho_0p'(\rho_0)\right]\over g'(\rho_0)\big\{\left[{g'}(\rho_0)\right]^2+b^2\big\}}\,
\label{nu}\\
%\eeqn
%\beqn
\lambda_{_K}&=&-\mu^\prime\,.
%-\mu?- {e_1p'(\rho_0)\over 2}+...
\label{lambda_K}
\eeqn
%where in the second equality in (\ref{K}) we have separated the chiral part (i.e., the piece proportional to $b$) from the achiral part.}

The noise in \rf{EOM:phi} is given by
\beq
f_\phi=\epsilon_{in}v_nf_i/v_0^2 \,.
\label{fp}
\eeq
Its statistics, after averaging over one cycle, are
\bew
\beq
\la f_\phi (\br,t) f_\phi(\br',t') \ra =\langle\epsilon_{in}v_nf_i\epsilon_{jm}v_mf_j\rangle/v_0^4=\epsilon_{in}\epsilon_{jm}\la f_i(\br,t) f_j(\br',t') \ra\langle v_nv_m\rangle_c /v_0^4=2D_\phi \delta^2(\br-\br')\delta(t-t')
\ ,
\label{fphicorr}
\eeq
\ew
where we`ve defined $D_\phi\equiv(D/v_0^2)$. In deriving \rf{fphicorr}, we`ve made  use of the identities \rf{epsid} and \rf{angaves} for the anti-symmetric tensor and the time averages of the velocity, respectively.

Eq. (\ref{EOM:phi}) is just the well-known (2+1)-dimension-KPZ equation.

This result is not a coincidence, but required by the following symmetry argument. Since the phase of the system  is isotropic in a time-averaged sense in the  long-time limit because of the constant rotation, the EOM for $\phi$ must be invariant under the rotation $\phi\to\phi+ {\rm constant}$, which implies the EOM can only involve derivatives of $\phi$. Also because of this isotropy, the EOM must be isotropic in space, which excludes all one-derivative terms, and only allows two-derivative terms like $\nabla^2\phi$ and $(\nabla\phi)^2$. Furthermore, the EOM (\ref{EOM:chiral_1}) is not invariant under the transformation $\phi\to -\phi$, because chiral symmetry is explicitly broken; the term violates this invariance is $(\nabla\phi)^2$. This explains why $(\nabla\phi)^2$ can only be generated from the chiral terms in the EOM for $\bv$ (\ref{EOM:chiral_1}).

The above symmetry argument proves that, even though we have neglected infinitely many possible terms in the starting model (\ref{EOM:chiral_1}), (\ref{EOM:phi})  is necessarily the ultimate form for the EOM for $\phi$. An alternative argument leading to the same conclusion is presented in appendix \rf{genmod}. This appendix shows  that all the possible one-derivative terms in (\ref{EOM:chiral_1}) have no contribution to (\ref{EOM:phi}), all the achiral two-derivative terms can only contribute to $\nabla^2\phi$, and all the chiral two-derivative terms can only contribute to $(\nabla\phi)^2$.
 Further, we don't need to consider higher derivative terms since their contribution to (\ref{EOM:phi}) is irrelevant.  This leaves the KPZ equation as the only possible EOM for $\phi$, in agreement with  the symmetry argument given above.
 
 \vspace{.2in}

\section{ The hydrodynamics of the density field}{\label{rho}}

Now we turn to the hydrodynamics of the local number density of flockers $\rho(\br,t)$.
In the achiral Malthusian case, we can include $\rho$, even though it is a non-hydrodynamic field, by adding to the usual continuity equation a source term $g(\rho(\br,t))$ which is
the difference between the local birth and death rate:
\beqn
\pp_t \rho +\vnab \cdot (\rho \bv) &=& g(\rho)\,.
\label{rhoEOMa}
\eeqn
We assume that
$g(\rho_0)=0$ at some $\rho_0\ne0$,  $g(\rho_0)<0$ for $\rho>\rho_0$, and $g(\rho_0)>0$ for $\rho<\rho_0$. This implies  that, in the absence of noise, $\rho$ will approach a steady state  value of $\rho_0$ exponentially in time, with a time constant
\beq
\tau_\rho=-{1\over g'(\rho_0)} \,.
\label{trho}
\eeq

In refs \cite{Toner_prl12, Chen_prl20,Chen_pre20}, it was shown that these equations imply that the fluctuation
\beq
\dr(\br,t)\equiv\rho(\br,t)-\rho_0 \,
\label{delrhodef}
\eeq

\noindent
of $\rho$ about its steady-state value $\rho_0$ becomes ``enslaved" to the local configuration of the velocity field $\bv(\br,t)$; that is, $\dr$ is determined at any instant of time by the spatial configuration of $\bv(\br,t)$ {\it at the same instant of time}. As a result, the slow hydrodynamic decay of velocity fluctuations leads to a slow decay of $\dr$ as well, even though $\dr$ is not, strictly speaking, a hydrodynamic variable.

We will now demonstrate that the same thing happens to the density fluctuations in the chiral problem, with $\dr$ becoming enslaved to our Goldstone mode $\phi$.

For the density EOM, we have 
\bew
\beqn
\pp_t \rho +\vnab \cdot {\bf J}&=& g(\rho, |\nabla\bv|^2, (\pp_iv_j)(\pp_jv_i), (\nabla\cdot\bv)^2, \epsilon_{ij}\pp_iv_j(\nabla\cdot\bv), |(\bv\cdot\nabla)\bv|^2)\,.\label{EOM:chiral_2}
\eeqn
\ew
where the number current ${\bf J}$ for the chiral case is, by the same logic we used earlier to be made up of vectors obtained from the velocity field $\bv(\br,t)$, the gradient operator, and, because our system is chiral, the anti-symmetric tensor $\beps$.

One might at first think that the gradient operator  inside the current ${\bf J}$ is unnecessary, since a term proportional to $\bv$ in   ${\bf J}$ (like the $\rho\bv$ term already present in the achiral model) would appear to dominate any terms involving gradients of the velocity in the long-wavelength limit.  However, that  term, and its chiral sibling $\beps\cdot\bv$,  both oscillate at  frequency $b$, while the dominant part of $\dr$ only involves frequencies much less than $b$, since it gets enslaved to $\phi(\br,t)$, as we`ll shortly show.

In order to make a term out of velocities that has components that do not oscillate rapidly, we need to make terms with an even number of velocity vectors $\bv$. The only way to make such terms that are not simply trivial powers of $|\bv|^2$ (which is just a constant $v_0^2$) is to include an odd number of gradient operators. The leading order terms in a gradient expansion therefore prove to be terms with two velocities and one gradient operator.  Including them all, we have 
\bew
\beqn
J_i=&&\rho_0v_i
+c_{1a}v_i\pp_jv_j+c_{2a}v_j\pp_jv_i\nonumber\\
&&+c_1\epsilon_{ij}v_j+c_2\epsilon_{ij}v_k\pp_kv_j+c_3\epsilon_{ij}v_j\pp_kv_k+c_4\epsilon_{jk}v_i\pp_jv_k
+c_5\epsilon_{jk}v_j\pp_iv_k+c_6\epsilon_{jk}v_j\pp_kv_i +c_7\epsilon_{j\ell}v_iv_\ell v_k\pp_kv_j \,.
\label{jrho}
\eeqn
\ew
The terms of the first line of this expression (\ref{jrho}) are achiral, while those on the second line are chiral.

%These can be rewritten in tensor notation as:

% in tensor form,
%\beqn
%\epsilon_{ij}v_j\pp_t \phi
%+\lambda_3 \bm{\epsilon} \cdot \left[(\bv \cdot {\bf \nabla})\bv\right]
%&=&-\lambda_1 \epsilon_{ik}v_kv_j\pp_j\phi-\lambda'_1\left[v_i(v_k\pp_k\phi)-v_0^2(\pp_i\phi)\right] -\lambda_2\epsilon_{jk}v_iv_k(\pp_j\phi)-p'(\rho_0)(\pp_i \delta\rho) +\mu_1\left[\epsilon_{ik}v_k\pp_j\pp_j\phi-v_i(\pp_j\phi)(\pp_j\phi)\right]\nonumber\\
%&&+\mu_2\left[\epsilon_{jk}v_k\pp_i\pp_j\phi-v_i(\pp_i\phi)(\pp_j\phi)\right]
%+\mu_3\left[\epsilon_{i\ell}v_\ell v_jv_k\pp_j\pp_k\phi-v_iv_kv_j(\pp_k\phi)(\pp_j\phi)\right]\nonumber\\
%&&-\mu'\left[v_i\pp_k\pp_k\phi+\epsilon_{im}v_m(\pp_k\phi)(\pp_k\phi)\right]+L(|v|)v_i +f_i
%\ ,\label{EOM:chiral_3}\\
%\pp_t (\delta\rho) &=&-\rho_0\pp_iv_i
%-v_i(\pp_i\delta\rho)
%-\pp_i\left(c_1\epsilon_{ij}v_j+c_2v_i\pp_jv_j+c_3\epsilon_{ij}v_j\pp_kv_k\right)
%-\vnab \cdot \left[c_2\bv(\nabla\cdot\bv)+c_3\bm{\epsilon}\cdot\bv(\nabla\cdot\bv)\right]
%-g?\rho_0)\delta\rho\,,\label{EOM:chiral_4}
%\eeqn
% {In deriving (\ref{EOM:chiral_3},\ref{EOM:chiral_4}) we have expanded $p(\rho,|\bv|=v_0)$ and $g(\rho)$ around $\rho=\rho_0$ to linear order in $\delta\rho$.}

 We have not included terms in $\bJ$  involving spatial derivatives of $\delta\rho$,   since they  always lead to terms in the EOM \rf{EOM:chiral_2} for $\dr$ which are   irrelevant compared to terms coming from  the birth minus death term $g(\rho, ...)$, to which we now turn.

 Expanding that term in powers of its arguments, we have 
 \bew
 \beqn
 g(\rho, |\nabla\bv|^2, (\pp_iv_j)(\pp_jv_i), (\nabla\cdot\bv)^2, \epsilon_{ij}\pp_iv_j(\nabla\cdot\bv), |(\bv\cdot\nabla)\bv|^2)=&&-\varpi\dr+f_1|\nabla\bv|^2+f_2(\pp_iv_j)(\pp_jv_i)+f_3 (\nabla\cdot\bv)^2\nonumber\\
 &&+f_4|(\bv\cdot\nabla)\bv|^2
+f_c \epsilon_{ij}\pp_iv_j(\nabla\cdot\bv) \,,
 \label{gexp}
 \eeqn
 \ew
 where $\varpi$, and $f_{1,2,3,4,c}$ are all constant expansion coefficients. Note that we must have $\varpi>0$ if the state with density $\rho_0$ is to be stable.

 We have only expanded to leading order in $|\nabla\bv|^2$, $(\pp_iv_j)(\pp_jv_i)$, $(\nabla\cdot\bv)^2$,  and $\epsilon_{ij}\pp_iv_j(\nabla\cdot\bv)$, $|(\bv\cdot\nabla)\bv|^2$, since higher order terms involve more derivatives, and hence prove to be irrelevant.

We can also drop the time derivative term $\pp_t\rho=\pp_t\delta\rho$ in the $\rho$ EOM \rf{EOM:chiral_2}, since it is irrelevant compared to the $\varpi\dr$ term (which has no derivatives).

Doing so, the resultant EOM for $\rho$ simply becomes a linear algebraic equation for $\dr$, whose solution is easily found to be
\bew
\beqn
\delta\rho&=&\left({1\over \varpi}\right)\left[\nabla\cdot{\bf J}+f_1|\nabla\bv|^2+f_2(\pp_iv_j)(\pp_jv_i)+f_3 (\nabla\cdot\bv)^2+f_4|(\bv\cdot\nabla)\bv|^2+f_c \epsilon_{ij}\pp_iv_j(\nabla\cdot\bv)\right]\nn\\
&=&\left({1\over \varpi}\right)\bigg[\pp_i\bigg(\rho_0v_i
+c_{1a}v_i\pp_jv_j+c_{2a}v_j\pp_jv_i%\right.
\nonumber\\
&&+c_1\epsilon_{ij}v_j+c_2\epsilon_{ij}v_k\pp_kv_j+c_3\epsilon_{ij}v_j\pp_kv_k+c_4\epsilon_{jk}v_i\pp_jv_k
+c_5\epsilon_{jk}v_j\pp_iv_k+c_6\epsilon_{jk}v_j\pp_kv_i +c_7\epsilon_{j\ell}v_iv_\ell v_k\pp_kv_j\bigg)%\right.
\nn\\
&&+f_1|\nabla\bv|^2+f_2(\pp_iv_j)(\pp_jv_i)+f_3 (\nabla\cdot\bv)^2+f_4|(\bv\cdot\nabla)\bv|^2+f_c \epsilon_{ij}\pp_iv_j(\nabla\cdot\bv)\bigg]
\,.\label{Eq:rho_1}
\eeqn
\ew
Using the second equation of  \rf{vderivs} to re-express the derivatives of the velocity in terms of derivatives of $\phi$ and velocities, and the
averaging the result (\ref{Eq:rho_1}) over short-time oscillation cycles using  \rf{angaves}, we find that $\delta\rho$ is enslaved to $\phi$:
\beqn
\delta\rho=e_3\nabla^2\phi+e_4|\nabla\phi|^2\,,\label{Eq:rho_2}
\eeqn
where 
\beqn
e_3={v_0^2\over 2\varpi}\left(c_3-c_2+c_4-2c_5-c_6-c_7v_0^2\right)\,,
\eeqn
and
\beqn
e_4=&&{v_0^2\over 2\varpi}\left(2f_1+f_2+f_3+f_4v_0^2\right) \,.
\label{e4}
\eeqn
We will use this result in the next section to show that the density fluctuations develop long-ranged spatio-temporal correlations, despite the fact that $\dr$ itself is {\it not} a hydrodynamic variable in this Malthusian flock, since birth and death violate conservation of flocker number.

\vspace{.2in}

\section{Phase, velocity, and density correlations in the Non-linear KPZ regime}{\label{kpzc}}

Having successfully mapped the 2D chiral Malthusian flock onto the (1+2)-dimensional-KPZ equation, we can now use the known \cite{KPZ, kpzexp1, kpzexp2,kpzexp3, kpzexp4,kpzexp5, kpzexp6}  scaling laws for the (1+2)-dimensional-KPZ equation to determine the scaling laws for our system.

The scaling laws of the KPZ equation are characterized by two universal exponents: a ``dynamical exponent" $z$ relating length scales $L$  to time scales $t$ via
$t\propto L^z$, and a ``roughness exponent" $\chi$ relating length scales to the magnitude of the fluctuations of $\phi$ via $\phi\propto L^\chi$.

This implies that the mean squared fluctuations of the difference between $\phi$'s at widely separated points in space and time obey:
%{The EOM (\ref{EOM:phi}) implies}
\beq
\langle[\phi(\br,t)-\phi(\br',t')]^2\rangle=A_\phi |\br-\br'|^{2\chi}\cG_\phi\left(|t-t^\prime|\over|\br-\br'|^z\right)\,,\label{phi_correl}
\eeq
where
%{$\chi$ is the roughness exponent for (2+1)-KPZ equation}
$A_\phi$ is a non-universal positive constant, and $\cG_\phi\left(X\right)$ has the following asymptotic scaling behavior:
\begin{eqnarray}
\cG_\phi\left(X\right)=\left\{
\begin{array}{ll}
1\,,&X\ll 1\,,\\
X^{2\chi\over z}\,,&X\gg1\,.
\end{array}
\right.
\end{eqnarray}
The current estimate of the values of
%m{ost accurate numerical result for}
$\chi$ and $z$ based on simulations is $\chi=0.388\pm.002$, $z=1.622\pm.002$\cite{kpzexp1, kpzexp2,kpzexp3, kpzexp4,kpzexp5, kpzexp6}.

Using (\ref{v_fluctuate}) the spatial-temporal correlation of the velocity can be expressed as
\bew
\beqn
%C(\br-\br^\prime, t-t')\equiv
\langle\bv(\br,t)\cdot\bv(\bm{\br^\prime},t')\rangle &=& v_0^2\Big\{\langle\cos[bt+\phi(\br,t)]\cos[bt'+\phi(\br',t')]\rangle
+\langle\sin[bt+\phi(\br,t)\sin[bt'+\phi(\br',t')]\rangle\Big\}
\nn\\
&=&v_0^2\big\langle\exp\big\{\ii\left[\phi(\br,t)-\phi(\br',t')\right]\big\}\big\rangle \cos\left[b(t-t')\right]
\,.\label{v_correl}
\eeqn
\ew
 % The velocity correlations between different times are hard to measure due to the fast oscillating cosine factor. Fortunately, the order of system can be fully captured by the equal-time velocity correlations.}

If the fluctuations of the field $\phi$ were Gaussian, then we could say
\bew
\beqn
\big\langle\exp\big\{\ii\left[\phi(\br,t)-\phi(\br^\prime,t')\right]\big\}\big\rangle  %\nonumber\\
&=&\exp\bigg\{-{1\over2}\bigg\langle\left[\bigg(\phi(\br,t)-\phi(\br^\prime%\br?
, t)\bigg)^2\right]\bigg\rangle\bigg\}\nonumber\\
&=&\exp\left[-\left({A_\phi\over2}\right)|\br-%\br'
\br^\prime|^{2\chi}\cG_\phi\left(|t-t^\prime|\over|\br-\br'|^z\right)\right]\,.
\label{gaussian}
\eeqn
\ew
Inserting (\ref{gaussian}) into (\ref{v_correl}),  we get velocity correlations for arbitrary spatial-temporal separation:
\bew
\beqn
\langle\bv(\br,t)\cdot\bv(\bm{\br^\prime},t')\rangle=v_0^2\exp\left[-\left({A_\phi\over2}\right)|\br-%\br'
\br^\prime|^{2\chi}\cG_\phi\left(|t-t^\prime|\over|\br-\br'|^z\right)\right]\cos\left[b(t-t')\right]\,.
\label{v_correl_g}
\eeqn
\ew

First we focus on the equal-time correlation. Setting $t'=t$ in (\ref{v_correl_g}) we find
\beq
\langle\bv(\br,t)\cdot\bv(\bm{\br^\prime},t)\rangle=v_0^2\exp\left[-\left({A_\phi\over2}\right)|\br-%\br'
\br^\prime|^{2\chi}\right]\,,~~~~~
\label{v_correl_et}
\eeq
which would imply (stretched) exponentially decaying velocity correlations.
However, since the $\phi$ fluctuations are non-Gaussian, due to the relevant nonlinearity $\lambda_{_K}$ in the KPZ equation, we can say very little, beyond noting that we expect velocity correlations to decay rapidly with increasing  separation $|\br-\br^\prime|$ once that separation is large enough that $\langle[\phi(\br,t)-\phi(\br',t)]^2\rangle\gtrsim O(1)$. Using \rf{phi_correl}, we see that this implies that correlations will decay rapidly for $|\br-\br^\prime|>\xi_v\equiv A_\phi^{-{1\over2\chi}}\approx A_\phi^{-1.29}$. 
However, as we mentioned in the introduction, in the limit of weak chirality, $\xi_v\ll L_{_{NL}}$, and so order will already be effectively lost before we even reach the non-linear regime. 
Hence, only for highly chiral systems will $\xi_v$ be the ``velocity correlation length". Even then, it must be kept in mind that the decay of velocity correlations is probably not exponential, nor even a simple stretched exponential like equation \rf{gaussian}.

 %{This shows an exponentially decaying correlation as a function of the distance between two locations. This implies}
 The essential point is that velocity correlations are ultimately short-ranged; that is, chirality has destroyed the long-ranged orientational order present in achiral Malthusian flocks. And it does so for arbitrarily weak noise (i.e., arbitrarily small noise strength $D$ and arbitrarily weak chirality). That is, if one takes a snapshot of the system at any instant of time,
 %{and computes}
 the average velocity  $\langle\bv\rangle$,
 %{the result,}
 in the limit of system size going to infinity, is $\langle\bv\rangle={\bm 0}$.

Combining (\ref{Eq:rho_2}) with (\ref{phi_correl}) we can predict the correlations of the density fluctuations. We start from the equal-time case:
\bew
\beqn
\langle\delta\rho(\br,t)\delta\rho(\bm{r}',t)\rangle&=&e_3^2\nabla^2{\nabla'}^2\langle\phi(\br,t)\phi(\br',t)\rangle + e_3e_4\bigg(\langle |\nabla\phi(\br,t)|^2{\nabla^\prime}^2\phi(\br^\prime,t)\rangle+\langle |\nabla^\prime\phi(\br^\prime,t)|^2\nabla^2\phi(\br,t)\rangle\bigg)
\nn\\
&+& e_4^2\langle |\nabla\phi(\br,t)|^2|\nabla^\prime\phi(\br^\prime,t)
|^2\rangle \,.
\label{rhocorKPZ1}
\eeqn
\ew
The first term in this expression is readily evaluated from the two point correlation function \rf{phi_correl}:
\beqn
&&e_3^2\nabla^2{\nabla'}^2\langle\phi(\br,t)\phi(\br',t)\rangle
\nonumber\\
&=&-{1\over 2}e_3^2\nabla^2{\nabla'}^2\langle[\phi(\br,t)-\phi(\br',t)]^2\rangle\nonumber\\
&=&-8A_\phi e_3^2\chi^2(\chi-1)^2|\br-\br'|^{2\chi-4}\nonumber\\
&\propto&|\br-\br'|^{2\chi-4}\approx|\br-\br'|^{-3.224}\,.\label{rho_corre1}
\eeqn

The remaining terms cannot be evaluated directly from the two-point correlation function \rf{phi_correl}, since they involve three and four-point correlation functions. These cannot be directly related to two point correlation functions because the field $\phi$ is non-Gaussian, due to the relevant $\lambda_{_K}$ non-linearity in the KPZ equation.  However, we {\it can} extract their behavior from scaling arguments: they contain either three or  four $\phi$`s,
%`s,
each of which contributes a factor of $|\br-\br'|^\chi$, while each of the four gradient contributes one factor of $|\br-\br'|^{-1}$. Hence we have
\bew
\beqn
&&\langle |\nabla\phi(\br,t)|^2{\nabla^\prime}^2\phi(\br^\prime,t)\rangle\propto |\br-\br'|^{3\chi-4}\approx |\br-\br'|^{-2.863}\propto\langle |\nabla^\prime\phi(\br^\prime,t)|^2\nabla^2\phi(\br,t)\rangle
\nn\\
&&\langle |\nabla\phi(\br,t)|^2|\nabla\phi(\br^\prime,t)
|^2\rangle\propto |\br-\br'|^{4\chi-4}\approx |\br-\br'|^{-2.448} \,,
\eeqn
\ew
where in the second, approximate, equality we have used the numerically obtained value $\chi=.38$ for $\chi$ \cite{kpzexp1, kpzexp2, kpzexp3, kpzexp4, kpzexp5, kpzexp6}.

Since $\chi>0$, the final term in  \rf{rhocorKPZ1} dominates, and at large $|\br-\br'|$. Thus we have
\bew
\beqn
\langle\dr(\br,t)\dr(\br^\prime,t)\rangle\approx e_4^2\langle |\nabla\phi(\br,t)|^2|\nabla\phi(\br^\prime,t)
|^2\rangle\propto |\br-\br'|^{4\chi-4}\approx |\br-\br'|^{-2.448} \,.
\label{rho_corre2}
\eeqn
\ew
Combining the scaling relation between the temporal and spatial dimensions $|t-t'|\sim|\br-\br'|^z$ and (\ref{rho_corre2}) implies the following scaling behavior for the two-point density correlation for arbitrary temporal and spatial separations:
\bew
\beqn
\langle\dr(\br,t)\dr(\br^\prime,t)\rangle
%\approx e_4^2\langle |\nabla\phi(\br,t)|^2|\nabla\phi(\br^\prime,t)
%|^2\rangle
\propto |\br-\br'|^{4\chi-4}\cG_\rho\left(|t-t^\prime|\over|\br-\br'|^z\right)\approx |\br-\br'|^{-2.448}\cG_\rho\left(|t-t^\prime|\over\,\,\,\,\,|\br-\br'|^{1.622}\right) \,,
\label{rho_corre_g}
\eeqn
\ew
where
\begin{eqnarray}
\cG_\rho\left(X\right)=\left\{
\begin{array}{ll}
1\,,&X\ll 1\,,\\
X^{4\chi-4\over z}\,,&X\gg1\,.
\end{array}
\right.
\end{eqnarray}

These results (\ref{v_correl_g},\ref{rho_corre_g}) hold only for length scales $|\br-\br^\prime|
\gg L_{_{NL}}$ or $|t-t'|\gg\tau_{_{NL}}$, where $L_{_{NL}}$ and $\tau_{_{NL}}$ are respectively the spatial and temporal length scales at which the non-linear $\lambda_{_K}$ term in the KPZ equation (\ref{EOM:phi}) becomes important.  We calculate these length scales in the next section.

There is also an {\it upper} bound on the length scales out to which these usual KPZ results hold for our problem. This is for a rather subtle but crucial reason: our mapping is not onto the standard KPZ equation, but onto the {\it compact} KPZ equation. What we mean by this is that, unlike, say, a growing interface, in which the KPZ variable $\phi$ is the height of the interface, and every value of the height represents a physically distinct local state, our KPZ variable is a {\it phase}. That is, a state with a given local value $\phi(\br,t)$ is locally physically indistinguishable from state $\phi(\br,t)+2\pi n$, for any integer $n$.

This is the same symmetry that is present in the 2D XY model. Here, as there, it allows   topological defects - i.e., vortices \cite{KT}.  These become important on length scales larger than the mean inter-vortex distance  $\lv$. We calculate this length scale in  section \rf{v},  in the limit of small chirality, as well.

\vspace{.2in}

\section{Achiral to chiral crossover lengths and times in the small chirality limit\label{sc}}

When the chirality is weak - that is, when $b$ is small - then for times
\beq
t\ll\tau_c\equiv b^{-1} \,,
\label{tauc}
\eeq
the direction of the mean velocity hardly changes, and   the system should behave like an achiral Malthusian flock for sufficiently small distances. To determine those distances, we can use the spatio-temporal scaling properties of achiral Malthusian flocks \cite{Toner_prl12, Chen_prl20,Chen_pre20, Chate_prl24}.

In an achiral Malthusian flock, Fourier modes with wavevector $\bq$ decay at a rate $\Gamma(\bq)$ given by 
the following scaling form \cite{Toner_prl12,Chen_prl20,Chen_pre20}:
\beq
\Gamma(\bq) = \Upsilon |q_y \xi_y|^{z_a-2}  q_y^2G_{\rm achiral}\left(\frac{|q_y \xi_y|^{\zeta_a}}{|q_x \xi_x|}\right)
\,,
\label{gammascale}
\eeq 
where $q_x$ and $q_y$ are the  components of $\bq$ along and perpendicular to the mean direction $\hx$ of flock motion, respectively. In addition, $\Upsilon$ is a non-universal constant with the dimensions of a diffusion constant, $z_a$ and $\zeta_a$ are the universal ``dynamic`` and ``anisotropy`` exponents of the achiral problem, respectively, $\xi_{x,y}$ are non-universal microscopic length scales for the achiral problem, 
and $G_{\rm achiral}$ is a scaling function,  with the limiting behaviors:
\beqn
G_{\rm achiral}(u \ll 1) &\propto & u^{-z_a/\zeta_a}
\\
G_{\rm achiral}(u \gg 1) &\approx & 1
\,.
\eeqn
Contrary to recent claims in the literature \cite{Chate_prl24,Ikeda_prl24}, there is no exact analytic expression for the universal exponents  $z_a$ and $\zeta_a$. Fortunately, these exponents {\it are} known through direct simulations of the noisy hydrodynamic equations of achiral Malthusian flocks\cite{Chate_prl24}, which find 
\beq
z_a\approx 5/4 \sep \zeta_a\approx 3/4 \,.
\label{achiralexps}
\eeq

In the {\it chiral} system, however, the direction of mean motion is rotating at a constant angular frequency $b$. This means that, in a time-averaged sense, all directions are equivalent. Therefore,we can without loss of generality take $\bq=(q,0)$. To obtain the {\it instantaneous} damping rate of such a mode, we need to project this wavevector  onto the {\it rotating} axis of mean motion $\hat{\bf x}(t)=(\cos(bt), \sin(bt))$, and the rotating  axis $\hat{\bf y}(t)=(-\sin(bt), \cos(bt))$ perpendicular to it, in order to obtain the $q_x$ and $q_y$ that enter \rf{gammascale}. Thus, we effectively have {\it time-dependent} components
\beq
q_x(t)=q\cos(bt) \sep q_y(t)=-q\sin(bt) \,.
\label{qrot}
\eeq

Inserting these expressions for the components of $\bq$ into our expression \rf{gammascale}, we see that  the damping also becomes time-dependent: 
\beq
\Gamma( \bq)  =\Upsilon q^{z_a} \xi_y^{z_a-2}  |\sin (bt)|^{z_a}  G_{\rm achiral}\left(u(q,t)\right) \,,
\eeq
where we`ve defined
\beq
u(q,t) = q^{\zeta_a-1} \frac{ |\xi_y \sin (bt)|^{\zeta_a}}{ |\xi_x \cos (bt)|}
\,.
\eeq

At long times, we are interested in the time average of this damping coefficient over a full cycle. This is
\beq
\la \Gamma( \bq) \ra_{\rm cycle} =\Upsilon q^{z_a} \xi_y^{z_a-2} \la |\sin bt|^{z_a}  G_{\rm achiral}\left(u(q,t)\right) \ra_{\rm cycle} \,.
\eeq
Since $\zeta_a$ has recently been estimated to be 3/4, which is less than 1, and we are interested in the small $q$ regime, $u\gg 1$  unless
\beq
\frac{ | \sin bt|^{\zeta_a}}{ |\cos bt|} \lesssim  q^{1-\zeta_a} \frac{ \xi_x}{ \xi_y^{\zeta_a}}
\eeq
which is a vanishing window within the cycle in the small $q$ limit. Therefore, we can approximate  $G_{\rm achiral}\left(u(q,t)\right)$ by 1 over most of the cycle. Hence,
\bew
\beq
\la \Gamma( \bq) \ra_{\rm cycle} =\Upsilon q^{z_a} \xi_y^{z_a-2} \la |\sin bt|^{z_a} \ra_{\rm cycle}=\Upsilon q^{z_a} \xi_y^{z_a-2} \frac{\Gamma(\frac{1+z_a}{2})}{\sqrt{\pi} \Gamma(1+z_a/2)}
\ .
\eeq
\ew
Using the recent estimate of 5/4 for $z_a$, we have
\beq
\la \Gamma( \bq) \ra_{\rm cycle} \simeq 7.13 \times \Upsilon q^{z_a} \xi_y^{z_a-2} 
\ .
\eeq

To obtain the achiral-chiral crossover length $L_c$, or, equivalently, the associated wavenumber $q_c=L_c^{-1}$,  we now equate the above expression with the corresponding one of the linearized KPZ, equation which is
\beq
\Gamma(\bq) =\nu q^2
\eeq
At the crossover length $L_c = \Lambda_c^{-1} =q_c^{-1}$, we have 
\beq
\nu \Lambda_c^2 = \Upsilon \Lambda_c^{z_a} \xi_y^{z_a-2} \times \cO(1)
\ .
\eeq

Further, we expect the crossover happens at a time scale when the relaxation time due to damping equals the rotation period. Therefore, we equate $\nu \Lambda_c^2$ with the rotation frequency $b$ to arrive at
\beq
\label{eq:crossoverL}
b =\nu \Lambda_c^2 \simeq \Upsilon \Lambda_c^{z_a} \xi_y^{z_a-2} 
 \ .
 \eeq
Equating the 1st and 3rd term in (\ref{eq:crossoverL}) leads to
\beq
\Lambda_c \propto b^{1/z_a} \ ,
\eeq
or 
\beq
L_c \propto b^{-1/z_a} \simeq b^{-4/5}
\label{lcsc_v2}
\ ,
\eeq
where we have used the recent numerical estimate\cite{Chate_prl24}  in the second relation of \rf{lcsc_v2}  above.

The fact that $\tau_c$ and $L_c$ both diverge as the chirality $b\to0$ leads to strong $b$ dependence of the parameters $\nu$ and $\lambda_{_K}$ of the ultimate KPZ model as well. We have already seen the strong dependence of $\lambda_{_K}$ on the chirality: since
$\lambda_{_K}$ is generated exclusively by the chiral terms in our model, it follows that it must vanish in the limit of zero chirality. Hence, by analyticity, it should vanish linearly with the chirality $b$ as
$b\to0$; that is
\beq
\lak\propto b \,.
\label{lak}
\eeq

The diffusivity $\nu$ in the ultimate KPZ equation, on the other hand, {\it  diverges} as $b\to0$. The origin of this divergence
lies in two phenomena: first, the $\tau_c\propto b^{-1}$ divergence of the time scale
$\tau_c$ for the crossover from achiral to chiral behavior, and second, the fact that the achiral system exhibits ``anomalous hydrodynamics'': that is, a divergence of the diffusion constants $\mu_{1,2,3}$ appearing in the EOM \rf{EOM:chiral_1} as a function of the length and time scales on which they are measured.

We can deduce this divergence from the known universal exponents $z_a$ and $\zeta_a$ by the following argument:

An  achiral system will reach a state in which only small fluctuations around a non-zero average velocity will occur. Taking the direction of this non-zero average velocity to be $\hx$, we can
expand  $\bv(\br,t)$ around the steady state:
\beqn
\bv(\br,t)=[v_0+u_x(\br,t)]\hat{{\bf x}}+u_y(\br,t)\hat{{\bf y}}\,,\label{v_achiral}
\eeqn
where $\hat{{\bf y}}$ denotes the direction perpendicular to $\hat{{\bf x}}$. Inserting (\ref{v_achiral}) into (\ref{eq:original_1}) and focusing on the $u_y$ part of the EOM since $u_y$ exhibits the largest
%{the dominant}
fluctuations, we get
\beqn
\pp_t u_y&=&
%-\lambda_1v_0\pp_x u_y +
\mu_x\pp_x^2 u_y + \mu_y\pp_y^2 u_y -\lambda_1u_y\pp_yu_y\nn\\
&&+{\lambda_1\over 2v_0}u_y^2\pp_xu_y+ f_y\,,
\label{EOM:achiral_1}
\eeqn
where
\beqn
\mu_x&=&\mu_1+\mu_3v_0^2\,,\label{mux}\\
\mu_y&=&\mu_1+\mu_2\,.\label{muy}
\eeqn
In deriving (\ref{EOM:achiral_1}), we have boosted to a moving frame via a Galilean transformation  $x\to x-\lambda v_0 t$ to eliminate the term $-\lambda_1v_0\pp_x u_y$ on the right hand side.

Previous work \cite{Toner_prl12, Chen_prl20, Chen_pre20} showed that the nonlinear term  in (\ref{EOM:achiral_1}) is ``relevant"  in the renormalization group sense of changing the long-wavelength behavior of the system. Indeed, it is responsible for the existence of long-ranged order in the 2D system; without it, the ``Mermin-Wagner" destruction of long-ranged order by fluctuations, which always occurs in equilibrium systems trying to break continuous rotation invariance in two dimensions, would occur here as well.

The effect of the non-linearity can be incorporated into an effective ``renormalization" of the diffusion constants $\mu_{x,y}$, making them
%{The diffusion coefficients $\mu_{x,y}$ are strongly renormalized by the nonlinearity, becoming frequency and wave}
length and time scale dependent, diverging as the length and time scales under consideration go to infinity. It is this length and time scale dependence that leads to the unusual, non-integer values of the dynamical and  anisotropy exponents $z_a$ and $\zeta_a$. To see this, note that in equation \rf{EOM:achiral_1} we can estimate
$\pp_tu_y\sim {u_y\over t}$ and $\mu_y\pp_y^2u_y\sim \mu_y{u_y\over y^2}$. Balancing these terms then gives
the scaling of the diffusivity $\mu_y(t)$:
\beq
\mu_y(t)\propto {y^2\over t}\propto t^{2/z_a-1}=t^{3/5} \,,
\label{muysc}
\eeq
where in the second proportionality we have used the scaling law $y\propto
t^{1/z_a}$ mentioned earlier,  while in the final equality we have used the values \rf{achiralexps} for  $\zeta_a$ and  $z_a$ obtained from the  numerical results  of \cite{Chate_prl24}.

A similar analysis for the diffusivity $\mu_x(t)$ gives 
\beq
\mu_x(t)\propto {x^2\over t}\propto t^{2\zeta_a/z_a-1}=t^{1/5} \,.
\label{muxsc}
\eeq

Comparing our expression \rf{nu} for the KPZ diffusivity $\nu$ with our expressions \rf{mux} and \rf{muy}, we see that
\beq
\nu={\mu_x+\mu_y\over2} \,.
\label{numuxy}
\eeq
Since $\mu_{x,y}$ are length and time scale dependent, we must decide at which time scale $t$ we should evaluate the right hand side of  \rf{numuxy}. But the answer is clear: it must be the crossover time $t=\tau_c=b^{-1}$ at which the system crosses over from achiral to chiral behavior. Thus,
\beq
\nu(b)={\mu_x(t=b^{-1})+\mu_y(t=b^{-1})\over2} \,
\label{numuxy2}
\eeq
Using our earlier results \rf{muysc} and \rf{muxsc}, we see that
\beqn
\begin{aligned}
\mu_y(t=b^{-1})&\propto& b^{1-2/z_a}\approx \,b^{-3/5}\,, \\ 
\mu_x(t=b^{-1})&\propto& b^{1-2\zeta_a/z_a}\approx\, b^{-1/5} \,.
\end{aligned}
\label{mub}
\eeqn

Inserting these results in \rf{numuxy}, we see that the contribution of $\mu_y$ dominates at small $b$, where we therefore have
\beq
\nu(b) \,\propto b^{1-2/z_a}\equiv b^{-\eta_\nu}  \,,
\label{nub}
\eeq
where the universal exponent 
\beq
\eta_\nu=2/z_a-1\approx3/5 \,.
\label{etanu}
\eeq
The numerical value given here is based on numerical simulations of the noisy hydrodynamic equation for achiral Malthusian flocks by \cite{Chate_prl24}.

Inserting \rf{nub} into the first equality of \rf{eq:crossoverL}, we (reassuringly) recover the value of $L_c$ obtained earlier in \rf{lcsc_v2}.

To summarize: small chirality $b$ only manifests itself on time scales larger than $\tau_c\propto b^{-1}$, or length scales  longer than $L_c\propto b^{-1/z_a} \simeq b^{-4/5}$. On those longer length and time scales, chiral Malthusian flocks are described by a KPZ equation with a very small non-linearity $\lambda_{_K}\propto b$, and a very large diffusivity 
$\nu\propto b^{-\eta_\nu}$, with $\eta_\nu\approx3/5$.

The final parameter of the KPZ equation is the noise strength
$D_\phi$ in equation \rf{fphicorr}. This will be independent of chirality $b$ for small chirality, because the noise strength $D_\phi$ of the chiral problem   is related to the velocity noise strength $D$ and the mean speed $v_0$ of the flockers by $D_\phi=D/v_0^2$, and neither $D$ nor  $v_0$  gets any large renormalization in the achiral problem \cite{Toner_prl12, Chen_prl20, Chen_pre20, Chate_prl24}.

We note that our result \rf{nub} for the $b$ dependence of $\nu(b)$ is consistent with the expectation that the time-space correlation function depends purely on the ratio
\beq
\frac{t/\tau_c}{(r/L_c)^2} \ .
\label{scale exp}
\eeq
This is because $\tau_c \sim b^{-1}$ and thus $L_c^2/\tau_c \simeq \nu$ (using the first relation in  Eq.~(\ref{eq:crossoverL})), leading to the ratio being of the form
\beq
\frac{\nu t}{r^2} \ ,
\label{lin kpz exp}
\eeq
which conforms to the expectation from the linearized KPZ equation.

The divergence of $\nu$ as $b\to0$ and the vanishing of the nonlinearity $\lak$ together imply that the asymptotic non-linear KPZ behavior described in the previous section, specifically, the scaling behavior of the correlations (\ref{phi_correl}), (\ref{v_correl_g}), and (\ref{rho_corre_g}) will only manifest itself on length scales $|\br-\br'|\gg L_{_{NL}}$ or time scales $|t-t'|\gg\tau_{_{NL}}$, and for small $b$, $L_{_{NL}}\gg L_c$, $\tau_{_{NL}}\gg \tau_c$.

We can calculate $L_{_{NL}}$ using the renormalization group recursion relations for  the dimensionless coupling
\beq
g\equiv
{\lambda_{_K}^2D_\phi\over 2\pi\nu^3} \,,
\label{gdef}
\eeq
which  provides a measure of the importance of  non-linear effects in the KPZ equation \cite{KPZ}.

The recursion relation for $g$ is \cite{KPZ}
\beq
{\dd g\over \dd\ell}={g^2\over4}+O(g^3)\,.
\label{g rr}
\eeq
Note that the ``bare" value $g_0$ of $g$, which in our problem is the value of $g$ at the length scales $L_c$ and $\tau_c$, vanishes quite rapidly as the chirality $b$ decreases. Specifically, using our earlier results \rf{lak} and \rf{nub} that $\lak\propto b$ and
$\nu\propto b^{-3/5}$, we see that
\beq
g_0\propto b^{2-3(1-2/z_a)}=b^{6/z_a-1} \equiv b^{\eta_g}\,,
\label{g0}
\eeq
where we`ve defined the universal exponent
\beq
\eta_g\equiv 6/z_a-1\approx19/5 \,,
\label{g0def}
\eeq
and, as with all numerical values based on the {\it achiral} exponents, we've used the values  $\zeta_a$ and$z_a$ obtained from the simulations of the noisy hydrodynamic equations for achiral flocks of \cite{Chate_prl24} in the last, approximate, equality.

The solution of the recursion relation \rf{g rr} for such an initially small value of $g_0$ is easily found:
\beq
g(\ell)={g_0\over1-\left({g_0\over4}\right)\ell} \,.
\label{gsol}\\
\eeq

It follows from this that $g(\ell)$ gets to be of $O(1)$, which signals the onset of important non-linear effects, when
\beq
\ell\approx4/g_0\equiv\ell_{_{NL}}\propto b^{-\eta_g}\ \,,
\label{ellnl}
\eeq
 where we remind the reader that the universal exponent $\eta_g\approx19/5$
.
As usual in the renormalization group (RG), we can associate with this RG time a ``non-linear length scale" $L_{_{NL}}$ given by
\beq
L_{_{NL}}=L_ce^{\ell_{_{NL}}}=L_c\exp\left[C_{_{NL}}b^{-\eta_g}\right] \,,
\label{lnl}
\eeq
and a ``non-linear time scale" $\tau_{_{NL}}$ given by
\beq
\tau_{_{NL}}=\tau_{c}e^{z_{_L}\ell_{_{NL}}}=\tau_c\exp\left[ 2C_{_{NL}}b^{-\eta_g}\right] \,,
\label{tnl}
\eeq
where $C_{_{NL}}$ is a non-universal, chirality $b$ independent constant, $z_{_L}=2$ is the dynamic exponent for the KPZ equation in the linear regime.

An equivalent alternative approach to the calculation  of $L_{_{NL}}$ and $\tau_{_{NL}}$ is to   perform  perturbation theory on (\ref{EOM:phi}) and calculate the lowest order correction to the noise strength $D_\phi$. The length and time scales at which this correction becomes comparable to the bare value of $D_\phi$ are $L_{_{NL}}$ and $\tau_{_{NL}}$ respectively. We have done this calculation, and find, reassuringly, that it reproduces the results \rf{lnl} and \rf{tnl} that we just obtained from the RG analysis.

Clearly, these length and time scales $L_{_{NL}}$ and $\tau_{_{NL}}$  given by \rf{lnl} and \rf{tnl}  can become much greater than the length and and time scales $L_c$ and $\tau_c$ at which the effects of chirality first become important, even for chiralities that are not particularly small.

To illustrate this, consider a system in which we'll call ``system 0", in which
%{To see how fast $L_{NL}$ diverges as $b$ decreases, let's assume}
$(L_{_{NL}})_{{\rm system~0}}=3L_c$ (which implies
%{[or equivalently}
$\exp\left(C_{_{NL}} b^{-{19\over 5}}\right)=3$) for some small $b\equiv b_0$. Now consider two other systems, in which all of the dynamical parameters are the same as those of system 0 except for the chirality $b$. The first system (``system 1") will have chirality $b_1=b_0/2$; the second (``system 2") will have chirality $b_2=b_0/3$. Then
%{for weaker chiralities, for instance, $b/2$ and $b/3$,}
\rf{lnl} predicts 
\bew
\beqn
&&\left(L_{_{NL}}\right)_{{\rm system~1}}=\left[2^{4\over 5}\cdot 3^{\left(2^{19\over 5}\right)-1}\right]\left(L_{_{NL}}\right)_{{\rm system~0}}\approx 2.57\times 10^6\left(L_{_{NL}}\right)_{{\rm system~0}}\,,\\
&&\left(L_{_{NL}}\right)_{{\rm system~2}}=3^{\left(3^{19\over 5}\right)-{1\over 5}}\left(L_{_{NL}}\right)_{{\rm system~0}}\approx  8.47 \times 10^{30}\left(L_{_{NL}}\right)_{{\rm system~0}}\,.
\eeqn
\ew
If $\left(L_{_{NL}}\right)_{\rm system~0}=1\mu m$, then $\left(L_{_{NL}}\right)_{\rm system~1}\sim  2.6m$, $\left(L_{_{NL}}\right)_{\rm system~2}\sim 8.47 \times10^{24}m$,  which is a cosmological length scale (roughly  $0.9$ {\it billion light years}, or about  $2\%$ of  the size of the observable universe)! Similarly, $\tau_{_{NL}}$ also exhibits  a similarly dramatic  divergence in the limit of weak chirality.

So
for weak chirality the nonlinear crossover length $L_{_{NL}}$ and time $\tau_{_{NL}}$ 
are certainly much larger than the chiral crossover length $L_c$ and time $\tau_c$.

In any event, at small chirality there will certainly be an appreciable range of length and time scales, specifically $L_c\ll|\br-\br'| \ll L_{_{NL}}$,  $|t-t'|\ll\tau_{_{NL}}$, or $\tau_c\ll |t-t'| \ll \tau_{_{NL}}$, $|\br-\br'|\ll L_{_{NL}}$, over which the phase $\phi$ correlations can be evaluated based on  the KPZ equation with the non-linearity neglected.

It therefore behooves us to consider the behavior of our flock in this linear regime $L_c\ll L\ll L_{_{NL}}$. We do so in the next section.

\section{Phase and velocity correlations in the Linear KPZ regime \label{pc}}
We begin by noting that in the linear regime, we can drop the $\lambda_{_K}$ non-linearity in the KPZ equation. This turns the EOM for $\phi$ into the ``Edwards-Wilkinson" equation \cite{EA}:
\beqn
\pp_t\phi=\nu\nabla^2\phi+f_\phi\,.\label{EOM:phi_v1}
\eeqn
This EOM is identical to a purely relaxational model for an equilibrium XY model:
\beqn
\pp_t\phi=-\Gamma{\delta H\over\delta\phi}+f_\phi\,,\label{EOM:phi_v2}
\eeqn
with the Hamiltonian
\beq
H={K\over2}\int d^dr |\nabla\phi(\br)|^2
\label{Hxy}
\eeq
In this mapping, we have
\beq
\nu=\Gamma K \,.
\label{nuK)}
\eeq
\beq
D_\phi=\Gamma k_BT \,.
\label{FDT}
\eeq

After spatio-temporal Fourier-transforming Eq.~(\ref{EOM:phi_v1}) and solving it for $\phi$, we obtain 
%\beq
%-i\omega \phi(\bq,\omega)=-\nu q^2 \phi(\bq,\omega)+f(\bq, \omega) \, ,
%\label{XYlinFT}
%\eeq
\beq
\phi(\bq, \omega) ={f(\bq, \omega) \over -i\omega+\nu q^2} \,,
\label{phisol}
\eeq
where
\beqn
&&\phi(\bq,\omega)=\int {\dd t\over \sqrt{2\pi}}\int{\dd^2\br\over\sqrt{A}}\,\phi(\br,t)\ee^{-\ii(\omega t-\bq\cdot\br)}\,,~~~~~~~~\\
&&f_\phi(\bq,\omega)=\int {\dd t\over \sqrt{2\pi}}\int{\dd^2\br\over\sqrt{A}}\,f_\phi(\br,t)\ee^{-\ii(\omega t-\bq\cdot\br)}\,,
\nn\\
\eeqn
where $A$ is the area of the system.
Autocorrelating this with itself then implies
\bew
\beqn
\langle \phi(\bq,\omega)\phi(\bq',\omega')\rangle&=&{2D_\phi\delta^K_{\bq+\bq'}\delta(\omega+\omega') \over \omega^2+\nu^2 q^4}  \equiv C_\phi(\bq,\omega)\delta^K_{\bq+\bq^\prime}
%\bq'}
\delta(\omega+\omega^\prime)
%%a?
\,,
\label{phicorFT}
\eeqn
\ew
where we've used the fact that
\beq
\langle f_\phi(\bq,\omega)f_\phi(\bq',\omega') \rangle= {2D_\phi\delta^K_{\bq+\bq'}} \delta(\omega+\omega') \,,~~~~~
\eeq
which follows from \rf{fphicorr} upon Fourier transforming.  Here $\delta^K_\bp$ is a Kronecker delta, which is equal to $1$ when $\bp={\bm 0}$, and equal to $0$ when $\bp\ne{\bm 0}$.
 The implies that this real-space, real-time correlation function is given by
\bew
\beqn
C_\phi(\br,t)\equiv\langle[\phi(\br+\bR,t+T)-\phi(\bR,T)
]^2\rangle&=&2\int\,{d\omega\over 2\pi}\int\,{d^2q\over(2\pi)^2}C_\phi(\bq,\omega)\bigg(1-\cos[\omega t-\bq\cdot\br]\bigg)e^{-(qL_c)^2}\nonumber\\
&=&4D_\phi\int\,{d\omega\over 2\pi}\int\,{d^2q\over(2\pi)^2}\left[{\bigg(1-\cos[\omega (t
-t')-\bq\cdot\br]\bigg)\over\omega^2+\nu^2q^4}\right]e^{-(qL_c)^2}\label{phi_correl2_1} \,,\nn\\
\eeqn
\ew
where we
have introduced a Gaussian ultraviolet cutoff at wavevectors comparable to $L_c^{-1}$, where $L_c$ is the achiral to chiral crossover length given by equation (\ref{lcsc_v2}). Since, as we shall see, the form of our final result only depends logarithmically on the ultraviolet cutoff, the precise form of the cutoff we use is unimportant. More precisely, changing the cutoff  will only introduce changes of additive constants of $\cO(1)$ to $C_\phi(\br,t)$. 

The integral in \rf{phi_correl2_1} is evaluated in appendix \rf{cphi}; the result is 
\beq
C_\phi(\br,t)=\alpha\bigg[2\ln\left({r\over2L_c}\right)-\re\left(-{r^2\over4(\nu |t|+L_c^2)}\right)+\gamma\bigg] \,.
\label{phi_corr_fin_1}
\eeq
where we've used our earlier definition $\alpha\equiv{D_\phi\over2\pi\nu}$.
This recovers the result of \rf{phi_corre2I} of the introduction; the limiting behaviors obtained there follow straightforwardly from the large and small argument limits of the exponential integral function $\re\left(-{r^2\over4(\nu |t|+L_c^2)}\right)$.
 
We can use this result to calculate the velocity correlation function from the linear theory. In the linear theory, fluctuations  are Gaussian, so we can write 
\bew
\beq
\big\langle\exp\big\{\ii\left[\phi(\br+\bR,t+T)-\phi(\br, t) \right]\big\}\big\rangle=\exp\bigg\{-{1\over2}\bigg\langle\left[\bigg(\phi(\br +\bR,t+T)-\phi(\br, t)\bigg)^2\right]\bigg\rangle\bigg\} =\exp\left(-{1\over2} C_\phi(\bR, T)\right)\,,
\label{gaussian exp_1}
\eeq
\ew
where in the first equality we have used the well-known identity for zero-mean Gaussian random variables 
$\langle\ee^{\ii x}\rangle=\ee^{-\langle x^2\rangle/2}$.

Inserting \rf{phi_corr_fin_1} into  \rf{gaussian exp_1}, and using the asymptotic limiting behaviors of the exponential integral function for large and small argument,  we get
\bew
\beq
\big\langle\exp\big\{\ii\left[\phi(\br+\bR,t+T)-\phi(\br, t) \right]\big\}\big\rangle \approx \left\{
\begin{array}{ll}
 \ee^{-\alpha \gamma/2} 
%\sqrt{\ee^{\alpha\gamma}}
\left(|\br-\br'|\over  2 L_c\right)^{-\alpha}\,,
&\frac{|t-t'|}{\tau_c}\ll\left( \frac{|\br-\br'|}{L_c}\right)^2\,,
\\
\left(|t-t'|\over \tau_c\right)^{-{\alpha\over 2}}\,,
&\frac{|t-t'|}{\tau_c}\gg\left( \frac{|\br-\br'|}{L_c}\right)^2\,,
\,,
\end{array}
\right.
\label{Exponential_phi}
\eeq
\ew
Inserting \rf{Exponential_phi} into \rf{v_correl} leads to the velocity correlation \rf{v_corre2I} given in the introduction.

\vspace{.2in}

\section{the  Purely temporal Fourier transform of the velocity correlation function}{\label{F}}
The equal-time velocity correlations are readily measurable. However,  non-equal-time velocity correlations  are more subtle, since they carry a frequency $b$, i.e., the cosine factor which oscillates in the time difference $t-t'$ with period $2\pi/b$.
%Setting $\br'=\br$ in (\ref{v_correl_g}) we get
%\beqn
%\langle\bv(\br,t)\cdot\bv(\bm{\br},t')\rangle=v_0^2\exp\left[-\left({A_\phi\over2}\right)|t-t'|^{2\chi\over z}\right]\cos\left[b(t-t')\right]\,.\label{v_correl_ep}
%\eeqn
This  temporal oscillation  is reminiscent of the  spatial oscillations of density correlations in smectics\cite{deGennes' book}. Those correlations in smectics also  oscillate rapidly, in that case in {\it space}, with period $a$ (the layer spacing). In smectics, the signature of this spatially rapid oscillation is the appearance of ``Bragg peaks`` in the Fourier component of the  density correlation function at $\bq=(2 m \pi/ a)\hat{n}$ ($m=1,2,...$). These peaks can be observed experimentally via X-ray scattering\cite{deGennes' book}.

Like our 2D system in the linear regime, three-dimensional smectics also exhibit
logarithmically divergent phase fluctuations\cite{deGennes' book, Caille}. These manifest themselves in the shape of the X-ray scattering, which exhibit algebraic divergences as the peaks are approached\cite{deGennes' book, Caille}.

 Here we make a similar prediction for the temporal Fourier transform of the  velocity correlation function near the ``Bragg %peaks?
 peaks`` at frequencies $\omega=\pm b$ in frequency\cite{only 2}. This Fourier transform is given by
\bew
\beqn
I(\bR, \omega)={1\over 2\pi}\int_{-\infty}^{\infty}\langle\bv(\br+\bR,t+T)\cdot\bv(\bm{\br}, t) 
\rangle\,\ee^{\ii\omega 
T}\,\dd T \,.
\label{Bragg1}
\eeqn
\ew

Using the relation \rf{v_fluctuate} between the velocity $\bv(\br,t)$ and the phase variable $\phi(\br, t)$,  the velocity correlations can be rewritten in terms of the
phase correlations:
\bew
\beqn
\langle\bv(\br+\bR,t+T)\cdot\bv(\bm{\br}, t) \rangle
&=&v_0^2\cos(bT)
%t?
\big\langle\exp\big\{\ii\left[\phi(\br+\bR,t+T)-\phi(\br, t)
%\br',t')
\right]\big\}\big\rangle \,.
%\nonumber\\
%&=&v_0^2\cos\left[b(t-t')\right]\exp\bigg\{-{1\over2}\bigg\langle\left[\bigg(\phi(\br,t)-\phi(\br^\prime%\br?, t')\bigg)^2\right]\bigg\rangle\bigg\}\nonumber\\&=&v_0^2\cos\left[b(t-t')\right]\left(|\br-\br'|\over L_c\right)^{-\alpha}{  \exp\bigg[{\alpha\over2}{\rm Ei}\left(-{|\br-\br'|^2\over4(\nu |t-t^\prime|+L_c^{2})}\right)\bigg]}\,,
\label{v_corre4I}
\eeqn
\ew
Inserting this into our expression \rf{Bragg1} for the temporal Fourier transform of the  velocity correlation function gives
\bew
\beqn
I(\bR, \omega)&=&{1\over 2\pi}\int_{-\infty}^{\infty}v_0^2\cos(bT)
%\left[b(t-t^\prime)
%t?\right]\
\big\langle\exp\big\{\ii\left[\phi(\br+\bR,t+T)-\phi(\br, t)
\right]\big\}\big\rangle\,\ee^{\ii\omega T}\,\dd T
\nn\\
&=&
{v_0^2\over 4\pi}\int_{-\infty}^{\infty}
%\left[b(t-t')\right]
\big\langle\exp\big\{\ii\left[\phi(\br+\bR,t+T)-\phi(\br, t)
\right]\big\}\big\rangle\, (\ee^{\ii b T}+\ee^{-\ii b T})\ee^{\ii\omega T}\,\dd T
\nn\\
&=&
{v_0^2\over 4\pi}\int_{-\infty}^{\infty}\big\langle\exp\big\{\ii\left[\phi(\br+\bR,t+T)-\phi(\br, t)
\right]\big\}\big\rangle\,(\ee^{\ii\dw_+ T}+\ee^{\ii\dw_- T})\,\dd T 
\nn\\
&=&G(\bR, \dw_+)+G(\bR, \dw_-)
\,,
\label{Bragg2}
\eeqn
\ew
where we`ve defined 
\beq
\dw_\pm\equiv\omega\pm b 
\label{opmdef}
\eeq
and
\vspace{.1in}
\bew
\beq
G(\bR, \dw)\equiv{v_0^2\over 4\pi}\int_{-\infty}^{\infty}\big\langle\exp\big\{\ii\left[\phi(\br+\bR,t+T)-\phi(\br, t)
\right]\big\}\big\rangle\,\ee^{\ii\dw T}\,\dd T  \,.
\label{gdef}
\eeq
\ew
Note that  $\dw_+$ measures the distance in frequency from the peak at $\omega=b$, while $\dw_-$  similarly measures the distance in frequency from the peak at $\omega=-b$.

If there were no fluctuations in the phase $\phi$, the average 
\beq
\big\langle\exp\big\{\ii\left[\phi(\br+\bR,t+T)-\phi(\br, t) \right]\big\}\big\rangle=1 \,\,\, ,\  {\rm no} \,\, {\rm fluctuations.}
\label{eiphinofluc}
\eeq
Inserting this into  \rf{Bragg2} reveals that the temporal Fourier transform of the  velocity correlation function in the abscence of fluctuations would consist of two delta-function Bragg peaks, at $\omega=\pm b$:
\beqn
I(\bR, \omega)&=&{v_0^2\over 2}[\delta(\dw_+)+\delta(\dw_-)]
\nn\\
&=&{v_0^2\over 2}[\delta(\omega+b)+\delta(\omega-b)]\,,\  {\rm no} \, \ {\rm fluctuations.}\nn\\
\label{nofluc}
\eeqn

In the presence of fluctuations in $\phi$, these peaks broaden. In light of the fact that the fluctuations of $\phi$ are slow (since $\phi$ is a hydrodynamic variable), $G(\dw)$ will only be appreciable when $\dw$ is small. Hence, in light of \rf{Bragg2}, $I(\bR,\omega)$ will only be large when either $\dw_+$ or $\dw_-$ is small; that is, when $\omega$ is near either $+b$ or $-b$.  When $\omega$ is near $+b$, $I(\bR,\omega)$ will be dominated by the contribution from $G(\dw_-)$, since only $\dw_-$ is small in that range of $\omega$; likewise, $I(\omega)$ will be dominated by the contribution from $G(\dw_+)$
when $\omega$ is near $-b$.

Furthermore, the integral over the time difference $T$ in \rf{gdef} is 
dominated by $T\sim|\dw|^{-1}$. Therefore, for $|\dw|$ in the range $b^{-1}\gg|\dw|\gg\tau_{_{NL}}^{-1}$,
we can calculate the velocity correlation function in \rf{gdef} from the linear theory. In that linear theory, fluctuations  are Gaussian, so we can write 
\bew
\beq
\big\langle\exp\big\{\ii\left[\phi(\br+\bR,t+T)-\phi(\br, t) \right]\big\}\big\rangle=\exp\bigg\{-{1\over2}\bigg\langle\left[\phi(\br +\bR,t +T)-\phi(\br, t)\right]^2\bigg\rangle\bigg\} =\exp\left(-{1\over2} C_\phi(\bR, T)\right)\,,
\label{gaussian exp}
\eeq
\ew
where we have used the well-known identity for zero-mean Gaussian random variables 
$\langle\ee^{\ii x}\rangle=\ee^{-\langle x^2\rangle/2}$.

Using this  and \rf{phi_corr_fin_1}, we can rewrite our expression \rf{gdef} for $G(\bR, \dw)$ as
\beq
G(\bR,\dw)={v_0^2\over 4\pi}\left({R \ee^{\gamma/2} \over2L_c}\right)^{-\alpha} h(\bR, \dw) \,,\label{I1}
\eeq
where we`ve defined
\beq
h(\bR, \dw)\equiv\int_{-\infty}^\infty \,  \exp\bigg[\ii\dw T+{\alpha\over2}{\rm Ei}\left(-{R^2\over4\nu |T|+L_c^2}\right)\bigg] \, \dd T \,.
\label{hdef_1}
\eeq
The integral in \rf{hdef_1} is dominated by  the range $T\sim |\delta\omega|^{-1}$. Since $b^{-1}\gg|\delta\omega|\gg\tau_{_{NL}}^{-1}$ and $L_c^2\sim\nu b$, we can neglect the $L_c^2$ in the argument of the Ei function is negligible. Therefore, we   approximate \rf{hdef_1} as
\beq
h(\bR, \dw)\equiv\int_{-\infty}^\infty \,  \exp\bigg[\ii\dw T+{\alpha\over2}{\rm Ei}\left(-{R^2\over4\nu |T|}\right)\bigg] \, \dd T \,.
\label{hdef}
\eeq

Making the change of variables of integration from $T$ to $u$, where
\beq
u\equiv{4\nu T\over R^2}
\label{udef}
\eeq
 in \rf{hdef}, we can write $h(R, \dw)$ in a scaling form:
 \beq
 h(\bR, \dw)={R^2\over4\nu}F_h\left({\dw R^2\over4\nu}\right) \,,
 \label{hscale}
 \eeq
 where the scaling function $F_h(S)$ is given by
 \beq
 F_h(S)=\int_{-\infty}^\infty \,  \exp\bigg[\ii S u+{\alpha\over2}{\rm Ei}\left(-{1\over |u|}\right)\bigg] \, \dd u \,,
 %\nn\\
% &=& \int_0^\infty \,  \ee^{g(u)} \, \dd u+{\rm C.C.}
% \,,}
\label{Fhdef}
\eeq
% where
%\beq
%g(u)=\ii Su+{\alpha\over 2}\re\left(-{1\over u}\right)\,.
%\eeq
%We show in the appendix that the integral is invariant under the rotation of the contour by any angle $\theta_0$. That is,
%\beqn
% F_h(S) =\int_0^{\infty\ee^{\ii\theta_0}} \,  \ee^{g(u)} \, \dd u+{\rm C.C.}
% \,.
%\label{}
%\eeqn}

We`ll first derive the limiting behavior of this scaling function for  small argument $S$, which will give us the behavior of $I(R, \dw)$ for $\dw\ll{\nu\over R^2}$.

In this limit, the integral in \rf{Fhdef} is dominated by $|u|\sim1/|S|\gg1$. In this limit, $1/|u|\ll1$, so we can use the small argument limit of the exponential integral function ${\rm Ei}\left(-{1\over |u|}\right)\approx-\ln|u|+\gamma$. Inserting this into \rf{Fhdef}, we obtain
\beqn
F_h(S)&\approx&\int_{-\infty}^\infty \,  e^{\ii Su}|u|^{-\alpha/2}e^{\alpha\gamma/2} \, \dd u 
\nn\\
&=&\int_{0}^\infty \,  e^{\ii |S|u}|u|^{-\alpha/2}e^{\alpha\gamma/2} \, \dd u +{\rm complex} \,\, {\rm conjugate}
\nn\\
\label{smallS}
\eeqn

The integral in this expression is easily done by rotating the contour of integration over $u$ in the complex plane to the imaginary axis. The result is
\beq
F_h(S)\approx C_h|S|^{{\alpha\over2} -1} \,.
\label{smallSfin}
\eeq
\\
with the constant $C_h(\alpha)$ (which is ``constant" in the sense of being independent of $S$) given by
\beq
C_h(\alpha)=
2e^{\alpha\gamma\over2}\Gamma\left(1-{\alpha\over2}\right)
\sin\left(\alpha\pi\over 4\right) \,.
\label{ch}
\eeq

Inserting the result \rf{smallSfin} into \rf{hscale} for $h(\bR,\dw)$, and inserting {\it that} result into our expression \rf{I1}  for $G(\bR, \dw)$, 
we see that the separation $R$ between the two spatial points being correlated cancels, leaving us with 
\beq
G(\bR,\dw)={C_h v_0^2\over 4\pi}\left({L_c^\alpha|\dw|^{\alpha/2-1}\over(\nu  \ee^\gamma)^{\alpha/2}}\right)\sep 
|\dw|\ll{\nu/R^2} \,.
\label{If}
\eeq
The fact that this result is independent of the spatial separation $R$ in this limit is unsurprising. It is because the time scales being probed in this range of $\dw$ are 
$\gg R^2/\nu$, which is simply the time it takes distortions of the phase field $\phi$ to diffuse a distance $R$. Thus, for longer times, we expect the correlations between these two points to be the same as if they were at the {\it same} point in space. Hence, $R$ should drop out, as we just found it does.

Now recalling that  the temporally Fourier transformed correlation function is dominated by the $G(\bR, \dw_-)$ term for $\omega$ near $b$, and by the  
$G(\bR, \dw_+)$ term for $\omega$ near $-b$,
we see that 
\beq
I(\bR,\omega)\approx 
{C_h v_0^2\over 4\pi}\left({|\dw|^{\alpha/2-1}\over(\nu  L_c^2%\Lambda^2_ c}
\ee^\gamma)^{\alpha/2}}\right)\sep 
|\dw|\ll{\nu/R^2}
%C_h v_0^2\left({|\dw|^{\alpha/2-1}\over(\nu\Lambda^2)^{\alpha/2}}\right)\sep 
%|\dw|\ll{\nu/R^2} \,,
\label{If}
\eeq
with $\dw$ the smaller of $\dw_\pm$; i.e., the distance in frequency from the nearer of the Bragg peak positions.

%\vspace{.2in}
In the opposite limit $|S|\gg1$, which will determine the behavior of $I(\bR, \omega)$ for 
$|\dw|\gg{\nu\over R^2}$, the integral in \rf{Fhdef} is dominated by $|u|\sim1/|S|\ll1$, which means $1/|u|\gg1$. Hence, we can use the large argument approximation to the exponential integral function ${\rm Ei}\left(-{1\over |u|}\right)\approx-|u|\exp\left[-{1\over|u|}\right]$. 

Since $|u|$, and, therefore, $e^{-1/|u|}$, is small, ${\rm Ei}\left(-{1\over |u|}\right)$ is also small, and so we can expand
\beq
{\rm exp}\left[-\frac{\alpha }{2}|u| {\rm exp} \left(-\frac{1}{|u|} \right)\right]\approx 1-\frac{\alpha }{2}|u| {\rm exp} \left(-\frac{1}{|u|} \right)
\eeq
The Fourier transform of the first term is zero (strictly speaking, it`s $\delta(S)$, but that of course is zero for $S\ne0$, which is what we're considering here). So $F_h(S)$ is dominated by the Fourier transform of the second term, which gives
\bew
\beq
F_h(S)\approx-\int_{-\infty}^\infty \,  \ee^{\ii Su}\frac{\alpha }{2}|u| {\rm exp} \left(-\frac{1}{|u|} \right)
 \, \dd u = - \frac{\alpha }{2}\int_{-\infty}^\infty \,  \ee^{\ii Su+\ln(|u|)-1/|u| } \, \dd u\,.
\label{bigS2}
\eeq
\ew

This equation can be rewritten
\beqn
F_h(S)\approx- \frac{\alpha}{2}\bigg(\int_{0}^\infty \,  e^{\Phi(u)} \, \dd u \,  +{\rm complex} \,\, {\rm conjugate}\bigg)
\nn\\
\label{bigS2}
\eeqn
with 
\beq
\Phi(u)=\ii |S|u+\ln(u)-1/u \,.
\label{phidef}
\eeq

The integral in \rf{bigS2} can be evaluated using the method of steepest descents. Details are given in appendix \ref{Steep_Descent_I}. The result is 
\beqn
\int_{0}^\infty \,  e^{\Phi(u)} \, \dd u\approx {\sqrt{\pi}\over |S|^{5/4}} \exp\left[2e^{{3\ii\pi\over4}}\sqrt{|S|} +{\ii 5\pi\over8}\right] \,.
\nn\\
\label{intfin}
\eeqn
Inserting this into \rf{bigS2}, we obtain 
\beq
F_h(S)\approx{\alpha\sqrt{\pi}\over |S|^{5/4}}\sin\left(\sqrt{2|S|} +{\pi\over8}\right)e^{-\sqrt{2S}} \,.
\label{bigSfin}
\eeq
This decays rapidly with the scaling variable $S$; hence, the  temporally Fourier transformed correlation function $I(\bR, \omega)$  decays rapidly for  $|\dw|\gg\nu/R^2$, where by $\dw$ we mean, as before, the distance of the frequency $\omega$ from the nearest Bragg peak at $\pm b$.

Now we turn to the region extremely close to the peaks; that is, 
$|\delta\omega|\ll\tau_{_{NL}}^{-1}$. On these time scales, the distribution of the field $\phi$ is no longer Gaussian. Therefore, even though we know the behavior of the mean squared phase difference $\bigg\langle\left[\bigg(\phi(\br,t)-\phi(\br+\bR, t+T)\bigg)^2\right]\bigg\rangle =C_\phi(\bR, T)$, we cannot use \rf{gaussian exp}
to determine the behavior of  $\big\langle\exp\big\{\ii\left[\phi(\br+\bR,t+T)-\phi(\br, t) \right]\big\}\big\rangle$, since \rf{gaussian exp} only holds for Gaussian distributions. 

However, it seems reasonable to assume that the much bigger fluctuations in $\phi$ that occur in the non-linear regime of the KPZ equation imply that the correlation function
$\big\langle\exp\big\{\ii\left[\phi(\br+\bR,t+T)-\phi(\br, t) \right]\big\}\big\rangle$ decays more rapidly in the non-linear regime than it does in {\it any} linear Edwards-Wilkinson model. Since $\big\langle\exp\big\{\ii\left[\phi(\br+\bR,t+T)-\phi(\br, t) \right]\big\}\big\rangle$  decays like $|T|^{-\alpha/2}$ in the limit $|T|\to \infty$ in the  Edwards-Wilkinson model (see \rf{Exponential_phi} from the linear section), and in principle $\alpha$ can be made arbitrarily large, this would imply that  the correlation function
$\big\langle\exp\big\{\ii\left[\phi(\br+\bR,t+T)-\phi(\br, t) \right]\big\}\big\rangle$ decays more rapidly than any power law in the non-linear regime.

If this is correct, then, even when $\dw=0$, $G(\bR, \dw=0)$ is finite, since the integral over all time of $\big\langle\exp\big\{\ii\left[\phi(\br+\bR,t+T)-\phi(\br, t) \right]\big\}\big\rangle$ will be finite.

Hence, the  temporal Fourier transform of the  velocity correlation function 
$I(\bR, \omega)$ remains finite even right {\it at} the Bragg peaks at $\omega=\pm b$. 
This implies that the $|\dw|^{\alpha/2-1}$ divergence of $I(\bR, \omega)$ must round off for $|\dw|\lesssim\tau_{_{NL}}^{-1}$.

The resultant behavior of $I(\bR, \omega)$ is summarized in Fig. \ref{iw}. Note that the oscillating and exponentially decaying tail implied by equation \rf{bigSfin} is too small to be visible in this figure.

\section{Density correlations in the Linear KPZ regime \label{DC}}

We can now use equation \rf{rhocorKPZ1} to obtain the density correlations  in the linear regime. We will focus on the equal-time correlation functions for large spatial separations; specifically $|\br-\br'| \gg L_c$. In this range, the exponential integral in equation \rf{phi_corr_fin_1} is negligible (since its argument is large), and so we have
\beq
C_\phi(\br-\br',t=0)={D_\phi\over\pi\nu}\bigg[\ln\left({|\br-\br'|\over2L_c}\right)+\gamma\bigg] \,.
\label{phi_corr_et}
\eeq
This implies that
 the first term in \rf{rhocorKPZ1} is short ranged:
\beqn
&&e_3^2\nabla^2{\nabla'}^2\langle\phi(\br,t)\phi(\br',t)\rangle\nn\\
&=&-2e_3^2\nabla^2{\nabla'}^2\langle[\phi(\br,t)-\phi(\br',t)]^2\rangle\nn\\
&=& -2e_3^2 \left(D_\phi\over\pi\nu\right)\nabla^2{\nabla'}^2\ln\left({|\br-\br^\prime|\over2L_c}\right)\nn\\
%&=&-4\pi e_3^2 \left(D_\phi\over\pi\nu\right)\nabla^2\delta(\br-\br^\prime)\nn\\
&=&-\left(4e_3^2D_\phi\over\nu\right)\nabla^2\delta(\br-\br^\prime)\,.
%\propto |\br-\br'|^{-4}\,.\label{rho_corre2}
\eeqn
The second term in \rf{rhocorKPZ1} vanishes in the linear regime, since it is odd in $\phi$, while the distribution of $\phi$ is even (indeed, a zero-mean Gaussian) in the linear regime.

However, the remaining  term in the our expression  \rf{rhocorKPZ1} for $\dr$ gives a  long-ranged $|\br-\br'|^{-4}$ contribution to $\langle\dr(\br,t)\dr(\br^\prime,t)\rangle$, as
we`ll now show.

%\beqn\langle\dr(\br,t)\dr(\br^\prime,t)\rangle=e_4^2\langle |\nabla\phi(\br,t)|^2|\nabla\phi(\br^\prime,t)|^2\rangle\eeqn
Fourier transforming, we have 
\bew
\beqn
\langle |\nabla\phi(\br,t)|^2|\nabla\phi(\br^\prime,t)|^2\rangle={1\over A^2}\sum_{\bp_{_{1,2,3,4}}}e^{i(\bp_1+\bp_2)\cdot(\br-\br^\prime)}(\bp_1\cdot\bp_2) (\bp_3\cdot\bp_4)\langle\phi(\bp_1,t)\phi(\bp_2,t)\phi(\bp_3,t)\phi(\bp_4,t)\rangle\,.
\label{4pt}
\eeqn
\ew
Since the distribution of $\phi$ is Gaussian in the linear regime, we can
Wick decompose the four-point average in \rf{4pt}, obtaining
\bew
\beqn
\langle\phi(\bp_1,t)\phi(\bp_2,t)\phi(\bp_3,t)\phi(\bp_4,t)\rangle
=&&\langle\phi(\bp_1,t)\phi(\bp_2,t)\rangle\langle\phi(\bp_3,t)\phi(\bp_4,t)\rangle+\langle\phi(\bp_1,t)\phi(\bp_3,t)\rangle\langle\phi(\bp_2,t)\phi(\bp_4,t)\rangle\nn\\
&&+\langle\phi(\bp_1,t)\phi(\bp_4,t)\rangle\langle\phi(\bp_2,t)\phi(\bp_3,t)\rangle\label{wick}
\eeqn
\ew
We'll now use our earlier result for the two-point correlations of $\phi$ in the linear theory, which can be written
\beqn
\langle\phi(\bp_1,t)\phi(\bp_3,t)\rangle={D_\phi\delta^K_{\bp_1+\bp_3}\over\nu p_1^2} \,.
\eeqn
Using this in \rf{wick}, and using the result in \rf{4pt}, we get, after doing two sums using the Kronecker delta`s, and some relabelling,
\bew
\beqn
\langle |\nabla\phi(\br,t)|^2|\nabla\phi(\br^\prime,t)|^2\rangle={D_\phi^2\over \nu^2V^2}\sum_{\bp_1,\bp_2}\bigg(1+2{(\bp_1\cdot\bp_2)^2e^{i(\bp_1+\bp_2)\cdot(\br-\br^\prime)}\over p_1^2p_2^2}\bigg)\,.
\label{4pt2}
\eeqn
\ew

We see that the first term
in \rf{4pt2} just gives a constant. This constant is just $\langle\dr\rangle^2$, which is non-zero because, in the  presence of fluctuations, the mean value of the density is changed by the fluctuations, and is no longer the value at which $g(\rho, ...)$ vanishes.
Because it is constant, this piece of the density-density correlation function is trivial, and, indeed, will drop out of the connected density-density correlation function. In other words, if we define $\delta\rho(\br,t)$ to be the departure of $\rho(\br,t)$ from its true mean value, rather than (as we have up to now) defining it as the departure of $\rho(\br,t)$ from the value at which $g(\rho, ...)$ vanishes, then only the last two terms of \rf{4pt2} will contribute to $\langle\dr(\br,t)\dr(\br^\prime,t)\rangle$.

The last two terms depend on $\br-\br^\prime$, and survive in the connected correlation function. They are equal, and lead to the following expression for the connected $|\nabla\phi|^2$ correlation function $\langle |\nabla\phi(\br,t)|^2|\nabla\phi(\br^\prime,t)|^2\rangle_c$:
\bew
\beqn
\langle |\nabla\phi(\br,t)|^2|\nabla\phi(\br^\prime,t)|^2\rangle_c&=&2\left({D_\phi\over\nu}\right)^2\sum_{\bp_1,\bp_2} {(\bp_1\cdot\bp_2)^2e^{\ii(\bp_1+
\bp_2)\cdot(\br-\br^\prime)}\over V^2p_1^2 p_2^2}
\nn\\
&=&2\left({D_\phi\over\nu}\right)^2
\left(\sum_{\bp_1} {p_{1i}p_{1j}e^{\ii[\bp_1\cdot(\br-\br^\prime)]}\over V p_1^2 }\right)
\left(\sum_{\bp_2} {p_{2i}p_{2j}e^{-\ii[\bp_2\cdot(\br-\br^\prime)]}\over V p_2^2 }\right)
\nn\\
&\equiv&
2\left({D_\phi\over\nu}\right)^2\sigma_{ij}(\bR)\sigma^*_{ij}(\bR)
\label{gradphisig}
\eeqn
\ew
where
we`ve defined $\bR\equiv\br-\br^\prime$, and
\bew
\beqn
\sigma_{ij}(\bR)\equiv\left(\sum_{\bp} {p_{i}p_{j}e^{\ii\bp\cdot\bR}\over V p^2 }
\right)={\pp\over \pp R_i}{\pp\over \pp R_j}\left(\sum_{\bp} {(1-e^{\ii\bp\cdot\bR})\over V p^2 }\right)=
{\pp\over \pp R_i}{\pp\over \pp R_j} \left(\int {d^2p\over(2\pi)^2} ~ {(1-e^{\ii\bp\cdot\bR})\over  p^2 }\right)
\eeqn
\ew
where the $1$ has been added to the second and third equalities simply to make the integrals over $\bp$ converge as $\bp\to{\bf 0}$. (Its contribution to $\sigma_{ij}$ clearly vanishes once the derivatives are taken).  We have also introduced an  ultraviolet cutoff
$\Lambda_c\sim L_c^{-1}$ to avoid having to worry about any ultraviolet divergences.

Performing the integral over $\bp$ in the last equality in polar coordinates, we obtain, after the angular integration,
\beqn
\int {d^2p\over(2\pi)^2} ~ {(1-e^{\ii\bp\cdot\bR})\over  p^2}={1\over2\pi}\int_0^{\Lambda_c} {dp\over p}\bigg[ 1-J_0(pR)\bigg] \,,\nn\\
\label{Bess1}
\eeqn
where $J_0(x)$ is the zeroth-order Bessel function. Hence
\bew
\beqn
{\pp\over \pp R_j} \left(\int {d^2p\over(2\pi)^2} ~ {(1-e^{\ii\bp\cdot\bR})\over  p^2 }\right)
&=&{\pp\over \pp R_j}{1\over2\pi}\int_0^{\Lambda_c} {dp\over p}\bigg[ 1-J_0(pR)\bigg]=-\left({R_j\over R}\right)\bigg({1\over2\pi}\int_0^{\Lambda_c} dp ~J_0^\prime(pR)\bigg)
\nn\\
&=&\left({R_j\over 2\pi R^2}\right)(J_0(0)-J_0(\Lambda_c R))\approx{R_j\over 2\pi R^2} \,,
\label{1deriv}
\eeqn
\ew
where in the final approximate equality we have used the fact that $J_0(x\to\infty)\to0$ to drop $J_0(\Lambda_c R)$ in the large $R$ limit (specifically, the $R\gg\Lambda_c^{-1}$limit).
We've also used the fact that $J_0(0)=1$.

From \rf{1deriv}, it follows that
\beqn
\sigma_{ij}(\bR)={R^2\delta_{ij}-2R_iR_j\over 2\pi R^4}
\eeqn
which readily implies
\beq
\sigma_{ij}(\bR)\sigma_{ij}^*(\bR)={1\over2\pi^2R^4} \,.
\label{sig2}
\eeq
Using this in  \rf{rhocorKPZ1} and \rf{gradphisig}, we have
\beqn
\langle\dr(\br,t)\dr(\br^\prime,t)\rangle={e_4^2D_\phi^2\over\pi^2\nu^2|\br-\br^\prime|^4} \,.
\label{rho_corr_lin_ET}
\eeqn

Performing a similar calculation for the equal-position case, we find
\beqn
\langle\dr(\br,t)\dr(\br,t')\rangle={e_4^2D_\phi^2\over 16\pi^2\nu^4|t-t^\prime|^2}\left(1-\ee^{-\nu\Lambda_c^2|t-t'|}\right)\,. \nn\\
%\approx{e_4^2D_\phi^2\over 16\pi^2\nu^4|t-t^\prime|^2}\,.
\label{rho_corr_lin_ep1}
\eeqn
Since we are now in the regime $|t-t'|\gg\tau_c$, and since $\tau_c\gg \nu\Lambda^2$ since $\tau\sim b^{-1}$, $\nu\sim b^{-3/5}$, and $\Lambda_c\sim L_c^{-1}\sim b^{4/5}$, the exponential in (\ref{rho_corr_lin_ep1}) is much less than 1 and hence negligible. Then we have
\beqn
\langle\dr(\br,t)\dr(\br,t')\rangle={e_4^2D_\phi^2\over 16\pi^2\nu^4|t-t^\prime|^2}\,.
\label{rho_corr_lin_ED}
\eeqn

\vspace{.2in}
\section{Vortices}{\label{v}}

If we assume that the analogy between the linearized KPZ equation and the XY model holds for vortices as well as at the level of ``spin-wave theory", then the recursion relation for the vortex fugacity $y$ reads \cite{KT}:
\beq
{\dd y\over \dd\ell}=2\left(1-{1\over4\alpha}\right)y \,,
\label{yrr}
\eeq
where $\alpha$ is defined in (\ref{aI}).  Since $\alpha$ vanishes like $b^{\eta_\nu}$  with $\eta_\nu\approx3/5$  for small chirality $b$, the second
$\left({1\over4\alpha}\right)$ term in (\ref{yrr}) will dominate in that limit.
Furthermore,
%} Now,} 
$\alpha$ is essentially constant for $\ell<\ell^*$, where
\beq
\ls\equiv\ln\left({\lnl\over L_c%\lch
}\right)= C_{_{NL}}b^{-{\eta_g}} 
\label{ell*}
\eeq
is the RG ``time" at which the $\lk$
non-linearity becomes important and we have used (\ref{lnl}) in the second equality.
Hence, solving (\ref{yrr}) gives
\beq
y(\ls)=y_0\exp\left[2\ls\left(1-{1\over4\alpha}\right)\right]\ .
\label{yls}
\eeq
For small chirality $b$, the argument of this exponential is dominated by the ${1\over4\alpha}$ term, which scales with $b$ like ${\ls\over\alpha}\propto b^{-\eta_y}$, where we`ve defined  
\beq
\eta_y\equiv\eta_g+\eta_\nu\approx22/5 \,,
\label{etay}
\eeq
 a result obtained by using  our earlier results $\alpha=C_\alpha b^{\eta_\nu}$ and $\ls= C_{_{NL}}b^{-\eta_g}$.

Thus we see that
\beq
y(\ls)\propto\exp\left(-E b^{-\eta_y}\right) \,,
\label{ylsscale}
\eeq
with $E={C_{_{NL}}\over2C_\alpha}$ an $O(1)$ constant.
%\beq
%\eta_y=\eta_g+\eta_\nu\approx22/5 \,.
%\label{etay}
%\eeq 
This fugacity $y(\ls)$  will be extremely small for small chirality $b$.

With this result  (\ref{ylsscale}) in hand, we can now ask for the length scale $\lv$ at which vortices actually become important. There are two equivalent ways to calculate this, which (reassuringly!) give the same result.

The first approach  is to say that the ``free" vortex density $n_f$  at length scales longer that $L_{_{NL}}$ is proportional to $y(\ls)$. One then estimates $\lv$ as the typical distance  between vortices of density $n_f$ per unit area, which is, of course, simply
\beq
\lv\approx n_f^{-1/2}\propto\exp\left(C_{vL}b^{-\eta_y}\right) \,.
\label{lv}
\eeq
where
\beqn
C_{vL}\equiv{E\over 2}={C_{_{NL}}\over 4C_\alpha}\,.
\eeqn

The second approach is to continue to use the recursion relation (\ref{yrr}) for RG times
$\ell>\ls$, but assume that $\alpha$ has now been driven to infinity by the non-linear term. Hence that recursion relation reduces to
\beq
{\dd y\over \dd\ell}=2y \,.
\label{yrr2}
\eeq
whose solution is trivially
\beq
y(\ell)=y(\ls)e^{2(\ell-\ls)} \,,
\label{ysolbigl}
\eeq
where we have matched this solution onto the solution \rf{ylsscale} that we found earlier for
$\ell<\ls$. We can now ask for the RG time $\ell_v$ at which this renormalized fugacity becomes $O(1)$. This is readily found to be
\beq
\ell_v=\ls - {\ln\left[y(\ls)\right]\over 2}\approx C_{vL}b^{-\eta_y} \,,
\label{ellv}
\eeq
where in the last, approximate equality, we have used the fact that $\left({\ln(y(\ls))\over2}\right)\approx C_{vL}b^{-\eta_y}\gg\ls\propto b^{-\eta_g}$, since $\eta_y-\eta_g=\eta_\nu>0$.

The length scale $L_v$ at which vortices become important is simply proportional to $e^{\ell_v}$; this recovers (\ref{lv}).

Note that this value  (\ref{lv}) of $\lv$ is, for small chirality $b$, much greater than the non-linear length scale $L_{_{NL}}$. Hence, there will be a large range of length scales over which the results
%\rf{phi_corre2},
we derive based on the KPZ equation in the linear and non-linear regimes will hold, before the unbinding of vortices destroys all remnants of order.

\section{Summary}{\label{s}}

We have shown that   the hydrodynamic behavior of a generic planar chiral Malthusian flock  in the rotating phase can be mapped onto the Kardar-Parisi-Zhang surface growth model in (2+1) dimensions. Specifically, the fluctuating angular field in the chiral flocks corresponds to the height function in the KPZ model. Surprisingly, although {\it achiral} Malthusian flocks {\it can} order in two dimensions, {\it  any} degree of chirality will destroy the  long-ranged orientational order. We then focus on the small chirality limit and elucidate the distinct regimes that depend on the length  and time scales of interest (Fig.~\ref{ltsc}).
Our work thus highlights the intricate interplay between chirality and motility.
Looking ahead, one interesting direction would be to relate our findings here to multi-component systems with nonreciprocal interactions where a chiral phase can also be present \cite{fruchart_nature21}.

\section{Acknowledgements}{\label{Acknowledgements}
 We thank the Max Planck Institute for the Physics of Complex Systems, where the early stage of this work was performed, for their support. L.C. acknowledges support by the National Science Foundation
of China (under Grants No.12274452). 

We also thank Tim Halpin-Healy and Mehran Kardar for useful discussions about the current state of our understanding of the KPZ equation, and for calling references \cite{kpzexp4}  and \cite{kpzexp5} to our attention. We also thank Ananyo Maitra for useful discussions about prior work on chirality in Malthusian flocks (see "Note added" below).  

\section{Note added} After this work was completed, we learned that the conclusion that chiral malthusian flocks are described by the KPZ equation had been independently reached by Ananyo Maitra\cite{scoop}, whom we thank  for calling this to our attention.

\appendix

\section{ Demonstration that the most general model still leads to the KPZ equation}{\label{genmod}

In this appendix, we demonstrate that our fundamental result, which is that the dynamics of a chiral Malthusian flock is described by the KPZ equation, is not an artifact of the simplifications we made to our  truncated dynamical EOM \rf{EOM:chiral_1}, but in fact holds even if all relevant terms in the EOM are included.

We will do this by considering all possible terms in  the EOM involving one and two derivatives of the velocity fields, and showing that all such terms either average to zero when averaged over one cycle, or lead to  terms of the same form as the two $\phi$-dependent terms in the KPZ equation; that is, either  $\nabla^2\phi$ or $|\nabla\phi|^2$. Terms involving more than two spatial derivatives are clearly irrelevant in the long-wavelength limit relative to these two terms, and so can be ignored.

\subsection{One-derivative terms}

\subsubsection{achiral one derivative terms}
The most general {\it achiral}  one-derivative term in the EOM (that is, the expression for $\pp_tv_i$)
can be written\cite{product rule} as
\beq
\lambda_\alpha^ n T_{ijkl...mnp}v_jv_k.v_l...\pp_ nv_p \,,
\label{1dgen}
\eeq
where the $2n$ index tensor ${\bf T}$ (with $n$ an integer) is composed of the product of $n$ Kronecker delta`s, which contract all but one of the indices in the product $v_jv_kv_l....\pp_ nv_p$ of velocities and the derivative of the  velocity, so as to leave a single free index $i$ to match the free index in $\pp_tv_i$.

Hence, the total number of indices in $v_jv_kv_l....\pp_ nv_p$ must be $2n+1$, and so the total number of velocities to the left of the derivative in \rf{1dgen} must be $2n-1$. Most of these must be contracted with each other, rather than with the indices $n$ and $p$ of the derivative of the velocity. Indeed, the only possibilities are that there is one velocity on the left ``left over``, with all of the others contracted with others on the left, or three velocities ``left over``.

We'll now consider these cases in turn.

\noindent One velocity ``left over". There are three possibilities here:

\noindent 1)   the remaining velocity component on the left is contracted with the $n$  of the $\pp_nv_p$, 

\noindent 2) the remaining velocity component on the left is contracted with the  $p$ of the $\pp_nv_p$, 

\noindent 3) the remaining velocity component on the left is the ``free" index $i$. 

In case 1), the remaining index ``$p$" must be $i$, and
 \rf{1dgen} reduces to
\beq
\lambda_\alpha^n v_0^{2n-2}v_j \pp_jv_i= \lambda_\alpha^nnv_0^{2n-2}(\bv\cdot\nabla\bv)_i \,.
\label{1vlcase1}
\eeq

Since $v_0$ is a constant, the sum of all such possible terms is simply
\beq
\lambda_1(\bv\cdot\nabla\bv)_i \,;
\eeq
with $\lambda_1=\sum_{n=1}^\infty \lambda^n_\alpha v_0^n2n-2$.
This is exactly the $\lambda_1$ term we had in our EOM \rf{EOM:chiral_1}.

In case 2), we have \beq
\lambda_\alpha^n v_0^{2n-2}v_j \pp_iv_j ={1\over2}\lambda_\alpha^n  v_0^{2n-2} \pp_i(v_jv_j)
%={1/2}\lambda_\alpha  v_0^{2n-2} \pp_i(v_0^2)
=0 \,,
\label{1vlcase2}
\eeq
So all such terms vanish identically.

Finally, for case 3), we have 
\beq
\lambda_\alpha^n v_0^{2n-2}v_i \pp_j v_j=\lambda_\alpha^n  v_0^{2n-2}[\bv (\nabla\cdot\bv)]_i
\eeq

All such terms are along the velocity. Hence, when we project perpendicular to the velocity by acting on the EOM with $\epsilon_{ik}v_k$, they vanish.

We now turn to the terms in which three of the velocities to the left of the derivative are left to be contracted with the derivative of the velocity term. We first note that if any of the three indices to the left contract with the index of the velocity to the right, we will get a term proportional to
\beq
v_j\pp_wv_j=\pp_w(v_jv_j)=
%{{1/2}\lambda_\alpha  v_0^{2n-2}}} 
\pp_w(v_0^2)=0 \,,
\label{vcont}
\eeq
where $w$ is some index, and it doesn't
matter which. Hence all of these terms vanish as well.

The only other possibility is that one of the three velocities to the left of the derivative contracts with the derivative, and the other two contract with each other. It is easy to see that this simply leads to another contribution to the $\lambda_1$ term equation \rf{EOM:chiral_1}.

So all possible achiral first derivative terms can be absorbed into the $\lambda_1$ term that we already had in equation \rf{EOM:chiral_1}.

\subsubsection{chiral one derivative  terms}

The chiral one-derivative terms can be reduced to terms involving only one $\bm{\epsilon}$ matrix since the product of two of them annihilate into Kronecker deltas:
\beqn
\epsilon_{jk}\epsilon_{\ell m} = \delta_{j\ell}\delta_{km}-\delta_{jm}\delta_{k\ell}\,.
\label{eps^2}
\eeqn
In addition,  any such  term must have an odd number of indices, so that after contraction one index remains to be $i$. So in general, the chiral one-derivative term is a combination of one matrix $\bm{\epsilon}$, one derivative, and even number of $v$ components. Again, once we act on such a term with the  ``projection operator`` $\left(\epsilon_{ik}v_k\over v_0\right)$  the resultant piece vanishes upon averaging over short time oscillation cycles of the rotating $\bv$ since it involves an odd number of $v$ components.

In conclusion, the only  one-derivative  term in the EOM for $\bv$ is the $\lambda_1(\bv\cdot\nabla\bv)$ term which we had in our original EOM \rf{EOM:chiral_1}. As we showed in our analysis in the main text, section \rf{kpz}, this term makes no contribution to the EOM for the phase field $\phi$. This is not a coincidence, but determined by the symmetry. Since the 2D chiral state itself is isotropic in the large scale time limit, there is no way to create one-derivative linear terms in the hydrodynamic equation for $\phi$  without violating this isotropy.

\subsection{terms with two derivatives}

\subsubsection{achiral two derivative terms}

There are two types of such terms:
\beq
\lambda^{(2.1)}_\alpha T^{(2.1)}_{ijkl...mnps}v_nv_pv_s...(\pp_ jv_k)(\pp_lv_m) \,,
\label{2dgen1}
\eeq
and
\beq
\lambda^{(2.2)}_\alpha T^{(2.2)}_{ijkl...mnps}v_nv_pv_s...(\pp_ j\pp_lv_m) \,,
\label{2dgen2}
\eeq
where, as before the tensors ${\bf T}^{(2.1)}$ and ${\bf T}^{(2.2)}$ are products of Kronecker deltas that contract the indices of the velocities and partial derivatives in pairs.

%\subsubsection{
We`ll consider first the $\lambda^{(2.1)}_\alpha$ terms.

After contracting all the velocity components to the left of the derivatives that must be contracted with each other, we are left only with terms of the form
\beq
\gamma^{(2.1)}_\alpha M^{(2.1)}_{ijklmnps}v_nv_pv_s(\pp_ jv_k)(\pp_lv_m) \,,
\label{2dgen3}
\eeq
where the coefficient $\gamma^{(2.1)}_\alpha$ absorbs all of the powers of $v_0^2$ that arose from contracting velocity components with themselves, and the  tensor ${\bf M}^{(2.1)}$ is again a product of Kronecker deltas which contract the free indices.

Now consider the possibilities for those contractions. If any of the indices $n$, $p$, or $s$ of the velocity components outside the derivatives contracts with either of the velocity components $k$ or $m$ inside the derivatives, then this term vanishes, by equation
\rf{vcont}. Hence, the velocities outside the derivatives can only contract either with the derivatives, or each other. This leaves two possibilities:

\noindent1) two of the outside velocities contract with the derivatives, and the remaining one is the free index $i$ (remember that we are looking at a term in the expression for $\pp_tv_i$). In this case, when we ``project" this term orthogonal to $\bv$ by acting on it with $\epsilon_{is} v_s$, it will vanish.

\noindent2) two of the outside indices contract with each other, and the remaining index contracts with one of the derivatives. This leaves three possibilities:
\beq
v_j(\pp_jv_i)(\pp_lv_l) \sep {\rm (1.2.1)} \,,
\label{1.2.1}
\eeq

\beq
v_k(\pp_kv_j)(\pp_jv_i) \sep {\rm (1.2.2)}  \,,
\label{1.2.2}
\eeq
and
\beq
v_k(\pp_kv_j)(\pp_iv_j) \sep {\rm (1.2.3)} \,.
\label{1.2.3}
\eeq

Consider first (1.2.1). Using
\beq
\pp_jv_i=\epsilon_{ik}v_k \pp_j\phi
\label{vderiv}
\eeq
we can rewrite that term as
\beq
v_j(\pp_jv_i)(\pp_lv_l)=\epsilon_{ik}\epsilon_{lm}v_jv_kv_m(\pp_j\phi)(\pp_l\phi) \,.
\label{1.2.1.1}
\eeq
Using the identity \rf{eps^2} for the product of two ${\bf \epsilon}$ matrices, this becomes
\bew
\beqn
v_j(\pp_jv_i)(\pp_lv_l)=(\delta_{il}\delta_{km}-\delta_{im}\delta_{kl})v_jv_kv_m(\pp_j\phi)(\pp_l\phi) =v_0^2v_j(\pp_j\phi)(\pp_i\phi)-v_iv_jv_l(\pp_j\phi)(\pp_l\phi)\,.
\label{1.2.1.2}
\eeqn
\ew
The second term
is proportional to $v_i$, and so it will vanish when we  act on it with $\epsilon_{is} v_s$. Acting on the first term with
$\epsilon_{is} v_s$
gives
\beq
\epsilon_{is} v_sv_j(\pp_jv_i)(\pp_lv_l)=v_0^2\epsilon_{is}v_jv_s(\pp_j\phi)(\pp_i\phi) \,.
\label{1.2.1.3}
\eeq
Taking the average of this over one cycle using (\ref{angaves}) gives
\bew
\beq
\langle\epsilon_{is} v_sv_j(\pp_jv_i)(\pp_lv_l)\rangle_c={v_0^4\over2}\epsilon_{is}\delta_{js}(\pp_j\phi)(\pp_i\phi)={v_0^4\over2}\epsilon_{ij}(\pp_j\phi)(\pp_i\phi)=0 \,,
\label{1.2.1.4}
\eeq
\ew
where the last equality holds because $\epsilon_{ij}$ is anti-symmetric under interchange of $i$ and $j$, while the product $(\pp_j\phi)(\pp_i\phi)$ is symmetric.

So (1.2.1) makes no contribution to the EOM for $\phi$.

Now we turn to (1.2.2). Using \rf{vderiv} again, this becomes
\beq
v_k(\pp_kv_j)(\pp_jv_i) =\epsilon_{im}\epsilon_{jl}v_kv_lv_m(\pp_j\phi)(\pp_k\phi)
%=(\delta_{ij}\delta_{ml}-\delta_{il}\delta_{mj})v_kv_lv_m(\pp_j\phi)(\pp_k\phi)\,,
\label{1.2.2.1}
\eeq
which is readily seen to be identical to (1.2.1) (see (\ref{1.2.1.1})).

Finally, (1.2.3) gives, upon using \rf{vderiv},
\bew
\beq
v_k(\pp_kv_j)(\pp_iv_j)=\epsilon_{jl}\epsilon_{jm}v_kv_lv_m(\pp_i\phi)(\pp_k\phi)
=\delta_{lm}v_kv_lv_m(\pp_i\phi)(\pp_k\phi)=v_0^2v_k(\pp_i\phi)(\pp_k\phi) \,,
\label{1.2.3.1}
\eeq
\ew
where in the second equality we have used the identity (\ref{eps^2}) for the ${\beps}$ tensor.

Multiplying this term by $\eps_{is}v_s$ and avergaing over one cycle using (\ref{angaves}) gives
\bew
\beq
\langle\eps_{is}v_sv_k(\pp_kv_j)(\pp_iv_j)\rangle_c={v_0^4\over2}\eps_{is}\delta_{sk}(\pp_i\phi)(\pp_k\phi)= {v_0^4\over2}\eps_{ik}(\pp_i\phi)(\pp_k\phi)=0\,,
\label{1.2.3.2}
\eeq
\ew
where the final equality again follows because we are contracting a symmetric and an anti-symmetric tensor.

Now we turn to the final achiral two derivative terms, \rf{2dgen2}. Since we 
cannot have any of the velocities outside the derivatives carrying the index $i$,  for the reasons discussed above, there are only four possibilities here:

\noindent1) All velocities outside the derivatives contract with each other, and the two derivatives contract with each other, leaving the differentiated velocity to have index $i$. This gives us a term $v_0^{2n} \pp_j\pp_jv_i$, which is proportional to the $\mu_1\nabla^2v_i$ term we had in our original EOM (\ref{EOM:chiral_1}).

\noindent2) All velocities outside the derivatives contract with each other, and 
%{the two derivatives contract with each other,}} 
one of the derivative indices contracts with the differentiated velocity, and the other is $i$. This gives us a term proportional to the $\mu_2\pp_i\nabla\cdot\bv$ term we had in our original EOM (\ref{EOM:chiral_1}).

\noindent3) All undifferentiated velocities but two contract with each other, and the two remaining undifferentiated velocities contract with the derivatives, leaving the differentiated velocity index to be $i$. This gives us the $\mu_3(\bv\cdot\nabla)^2v_i$ term in our original EOM (\ref{EOM:chiral_1}).

\noindent4) All undifferentiated velocities but two contract with each other, and one of the two remaining undifferentiated velocities contracts with the differentiated velocity, and the other contracts with one of the derivatives, leaving the remaining derivative index to be $i$. This gives $v_jv_k\pp_i\pp_jv_k$, which is equivalent to (\ref{1.2.3}) since
\bew
\beqn
v_jv_k\pp_i\pp_jv_k=v_j\pp_i(v_k\pp_jv_k)-v_j(\pp_iv_k)(\pp_jv_k)=-v_j(\pp_iv_k)(\pp_jv_k)\,,
\eeqn
\ew
where in the second equality we have used $v_j\pp_i(v_k\pp_jv_k)={1\over 2}v_j(\pp_i\pp_jv_0^2)=0$.

So our original EOM (\ref{EOM:chiral_1}) contained all possible achiral two derivative terms.

\subsubsection{chiral two derivative terms}

As for the one-derivative terms,  chiral two-derivative terms can  be reduced to terms involving only one $\bm{\epsilon}$ matrix since the product of two of them annihilate into Kronecker deltas, as illustrated by  \rf{eps^2}.
These again fall into two classes:

\beq
\lambda^{(2.3)}_\alpha T^{(2.3)}_{ijkl...mnptw}v_jv_pv_s...\eps_{tw}(\pp_kv_l)(\pp_mv_n) \,,
\label{2dgen3}
\eeq
and
\beq
\lambda^{(2.4)}_\alpha T^{(2.4)}_{ijkl...mnptw}v_nv_pv_s...\eps_{tw}(\pp_ k\pp_lv_m) \,,
\label{2dgen4}
\eeq

Class \rf{2dgen3} can be considerably simplified by recognizing that only one of the non-differentiated $v$'s can contract with the $\beps$ tensor; if two contract, the term vanishes because that's an odd-even contraction. We also know from our earlier discussion that this term vanishes if any of the non-differentiated
$v$'s contracts with one of the {\it differentiated} $v$'s.
Therefore, there are only three available indices in the $\eps_{tw}(\pp_kv_l)(\pp_mv_n)$ factor with which the non-differentiated velocities
can contract. Therefore, either all but three of them, or all but one of them, must contract
with each other.

Consider first the terms  in \rf{2dgen3} with three ``outside" velocities that do not contract with other outside velocities. These are of the form 
\beq
\lambda^{(2.3)}_\alpha M^{(2.3.a)}_{imns}v_jv_kv_l\eps_{jm}(\pp_kv_n)(\pp_lv_s) \,,
\label{2.3.a.1}
\eeq
\beq
\nn
\eeq
where ${\bf M}^{(2.3.a)}$ is a product of two Kronecker deltas, which contract two of the three indices $m$, $n$, and $s$, and turn the remaining one into the free index $i$.
Using \rf{vderiv}, we can write
\bew
\beqn
v_jv_kv_l\eps_{jm}(\pp_kv_n)(\pp_lv_s)&=&\eps_{jm}\eps_{nt}\eps_{sw}v_jv_kv_lv_tv_w(\pp_k\phi)(\pp_l\phi)=(\delta_{ns}\delta_{tw}-\delta_{nw}\delta_{st})v_jv_kv_lv_tv_w\eps_{jm}(\pp_k\phi)(\pp_l\phi)
%(\pp_k\phi)(\pp_l\phi)
\nn\\
&=&(v_0^2\delta_{ns}-v_nv_s) v_jv_kv_l \eps_{jm}(\pp_k\phi)(\pp_l\phi) \,,
\label{2.3.a.2}
\eeqn
\ew
where in the second equality we have used the identity (\ref{eps^2}) for the product $\eps_{nt}\eps_{sw}$.

Now multiplying this term \rf{2.3.a.1} by $\eps_{i\alpha}v_\alpha$ to obtain the contribution to $\pp_t\phi$, we get
\bew
\beqn
&&\eps_{i\alpha}v_\alpha\eps_{jm}(\pp_k\phi)(\pp_l\phi)(v_0^2\delta_{ns}-v_nv_s) v_jv_kv_l
\lambda^{(2.3)}_\alpha M^{(2.3.a)}_{imns}
\nn\\
&=&(\delta_{ij}\delta_{\alpha m}-\delta_{im}\delta_{\alpha j})
(\pp_k\phi)(\pp_l\phi)(v_0^2\delta_{ns}-v_nv_s) v_jv_kv_lv_\alpha \lambda^{(2.3)}_\alpha
M^{(2.3.a)}_{imns}
\label{2.3.a.3}
\eeqn
\ew
where in the second equality we have used the identity (\ref{eps^2}) for the product $\eps_{i\alpha}\eps_{jm}$.

The Kronecker deltas explicitly diplayed in \rf{2.3.a.3} , and those implicit in
$M^{(2.3.a)}_{imns}$,
can only contract the indices on the velocity components with themselves; the only other indices are $k$ and $l$, which are already contracted. Thus this term (if not vanishing) is ultimately proportional to
\beq
v_kv_l(\pp_k\phi)(\pp_l\phi) \,.
\label{2.3.a.4}
\eeq
Averaging this over one cycle using equation (\ref{angaves}) gives
\beq
{v_0^2\over2}\delta_{kl}(\pp_k\phi)(\pp_l\phi)={v_0^2\over2}|\nabla\phi|^2 \,,
\label{2.3.a.5}
\eeq
which is a contribution to the KPZ non-linearity $\lambda_{_K}$ term in (\ref{EOM:phi}).

We now turn to terms with only one non-differentiated velocity
that's not contracted with other undifferentiated velocities. These terms are of the form 
\beq
% {M_{ijkmns}\eps_{jk}v_l(\pp_lv_m)(\pp_nv_s)}
M_{ijkplmns}\eps_{jk}v_p(\pp_lv_m)(\pp_nv_s)  \,,
\label{2.3.b.1}
\eeq
where again the ${\bf M}$ tensor is composed of Kronecker deltas that contract the indices.

Using our old standby \rf{vderiv}, we now get
\bew
\beqn
%
%{\eps_{jk}v_l(\pp_lv_m)(\pp_nv_s)=\eps_{jk}\eps_{mt}\eps_{sw}v_lv_tv_w(\pp_n\phi)(\pp_l\phi)=\eps_{jk}(\delta_{ms}\delta_{tw}-\delta_{mw}\delta_{st})v_lv_tv_w(\pp_n\phi)(\pp_l\phi) \,,}}\\
\eps_{jk}v_p(\pp_lv_m)(\pp_nv_s)=\eps_{jk}\eps_{mt}\eps_{sw}v_pv_tv_w(\pp_n\phi)(\pp_l\phi)=\eps_{jk}(\delta_{ms}\delta_{tw}-\delta_{mw}\delta_{st})v_pv_tv_w(\pp_n\phi)(\pp_l\phi) \,,
\label{2.3.b.2}
\eeqn
\ew
where in the second equality we have used the identity  (\ref{eps^2}) for the product $\eps_{mt}\eps_{sw}$.

Now multiplying this term \rf{2.3.a.1} by $\eps_{i\alpha}v_\alpha$ to obtain the contribution to $\pp_t\phi$, we get
\bew
\beqn
\eps_{i\alpha}v_\alpha\eps_{jk}v_p(\pp_lv_m)(\pp_nv_s)&=&\eps_{i\alpha}\eps_{jk}(\delta_{ms}\delta_{tw}-\delta_{mw}\delta_{st})v_\alpha v_pv_tv_w(\pp_n\phi)(\pp_l\phi)
\nn\\
&=& (\delta_{ij}\delta_{\alpha k}-\delta_{ik}\delta_{\alpha j})(\delta_{ms}\delta_{tw}-\delta_{mw}\delta_{st})v_\alpha v_pv_tv_w(\pp_n\phi)(\pp_l\phi)
\label{2.3.b.3}
\eeqn
\ew
where in the second equality we have used the identity  (\ref{eps^2}) for the product $\eps_{i\alpha}\eps_{jk}$.

When we act on this with 
%{$M_{ijkmns}$ 
$M_{ijkplmns}$, if the result does not vanish, there are two possibilities. %{the only} 
One possibility is that two of the velocities $v_\alpha v_pv_tv_w$ contract  with the two  $\pp$`s, and  the other two contract with each other. Thus we are left with a term proportional to
\beq
v_kv_l(\pp_k\phi)(\pp_l\phi) \,.
\label{2.3.b.4}
\eeq
We considered this term earlier (see (\ref{2.3.a.4})), and showed it contributes only to the $\lambda_{_K}$ term in (\ref{EOM:phi}).
So that is the form of the contribution from \rf{2.3.b.1} as well. The other possibility is that all the velocities $v_\alpha v_pv_tv_w$ contract with each other, and the two $\pp$`s contract with each other. Then we get a term proportional to $|\nabla\phi|^2$, which again  contributes only to the $\lambda_{_K}$ term in (\ref{EOM:phi}).

Finally, we turn to terms of type \rf{2dgen4}. Again only one of the non-differentiated $v$`s can contract with the $\beps$ tensor. Furthermore, the total number of undifferentiated $v$'s in \rf{2dgen4} must be even; if not, the piece will have odd number of $v$'s after multiplying $\epsilon_{i\alpha}v_\alpha$, and its time average over one cycle will vanish. Therefore, either all but four of non-differentiated $v$`s, or all but two of them, or all of them, must contract with each other.

First consider the case in which all of the non-differentiated $v$`s contract with each other. In this case  \rf{2dgen4} is proportional to
\beqn
M_{ijklmp}\eps_{jp}(\pp_ k\pp_lv_m)\,.\label{new1}
\eeqn

Using \rf{vderiv} twice,  we have
\bew
\beqn
\pp_k\pp_l v_m&=&\eps_{mn}\pp_k(v_n\pp_l\phi)=\eps_{mn}\bigg[(\pp_kv_n)\pp_l\phi+v_n\pp_k\pp_l\phi\bigg]=\eps_{mn}\bigg[\eps_{ns}v_s(\pp_k\phi)\pp_l\phi+v_n\pp_k\pp_l\phi\bigg]
\nn\\
&=&-\delta_{ms}v_s(\pp_k\phi)\pp_l\phi+\eps_{mn}v_n\pp_k\pp_l\phi=-v_m(\pp_k\phi)\pp_l\phi+\eps_{mn}v_n\pp_k\pp_l\phi
\label{2.4.1}
\eeqn
\ew
where we have also used the identity $\eps_{mn}\eps_{ns}=-\delta_{ms}$.

Inserting \rf{2.4.1} into \rf{new1} and multiplying the resultant piece by $\eps_{i\alpha}v_\alpha$ to obtain the contribution to $\pp_t\phi$, we get
\bew
\beqn
&&M_{ijklmp}\epsilon_{jp}\bigg(-\eps_{i\alpha}v_\alpha v_m(\pp_k\phi)\pp_l\phi+\eps_{i\alpha}\eps_{mn}v_\alpha v_n\pp_k\pp_l\phi\bigg)\nn\\
&=&M_{ijklmp}\bigg(-(\delta_{ij}\delta_{p\alpha}-\delta_{ip}\delta_{j\alpha})v_\alpha v_m(\pp_k\phi)\pp_l\phi+\epsilon_{jp}(\delta_{im}\delta_{\alpha n}-\delta_{in}\delta_{\alpha m})v_\alpha v_n\pp_k\pp_l\phi\bigg)\nn\\
&=&M_{ijklmp}\bigg(-(\delta_{ij}\delta_{p\alpha}-\delta_{ip}\delta_{j\alpha})v_\alpha v_m(\pp_k\phi)\pp_l\phi+\epsilon_{jp}(\delta_{im}v_0^2-v_iv_m)\pp_k\pp_l\phi\bigg)\,.
\label{new2}
\eeqn
\ew
Next we take the time average of this using (\ref{angaves}):
\bew
\beqn
&&\bigg\langle M_{ijklmp}\epsilon_{jp}\bigg(-\eps_{i\alpha}v_\alpha v_m(\pp_k\phi)\pp_l\phi+\eps_{i\alpha}\eps_{mn}v_\alpha v_n\pp_k\pp_l\phi\bigg)\bigg\rangle_c\nn\\
%&=&M_{ijklmp}\bigg(-(\delta_{ij}\delta_{p\alpha}-\delta_{ip}\delta_{j\alpha})v_\alpha v_m(\pp_k\phi)\pp_l\phi+\epsilon_{jp}(\delta_{im}\delta_{\alpha n}-\delta_{in}\delta_{\alpha m})v_\alpha v_n\pp_k\pp_l\phi\bigg)\nn\\
&=&M_{ijklmp}\bigg(-{1\over 2}
(\delta_{ij}\delta_{p\alpha}-\delta_{ip}\delta_{j\alpha})v_0^2\delta_{\alpha m}(\pp_k\phi)\pp_l\phi+\epsilon_{jp}\left(\delta_{im}v_0^2-{v_0^2\over 2}\delta_{im}\right)\pp_k\pp_l\phi\bigg)
\nn\\
&=&\left(v_0^2\over 2\right)M_{ijklmp}\bigg((-\delta_{ij}\delta_{pm}+\delta_{ip}\delta_{jm})(\pp_k\phi)\pp_l\phi+\epsilon_{jp}\delta_{im}\pp_k\pp_l\phi\bigg)\,.
\label{new3}
\eeqn
\ew
For the first and the second term on the last line of (\ref{new3}), the explicitly displayed Kronecker deltas and those implicitly in the tensor ${\mathbf M}$ must ultimately contract the indices $k$ and $l$, giving a piece proportional to $|\nabla  \phi|^2$, which is a contribution to the KPZ non-linearity $\lambda_{_K}$ in (\ref{EOM:phi}).
%Likewise, the second term gives ${v_0^2\over 2}|\nabla u|^2$, canceling the previous $-{v_0^2\over 2}|\nabla u|^2$.
For the third term, the Kronecker deltas contract the indices $j$, $p$, $k$, and $l$. There are two possibilities: one is that $j$ contracts with $p$ and $k$ contracts   with $l$, and the other is that $j$ and $p$ contract   with $k$ and $j$. Both of these vanish, the former because 
$\epsilon_{jj}=0$, the second because $\epsilon_{kl}\pp_k\pp_l\phi=0$.

Next consider the case in which all but two of the non-differentiated $v$`s in \rf{2dgen4} contract with each other. In this case \rf{2dgen4} is proportional to
\beqn
M_{ijklmpqw}\eps_{jp}v_qv_w(\pp_ k\pp_lv_m)\,.\label{new4}
\eeqn

Inserting \rf{2.4.1} into \rf{new4} and multiplying the resultant piece by $\eps_{i\alpha}v_\alpha$ to obtain the contribution to $\pp_t\phi$, we get
\bew
\beqn
&&M_{ijklmpqw}\epsilon_{jp}\bigg(-\eps_{i\alpha}v_\alpha v_qv_wv_m(\pp_k\phi)\pp_l\phi+\eps_{i\alpha}\eps_{mn}v_\alpha v_qv_wv_n\pp_k\pp_l\phi\bigg)\nn\\
&=&M_{ijklmpqw}\bigg(-(\delta_{ij}\delta_{p\alpha}-\delta_{ip}\delta_{j\alpha})v_\alpha v_qv_wv_m(\pp_k\phi)\pp_l\phi+\epsilon_{jp}(\delta_{im}\delta_{\alpha n}-\delta_{in}\delta_{\alpha m})v_\alpha v_qv_wv_n\pp_k\pp_l\phi\bigg)\nn\\
&=&M_{ijklmpqw}\bigg(-(\delta_{ij}\delta_{p\alpha}-\delta_{ip}\delta_{j\alpha})v_\alpha v_qv_wv_m(\pp_k\phi)\pp_l\phi+\epsilon_{jp}(\delta_{im}v_0^2-v_iv_m)v_qv_w\pp_k\pp_l\phi\bigg)\,.
\label{new5}
\eeqn
\ew

Again taking the  time average using (\ref{angaves}), we obtain:

\bew
\beqn
&&\bigg\langle M_{ijklmpqw}\epsilon_{jp}\bigg(-\eps_{i\alpha}v_\alpha v_qv_wv_m(\pp_k\phi)\pp_l\phi+\eps_{i\alpha}\eps_{mn}v_\alpha v_qv_wv_n\pp_k\pp_l\phi\bigg)\bigg\rangle_c\nn\\
&=&\left(v_0^4\over 8\right)M_{ijklmpqw}\bigg(-(\delta_{ij}\delta_{p\alpha}-\delta_{ip}\delta_{j\alpha})
(\delta_{\alpha q}\delta_{wm}+\delta_{\alpha w}\delta_{qm}+\delta_{\alpha m}\delta_{qw})(\pp_k\phi)\pp_l\phi\nonumber\\
&&+\epsilon_{jp}\left(3\delta_{im}\delta_{qw}-\delta_{iq}\delta_{mw}-\delta_{iw}\delta_{mq}\right)\pp_k\pp_l\phi\bigg)\,.
\label{new6}
\eeqn
\ew

The argument used on (\ref{new3}) also applies here. The conclusion is that if (\ref{new6}) gives something non-vanishing, it can only be a contribution to the KPZ non-linearity $\lambda_{_K}$ term in (\ref{EOM:phi})

Finally consider    the case in which all but four of the non-differentiated $v$`s in \rf{2dgen4} contract with each other. Since only one of the non-differentiated $v$`s can contract with the $\beps$ tensor, the other three must contract the two derivatives and the differentiated $v$, leaving one if index of  $\beps$ to be $i$. Therefore, in this case \rf{2dgen4} is proportional to
\beqn
\eps_{ij}v_jv_kv_lv_m(\pp_ k\pp_lv_m)\,.\label{new7}
\eeqn

Inserting \rf{2.4.1} into \rf{new7} and multiplying the resultant piece by $\eps_{i\alpha}v_\alpha$ to obtain the contribution to $\pp_t\phi$, we get
\bew
\beqn
\eps_{i\alpha}v_\alpha\eps_{ij}v_jv_kv_lv_m\Big(-v_m(\pp_k\phi)\pp_l\phi+\eps_{mn}v_n\pp_k\pp_l\phi\Big)
=-v_0^2\eps_{i\alpha}v_\alpha\eps_{ij}v_jv_kv_l(\pp_k\phi)\pp_l\phi
=-v_0^4v_kv_l(\pp_k\phi)\pp_l\phi\,.\label{new8}
\eeqn
\ew
We considered this term earlier (see (\ref{2.3.a.4})), and showed it contributes only to the $\lambda_{_K}$ term in (\ref{EOM:phi}).

Thus, we have exhaustively shown that all possible symmetry allowed terms in the velocity EOM for a chiral flock lead to one of the two terms in the KPZ equation for $\phi$. Furthermore, the non-linear $\lambda_{_K}$ term comes only from the chiral terms in the velocity equation, which implies that $\lambda_{_K}$ is linear in the chirality for small chirality, a result we needed to obtain the scaling laws relating the crossover length scales $L_c$, $L_{_{NL}}$, and $L_v$, and time scales $\tau_c$, $\tau_{_{NL}}$, and $\tau_v$ to the chirality $b$ for small chirality.

\section{Non-racemic small ``chirality`` systems}{\label{nonrac}

Our discussion of the small chirality case throughout this paper has imagined the chirality is tuned to zero in a racemic way (that is, by mixing chiral components that are mirror images of each other), which leads to all chiral parameters vanishing at the same time. However, as discussed in the introduction, when chirality is tuned by, e.g., making non-racemic mixtures (i.e., mixtures of different components which are not mirror images of each other), one would expect in general that the rotation rate $b$ and the chiral non-linear KPZ term $\lambda_{_K}$ would vanish at different concentrations. Therefore, in such systems, two possibilities arise:

\noindent1) $\lambda_{_K}\to0$ while $b$ remains finite, and

\noindent2) $\lambda_{_K}$ remains finite while $b\to0$. 

We`ll now discuss the behavior of each of these cases in turn.

\subsection{When $\lambda_{_K}\to 0$ with $b$ remaining finite}

In this case, the large value of $b$ makes the crossover to chiral behavior happen almost immediately; that is, at small length and time scales. Hence, there is no substantial renormalization of the diffusion constants  in the achiral regime, because that regime has essentially disappeared. Thus, the KPZ diffusion constant $\nu$ has no strong dependence on, e.g., the departure $\delta c\equiv c-c_*$ of the concentration of components of the competing chirality from the value $c_*$ at which $\lambda_{_K}$ vanishes. One can, of course, replace ``concentration'' in the definition of this departure with whatever parameter one is tuning to drive $\lambda_{_K}$ towards zero.

Hence, we expect the dimensionless KPZ coupling 
\beq
g\equiv
{\lambda_{_K}^2D_\phi\over 2\pi\nu^3} \,,
\label{gdefapp}
\eeq
which  provides a measure of the importance of  non-linear effects in the KPZ equation \cite{KPZ}, to acquire all of its dependence on $\delta c$ from $\lambda_{_K}$, which we would expect to vanish linearly with $\delta c$; i.e., 
\beq
\lambda_{_K}\propto\delta c \,.
\label{ldc}
\eeq
Using this in \rf{gdefapp}, and taking $\nu$ and $D_\phi$ to be independent of $\delta c$, we predict that the bare value $g_0$ of $g$ vanishes quadratically with $\delta c \,$:
\beq
g\propto\left(\delta c\right)^2 \,.
\label{gscaleapp}
\eeq
Using our result (\ref{ellnl}) relating $g_0$ to the length scale $L_{_{NL}}$, at which non-linear effects become important, we thereby obtain
\beq
L_{_{NL}}=L_c\exp\left[{C_n\over\left(\delta c\right)^2}\right] \,.
\label{Lcscale1}
\eeq
The argument of the exponential in this expression should be contrasted with the $b^{-\eta_g}$, with $\eta_g\approx19/5$, scaling of the corresponding argument in equation  (\ref{lnl}). 

\subsection{When $b\to0$ with $\lambda_{_K}$ remaining finite}

In this case, we still get the fluctuation induced enhancement of the diffusivity $\nu\propto b^{-\eta_\nu}$ given by 
equation  (\ref{nub}). But now, because the non-linearity $\lambda_{_K}$ remains finite as $b\to0$, the scaling law relating the bare value $g_0$ of $g$ to $b$ now changes to
\beq
g_0\propto\nu^{-3}\propto b^{3\eta_\nu}
\label{gscaleapp2}
\eeq
with the exponent $3\eta_\nu\approx9/5$, 
which in turn leads, after again using our result (\ref{ellnl}) relating the value of the RG ``time" at which non-linear effects become important to $g_0$, to
\beq
L_{_{NL}}=L_c\exp\left[{C_n\over b^{3\eta_\nu}}\right] \,.
\label{Lcscale2}
\eeq

\section{Evaluation of $C_\phi(\br,t)$}{\label{cphi}

The integral over $\omega$ in this expression can be done by elementary complex contour techniques. The result is
\bew
\beqn
C_\phi(\br,t)&=&{2D_\phi\over\nu}\int\,{\dd^2q\over(2\pi)^2}\bigg({1-\ee^{-\nu q^2|t|}(\ee^{\ii\bq\cdot\br}+\ee^{-\ii\bq\cdot\br})/2\over q^2}\bigg)\ee^{-(qL_c)^2} \,.
\label{phi_correl3}
\eeqn
\ew
Differentiating this with respect to $|t|$ gives
\beqn
{\pp C_\phi(\br,t)\over\pp |t|}&=&D_\phi\int\,{\dd^2q\over(2\pi)^2}\bigg(\ee^{\ii\bq\cdot\br-\nu q^2|t|}+\ee^{-\ii\bq\cdot\br-\nu q^2|t|}\bigg)\nn\\
&&\times \ee^{-(qL_c)^2} \,.
\label{dtc}
\eeqn
Both integrals in this expression are now simple Gaussian integrals. Evaluating them gives 
\beq
{\pp C_\phi(\br,t)\over\pp |t|}={D_\phi\over2\pi(\nu|t|+L_c^2)}\exp\left[-\left({r^2\over4(\nu|t|+L_c^2)}\right)\right] \,.
\label{dtc2}
\eeq
Integrating this expression gives
\bew
\beq
C_\phi(\br,t)=\int_0^{|t|}{D_\phi\over2\pi(\nu t^\prime+L_c^2)}\exp\left[-\left({r^2\over4(\nu t^\prime+L_c^2)}\right)\right]\dd t^\prime + C_\phi(\br,t=0) \,.
\label{phicorrel4}
\eeq
\ew
Changing variable of integration from $t^\prime$ to $u\equiv{r^2\over4(\nu t^\prime+L_c^2)}$ in the integral over $t^\prime$ gives
%\bew
\beqn
C_\phi(\br,t)&=&{D_\phi\over2\pi\nu}\int_{r^2\over4(\nu |t|+L_c^2)}^{{r^2\over4L_c^2}} \left({e^{-u}\over u}\right)\dd u + C_\phi(\br,t=0) 
\nn\\
&=&{D_\phi\over2\pi\nu}\bigg[{\rm Ei}\left(-{{r^2\over4L_c^2}}\right)-{\rm Ei}\left(-\left({r^2\over4(\nu |t|+L_c^2)}\right)\right)\bigg]\nn\\
&&+ C_\phi(\br,t=0)\,,
\label{phicorrel5}
\eeqn
%\ew
where ${\rm Ei}(x)$ is the Exponential Integral Function defined as
\beqn
{\rm Ei}(x)\equiv -\int_{-x}^\infty\,\left(\ee^{-u}\over u\right)\dd u\,.
\eeqn

We have thus reduced the problem to finding the equal-time correlation function.
This can be obtained from \rf{phi_correl3} by setting $t=0$, which gives
\bew
\beqn
C_\phi(\br,t=0)&=&{2D_\phi\over\nu}\int\,{\dd^2q\over(2\pi)^2}\bigg[{1-(\ee^{\ii\bq\cdot\br}+\ee^{-\ii\bq\cdot\br})/2\over q^2}\bigg]\ee^{-(L_cq)^2} \,.
\label{phi_corr_et_0}
\eeqn
\ew
Defining a related function of two variables $\br$ and $k$ via
\beqn
F(\br, k)&=&{2D_\phi\over\nu}\int\,{\dd^2q\over(2\pi)^2}\bigg[{1-(\ee^{\ii\bq\cdot\br}+\ee^{-\ii\bq\cdot\br})/2\over q^2}\bigg]\ee^{-kq^2} \,,\nn\\
\label{Fdef}
\eeqn
we see that
\beq
C_\phi(\br,t=0)=F(\br, k=L_c^2) \,.
\label{fccon}
\eeq
We can evaluate $F(\br, k)$ by first differentiating \rf{Fdef} with respect to $k$, which
gives
\beqn
{\pp F(\br,k)\over\pp k}&=&-{D_\phi\over2\pi^2\nu}\int\,\dd^2q\bigg[1-(\ee^{\ii\bq\cdot\br}+\ee^{-\ii\bq\cdot\br})/2\bigg]\ee^{-kq^2} \,.\nn\\
\label{dphi_corr_et}
\eeqn
Evaluating the  Gaussian integrals in \rf{dphi_corr_et}
gives
\beqn
{\pp F(\br,k)\over\pp k}&=&{D_\phi\over2\pi\nu}\left({\ee^{-{r^2\over4k}}-1\over k}\right)\,.
\label{dphi_corr_et_eval}
\eeqn
Integrating this from $k=L_c^2$ to $\infty$, and noting that $F(\br, k=\infty)=0$ for all $\br$, and using \rf{fccon} to relate the result to $C_\phi(\br,t)$, we get
\beqn
C_\phi(\br,t=0)&=&{D_\phi\over2\pi\nu}\lim_{W\to\infty}\int_{L_c^2}^{W^2}\left[{1\over k}-{\exp\left({-{r^2\over4k}}\right)\over k}\right]\,\dd k \,.\nn\\
\label{phi_corr_et_int}
\eeqn
The integral of the first term in parentheses is, of course, elementary, while the second can be evaluated with the change of variable of integration from $k$ to  $u\equiv{r^2\over4k}$. The result is
\bew
\beqn
\int_{L_c^2}^{W^2}\left({1\over k}-{\exp\left[{-{r^2\over4k}}\right]\over k}\right)\,\dd k &=&\ln\left({W^2\over L_c^2}\right)+\re\left(-{r^2\over4W^2}\right)-\re\left(-{r^2\over4L_c^2}\right) \,.
\label{et_int}
\eeqn
\ew
Using the well-known \cite{GR} limiting behavior of the exponential integral function for small argument $-u$
\beq
-\re(-u)\approx\ln\left({1\over u}\right)-\gamma 
\label{eilim}
\eeq
and taking the limit $W\to\infty$ in (\ref{et_int}),
we see that the dependence on $W$ cancels in this limit, and we get our final result for $C_\phi(\br,t=0)$:
\beqn
C_\phi(\br,t=0)&=&{D_\phi\over2\pi\nu}\bigg[2\ln\left({r\over2L_c}\right)-\re\left(-{r^2\over4L_c^2}\right)+\gamma\bigg] \,.\nn\\
\label{phi_corr_et_fin}
\eeqn
Inserting this result in our expression (\ref{phicorrel5}) for the full, position and time dependent correlation function gives
\beqn
C_\phi(\br,t)&=&{D_\phi\over2\pi\nu}\bigg[2\ln\left({r\over2L_c}\right)-\re\left(-{r^2\over4(\nu |t|+L_c^2)}\right)+\gamma\bigg] \,,\nn\\
\label{phi_corr_fin}
\eeqn
which is equation (\ref{phi_corre2I}) of the introduction.

\section{Steepest Descents calculation of $I(R, \omega)$ for  $\dw\gg\nu/R^2$\label{Steep_Descent_I}}

The scaling function $F_h(S)$  in \rf{Fhdef} is given by
\beq
F_h(S)=\int_0^{\infty}\,  \ee^{g(u)} \, \dd u +\rm{C. \,C.}
\label{F.1}
\eeq
where
\beq
g(u)=\ii |S|u+{\alpha\over 2}\re\left(-{1\over u}\right)\,.
\label{Def_g}
\eeq

Now consider the closed contour (see Fig. \ref{Contour}) and the integral
\beqn
I=\oint\,  \ee^{g(u)} \, \dd u \,.
\label{Closed_Contour}
\eeqn

\begin{figure}
		\begin{center}
	\includegraphics[scale=.25]{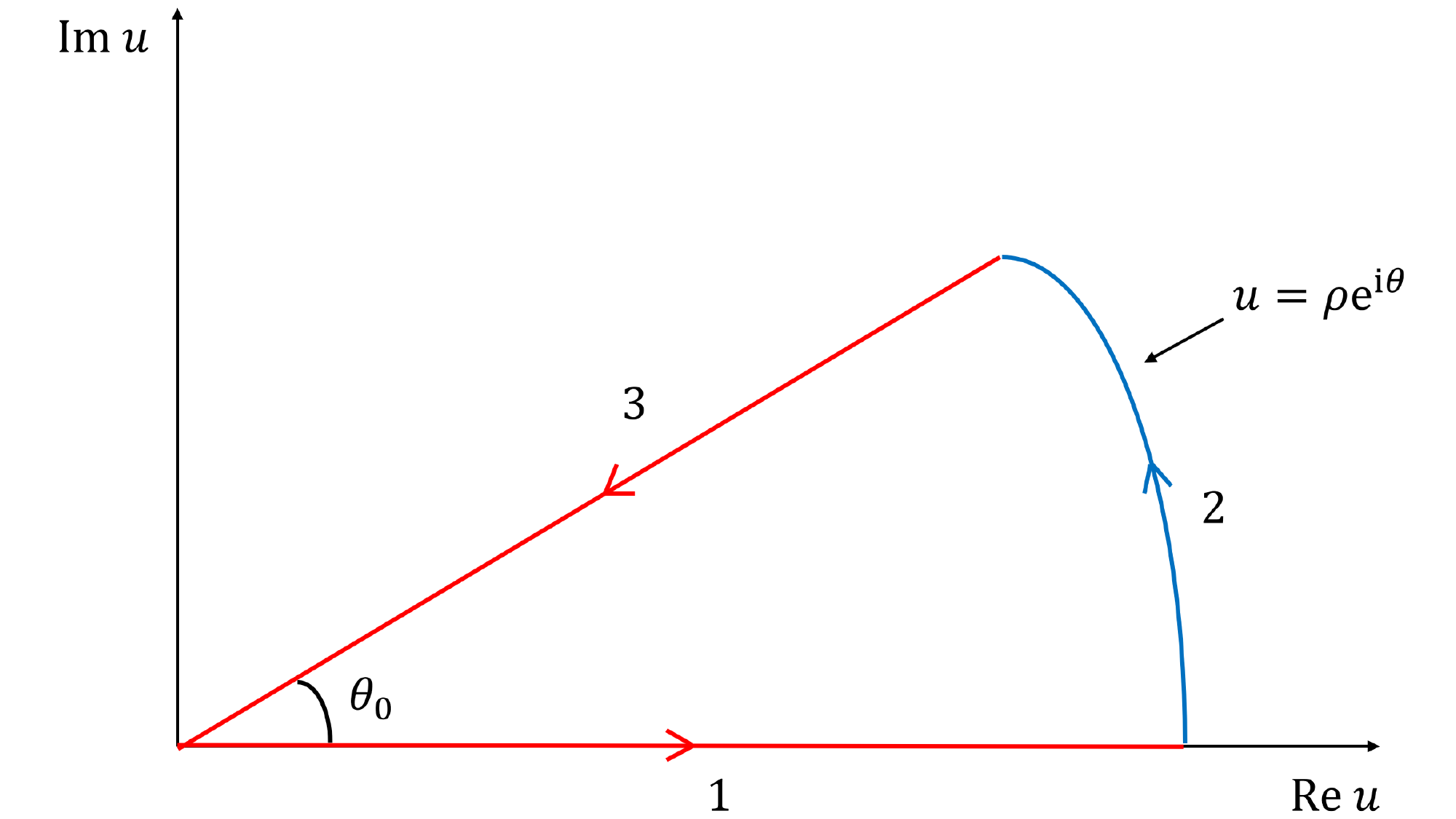}
		\end{center}
		\caption{The illustration of the closed contour of the contour integral
\rf{Closed_Contour}.}
\label{Contour}
\end{figure}

The integral around the closed contour vanishes since $\ee^{g(u)}$ is analytic in the region enclosed by the contour. The integral over the circular section 2, which is defined by $u=R \ee^{\ii\theta}$, with $R$ a positive real constant, which we will ultimately take to infinity, is given by
\beq
I_2=\ii\int_0^{\theta_0}\exp\left[\ii |S|R\ee^{\ii\theta}
+{\alpha\over 2}\re\left(-{1\over R\ee^{\ii\theta}}\right)\right]
R\ee^{\ii\theta}\dd\theta \,.
\eeq
For large $R$,
\beq
\re\left(-{1\over R\ee^{\ii\theta}}\right)
\approx -\ln\left(R\ee^{\ii\theta}\right)+\gamma\,.
\eeq
Hence,
\beq
I_2=\ii\ee^{\alpha\gamma\over 2}R^{-{\alpha\over 2}}\int_0^{\theta_0}\exp\left[\ii |S|R\ee^{\ii\theta}
-\ii{\alpha\theta\over 2}\right]
R\ee^{\ii\theta}\dd\theta\,.
\eeq
For large $R$, this integral is dominated by small $\theta$. Expanding the integrand for small $\theta$ and pushing the upper limit on $\theta$ from $\theta_0$ to infinity (since the integral over $\theta$ converges rapidly as $\theta$ grows from zero). We get
\beqn
I_2&\approx&\ii\ee^{{\alpha\gamma\over 2}+\ii |S|R}R^{1-{\alpha\over 2}}\int_0^{\infty}\exp\left[-\left(|S|R+\ii{\alpha\over 2}-\ii\right)\theta\right]\dd\theta\nn\\
&=&\ii\ee^{{\alpha\gamma\over 2}+\ii |S|R}R^{1-{\alpha\over 2}}\int_0^{\infty}\exp\left[-\left(|S|R+\ii{\alpha\over 2}-\ii\right)\theta\right]
\dd\theta\nn\\
&=&{\ii\ee^{{\alpha\gamma\over 2}+\ii |S|R}R^{1-{\alpha\over 2}}
\over |S|R+\ii{\alpha\over 2}-\ii}\,,
\eeqn

\vspace{.2in}
\noindent which vanishes in the limit $R\to\infty$. Hence, since the integral (\ref{Closed_Contour}) vanishes, we have
\beq
\int_0^{\infty}\,  \ee^{g(u)} \, \dd u=\int_0^{\infty\ee^{\ii\theta_0}}\,  \ee^{g(u)} \, \dd u 
\label{F_Rotate}
\eeq
for any $\theta_0$. That is, we can rotate the contour with complete impunity. 

Inserting (\ref{F_Rotate}) into (\ref{F.1}) gives
\beq
F_h(S)=\int_0^{\infty\ee^{\ii\theta_0}}\,  \ee^{g(u)} \, \dd u  +\rm{C. \,C.}
\label{}
\eeq

For small $S$, we choose $\theta_0={\pi\over 2}$. That is, we rotate the contour of integration to the imaginary axis, as we have done in the main text section \ref{F}. This gives us the result in the small $S$ limit.

For large $S$, we rotate the contour by $\theta_0={\pi\over 4}$ and change the variable by integration by writing $u$ as
\beq
u=v\ee^{\ii{\pi\over 4}}\,.
\eeq
This gives
\beq
F_h(S)=\ee^{\ii{\pi\over 4}}\int_0^{\infty}\,  \ee^{g(v\ee^{\ii{\pi\over 4}})} \, \dd v  +\rm{C. \,C.}
\label{Fh_large}
\eeq

Now we focus on the integral
\beq
I_3=\int_0^{\infty}\,  \ee^{g(v\ee^{\ii{\pi\over 4}})} \, \dd v \,.
\eeq
Let's separate this integral into small-$v$ and large-$v$ parts:
\beq
I_3=\int_0^{\Delta_2}\,  \ee^{g(v\ee^{\ii{\pi\over 4}})} \, \dd v 
+\int_{\Delta_2}^{\infty}\,  \ee^{g(v\ee^{\ii{\pi\over 4}})} \, \dd v\,.
\label{I_3}
\eeq
where we choose
\beq
\Delta_2=|S|^{-\sigma}
\label{Delta_2}
\eeq
with
\beq
0<\sigma<{1\over 2}\,.
\eeq
In this range of $\sigma$, $\Delta_2$ will be much less than $1$  for large $|S|$, so that, in the first integral of (\ref{I_3}), namely,
\beq
I_{3<}=\int_0^{\Delta_2}\,  \ee^{g(v\ee^{\ii{\pi\over 4}})} \, \dd v \,,
\label{I_3_1}
\eeq 
the argument of the exponential integral function will be large.  We can use the large argument expansion of $\rm{Ei}(-{1\over u})$ in the first integral. In this regime $\rm{Ei}(-{1\over u})$ will also be small, and we will use this to our advantage too.

For now, let's defer that discussion for later, and first show that the second integral in (\ref{I_3}), namely,
\beq
I_{3>}=\int_{\Delta_2}^\infty\,  \ee^{g(v\ee^{\ii{\pi\over 4}})} \, \dd v \,,
\label{I_3_2}
\eeq 
is negligible.

To do this, we will start by noting that the magnitude of $I_{3>}$ is bounded above:
\beq
|I_{3>}|\leq\int_{\Delta_2}^\infty\,  |\ee^{g(v\ee^{\ii{\pi\over 4}})}| \, \dd v 
=\int_{\Delta_2}^\infty\,  \ee^{{\rm Re}\left[{g(v\ee^{\ii{\pi\over 4}})}\right]} \, \dd v\,.
\label{I_3_3}
\eeq 
Now one can show that  
\bew
\beq
{\rm Re}\left[{\rm Ei}\left(-{1\over v\ee^{\ii{\pi\over 4}}}\right)\right]<{\rm Re}\left[{\rm Ei}\left(-{\pi\ee^{ -\ii{\pi\over 4}}\over \sqrt{2}}\right)\right]=.0415139\equiv C \,,
\label{Uplimit}
\eeq 
\ew
for $v>0$. To see this, we begin with the definition of the Ei function
\beqn
{\rm Ei}\left(-{1\over v\ee^{\ii{\pi\over 4}}}\right)\equiv -\int_{1\over v\ee^{\ii{\pi\over 4}}}^\infty\,\left(\ee^{-u}\over u\right)\dd u\,.
\label{Ei_1}
\eeqn 
By changing variable of integration from $u$ to $w\equiv u\ee^{\ii{\pi\over 4}}$, we can rewrite \rf{Ei_1} as\cite{upper limit}
\beqn
{\rm Ei}\left(-{1\over v\ee^{\ii{\pi\over 4}}}\right)\equiv -\int_{1\over v}^\infty\,\left[\ee^{{\sqrt{2}\over 2}w(-1+\ii)}\over w\right]\dd w\,.
\label{Ei_2}
\eeqn 
The real part of this  is given by
\bew
\beq
{\rm Re}\left[{\rm Ei}\left(-{1\over v\ee^{\ii{\pi\over 4}}}\right)\right]= -\int_{1\over v}^\infty\,\left[\ee^{-{\sqrt{2}\over 2}w}\cos\left({\sqrt{2}\over 2}w\right)\over w\right]\dd w  \, \equiv f(v)\,.
\label{Ei_3}
\eeq
\ew
It follows from this expression that the derivative 
\beq
f'(v)= -{\left[\ee^{-{1\over\sqrt{2}v}}\cos\left({1\over\sqrt{2}v}\right)\over v\right]} \,.
\label{dEi_1}
\eeq
Note first that this implies 
$f'(v)<0$ for $v>\sqrt{2}/\pi$. Thus the first maximum of $f(v)$ that we will encounter when decreasing $v$ from very large positive values will occur when
\beq
{1\over\sqrt{2}v}={\pi\over2} \,.
\label{firstmax}
\eeq

At this maximum, we have
\beq
f(v=\sqrt{2}/\pi)={\rm Re}\left[{\rm Ei}\left(-{\pi\ee^{ -\ii{\pi\over 4}}\over\sqrt{2}}\right)\right]=.0415139 \,,
\label{fmax}
\eeq
where the numerical value was obtained from Mathematica.

From \rf{dEi_1}, it is clear that there are infinitely many other extrema of $f(v)$ at smaller values of $v$, whose positions satisfy
\beq
{1\over\sqrt{2}v}=\left(m+{1\over2}\right)\pi
\eeq
with $m$ running over all positive integers. A moment's reflection reveals that the odd $m$'s correspond to minima, while the even $m$'s correspond to maxima. hence, the positions of the {\it maxima} are given by
\beq
v_n={\sqrt{2}\over\pi(4n+1)} \,.
\label{maxes}
\eeq
All of these maxima, and, hence, all values of the entire function $f(v)$, are smaller than the first maximum \rf{firstmax}. To prove this, we first note that $f(v)$ satisfies a rigorous upper bound, which can be derived by noting that, for real $v$,
$-\cos w\le1$ and, in the region of integration, $w>1/v$.  It follows that 
\bew
\beq
f(v)=\int_{1\over v}^\infty\,\left[\ee^{-{\sqrt{2}\over 2}w}\left(-\cos\left({\sqrt{2}\over 2}w\right)\right)\over w\right]\dd w <\int_{1\over v}^\infty\,\left[\ee^{-{\sqrt{2}\over 2}w}\over({1\over v}) \right]\dd w =v\sqrt{2}\ee^{-{\sqrt{2}\over 2v}}
\eeq
\ew

Of all of the maxima at $v_n$ with $n>0$ as given by equation \rf{maxes}, this upper bound is clearly largest
for $n=1$, at which point the bound is $f(v)<{2\over5\pi}\ee^{- 5\pi\over2}\approx 4.943\times 10^{-5}$, 
which is less {\it much less!} than the first ($n=0$) maximum given by \rf{fmax}.

Hence, that $n=0$ maximum \rf{fmax} is the absolute maximum of ${\rm Re}\left[{\rm Ei}\left(-{1\over v\ee^{\ii{\pi\over 4}}}\right)\right]$ over all positive $v$. This leads immediately to the bound \rf{Uplimit}.

It follows that
\beqn
{\rm Re}\left[{g(v\ee^{\ii{\pi\over 4}})}\right]&=&{\rm Re}\left(\ii |S|v\ee^{\ii{\pi\over 4}}\right)+{\alpha\over 2}{\rm Re}\left[{\rm Ei}\left(-{1\over v\ee^{\ii{\pi\over 4}}}\right)\right]\nn\\
&<&{\rm Re}\left(\ii |S|v\ee^{\ii{\pi\over 4}}\right) +{\alpha C\over 2}\nn\\
&<&-{|S|v\over\sqrt{2}} +{\alpha C\over 2}\,,
\label{Re_Less}
\eeqn
where in the equality we have used the definition of $g(u)$ \rf{Def_g}. Hence, using \rf{Re_Less} in \rf{I_3_3} gives
\beqn
|I_{3>}|&<&\ee^{\alpha C\over 2}\int_{\Delta_2}^\infty\,  \exp\left(-{|S|v\over\sqrt{2}}\right) \, \dd v \nn\\
&=& \ee^{\alpha C\over 2}{\sqrt{2}\over |S|}\exp\left(-{|S|\Delta_2\over\sqrt{2}}\right) \nn\\
&=& \ee^{\alpha C\over 2}{\sqrt{2}\over |S|}\exp\left(-{|S|^{1-\sigma}\over\sqrt{2}}\right) 
\label{I_3>_1}
\eeqn
where in the last equality we have used \rf{Delta_2}. Since $0<\sigma<{1\over 2}$, \rf{I_3>_1} implies
\beq
|I_{3>}|\ll {1\over |S|^{5\over 4}}\exp\left(-\sqrt{2|S|}\right)\,.
\label{I_3>_2}
\eeq
This is because for large $S$ and $0<\sigma<{1\over 2}$, the dominating exponential $\exp\left(-{|S|^{1-\sigma}\over\sqrt{2}}\right)$ on the right hand side of \rf{I_3>_1} is less than $\exp\left(-\sqrt{2|S|}\right)$ in \rf{I_3>_2} (as $\sigma$ is strictly less than $1/2$). Since the right hand side of \rf{I_3>_2} is the order of magnitude of the dominant contribution to $I_3$ coming from $I_{3<}$, as we will show in a while, $I_{3>}$ is negligible.

Now let's return to the first integral in \rf{I_3}, namely \rf{I_3_1}.
Since $\Delta_2\ll 1$, it follows that, throughout this entire region of integration over $v$, the magnitude of $v\ee^{\ii{\pi\over 4}}$ is $\ll 1$. Therefore, the argument of the Ei function in $g(v\ee^{\ii{\pi\over 4}})$: 
\beq
{g(v\ee^{\ii{\pi\over 4}})}=\ii |S|v\ee^{\ii{\pi\over 4}}+{\alpha\over 2}{\rm Ei}\left(-{1\over v\ee^{\ii{\pi\over 4}}}\right)
\label{}
\eeq
is large, which means the exponential integral itself is small. Therefore, we can expand
\beq
\ee^{g(v\ee^{\ii{\pi\over 4}})}\approx\ee^{\ii |S|v\ee^{\ii{\pi\over 4}}}\left[1+{\alpha\over 2}{\rm Ei}\left(-{1\over v\ee^{\ii{\pi\over 4}}}\right)\right]\,.
\label{Expansion}
\eeq
Inserting \rf{Expansion} into \rf{I_3_1} we get
\beq
I_{3<}\approx \int_0^{\Delta_2}\,  \ee^{\ii |S|v\ee^{\ii{\pi\over 4}}} \, \dd v
-{\alpha\over 2}\int_0^{\Delta_2}\,  \ee^{\Phi(v\ee^{\ii{\pi\over 4}})} \, \dd v
\label{I_3<_3}
\eeq
where
\beq
\Phi(u)=\ii |S|u+\ln{u}-{1\over u}
\label{Phi_u}
\eeq
as obtained by using the large argument expansion of the exponential integral. We note that the argument $\Phi(u)$ of the integral has a maximum in the complex plane at the complex values of $u$ at which
\beq
{\dd\Phi\over \dd u}=\ii |S|+{1\over u}+{1\over u^2}=0 \,.
\label{}
\eeq
For  $|S|\gg1$, the solution $u_m$ of this equation is well approximated by
\beq
u_m={\ee^{\ii\pi\over4}\over\sqrt{|S|}} \,.
\label{}
\eeq

The first integral in \rf{I_3<_3} is
\beq
\int_0^{\Delta_2}\,  \ee^{\ii |S|v\ee^{\ii{\pi\over 4}}} \, \dd v
={\ee^{\ii{\pi\over 4}}\over  |S|}\left(1-\ee^{\ii |S|\Delta_2\ee^{\ii{\pi\over 4}}} \right)\,.
\label{I_3<_4}
\eeq
The first term in \rf{I_3<_4} eventually leads to a contribution to $F_h(S)$ \rf{Fh_large}:
\beq
\ee^{\ii{\pi\over 4}}{\ee^{\ii{\pi\over 4}}\over  |S|}+{\rm C.\,C.} ={\ii+(-\ii)\over |S|}=0 \,.
\eeq
 The second term in \rf{I_3<_4} 
 can be written as
\beq
 -\frac{\ee^{-|S| \tri_2\sin \pi/4} \left(\ee^{\ii(|S|\tri_2 \cos \pi/4 + \pi/4)}\right)}{|S|} 
\ ,
\eeq
and its contribution to $F_h(S)$ is thus
\beq
-\frac{\ee^{-|S|^{1-\sigma}\sin \pi/4} \left(\ee^{\ii(|S|^{1-\sigma}\cos \pi/4 + \pi/2)}\right)}{|S|} +{\rm C.\,C.}
\ ,
\eeq
which, due to the fact that $\sigma$ is strictly less than 1/2, is sub-dominant to other contributions to this contour integral, as we will soon see.

So the contribution to $F_h(S)$ is entirely dominated by the 2nd term in \rf{I_3<_3}, the integral in which is given by 
\beq
I_4=\int_0^{\Delta_2}\,  \ee^{\Phi(v\ee^{\ii{\pi\over 4}})} \, \dd v\,.
\eeq
Now let's break this integral into three parts
\beqn
I_{4.1}&=&\int_0^{{1\over\sqrt{|S|}}-\Delta_3}\,  \ee^{\Phi(v\ee^{\ii{\pi\over 4}})} \, \dd v\,,\\
I_{4.2}&=&\int_{{1\over\sqrt{|S|}}-\Delta_3}^{{1\over\sqrt{|S|}}+\Delta_3}\,  \ee^{\Phi(v\ee^{\ii{\pi\over 4}})} \, \dd v\,,\label{4.2}\\
I_{4.3}&=&\int_{{1\over\sqrt{|S|}}+\Delta_3}^{\Delta_2}\,  \ee^{\Phi(v\ee^{\ii{\pi\over 4}})} \, \dd v\,,
\eeqn
where $\Delta_3$ is chosen to lie in the range:
\beq
|S|^{-{3\over 4}}\ll\Delta_3\ll |S|^{-{2\over 3}}\,.
\label{Range}
\eeq
Our choice of the lower bound on $\Delta_3$ is motivated by the fact that  $\ee^{\Phi(u)}$
is well approximated by a Gaussian near $u=u_m=\ee^{\ii {\pi\over 4}}/\sqrt{|S|}$; the width of this Gaussian is $O(|S|^{-{3\over 4}})$. We have chosen $\Delta_3$ to be much bigger than this width, which proves to be convenient later.

} 

%{ensures that $\Delta_3$ is much bigger than the width of the Gaussian which is a good approximation to $\ee^{\Phi(u)}$ near $u=u_m=\ee^{\ii {\pi\over 4}}/\sqrt{S}$.}

Now we will prove that $I_{4.2}$, which is extremely well approximated by a  Gaussian integral, is much larger than $I_{4.1}$ and $I_{4.3}$, which can therefore be neglected. Let's start with $I_{4.1}$. This can be written as
\beq
I_{4.1}=\ee^{\Phi(u_m)}I_{4.1.a}
%\int_0^{{1\over\sqrt{S}}-\Delta_3}\,
%\exp\left[\Phi(v\ee^{\ii{\pi\over 4}})-\Phi(u_m)\right]
\,.
\label{I_4.1_1}
\eeq 
where we have defined
\beq
I_{4.1.a}\equiv
\int_0^{{1\over\sqrt{|S|}}-\Delta_3}\,
\exp\left[\Phi(v\ee^{\ii{\pi\over 4}})-\Phi(u_m)\right]\, \dd v\,,
\label{I_4.1_2}
\eeq 
Now we show that the magnitude of this is bounded. 
It's clear that
\beqn
|I_{4.1.a}|&\leq&
\int_0^{{1\over\sqrt{|S|}}-\Delta_3}\,
\Big|\exp\left[\Phi(v\ee^{\ii{\pi\over 4}})-\Phi(u_m)\right]\Big|\, \dd v\nn\\
&=&\int_0^{{1\over\sqrt{|S|}}-\Delta_3}\,
\exp\left[{\rm Re}\left(\Phi(v\ee^{\ii{\pi\over 4}})-\Phi(u_m)\right)\right]\, \dd v\,.\nn\\
\label{I_4.1_3}
\eeqn
Recall that $\Phi(u)$ is given by \rf{Phi_u}. It follows that
\beq
{\dd\over\dd v}
\left[{\rm Re}\left(\Phi(v\ee^{\ii{\pi\over 4}})\right)\right]
=-{|S|\over\sqrt{2}}+{1\over v}+{1\over\sqrt{2}v^2}\,.
\eeq
For all $v<{1\over\sqrt{|S|}}$, it is clear that
\beq
{\dd\over\dd v}
\left[{\rm Re}\left(\Phi(v\ee^{\ii{\pi\over 4}})\right)\right]>0\,.
\eeq
Hence,
\beqn
&&{\rm Re}\left(\Phi(v\ee^{\ii{\pi\over 4}})-\Phi(u_m)\right)\nn\\
&<&{\rm Re}\left[\Phi\left(\left({1\over\sqrt{|S|}}-\Delta_3\right)\ee^{\ii{\pi\over 4}}\right)-\Phi(u_m)\right]
\label{Bond_1}
\eeqn
throughout the region of integration $0<v<{1\over\sqrt{|S|}}-\Delta_3$.
Furthermore, we can approximate ${\rm Re}\left[\Phi\left(\left({1\over\sqrt{|S|}}-\Delta_3\right)\ee^{\ii{\pi\over 4}}\right)\right]$ by expanding it around the maximum at $u_m$:
\beqn
&&{\rm Re}\left[\Phi\left(\left({1\over\sqrt{|S|}}-\Delta_3\right)\ee^{\ii{\pi\over 4}}\right)-\Phi(u_m)\right]\nn\\
&\approx& {1\over 2}{\dd^2 \left[{\rm Re}\left(\Phi\left(v\ee^{\ii{\pi\over 4}}\right)\right)\right]\over\dd v^2}\Bigg|_{v={1\over\sqrt{|S|}}}
\Delta_3^2\nn\\
&=&-{1\over 2}\sqrt{2|S|^3}\Delta_3^2\,.
\label{Expan_1}
\eeqn
Note that this expansion works provided the $O(\Delta_3^3)$ term in the expansion is $\ll 1$. 

That term is
\beq
{1\over 6}{\dd^3 \left[{\rm Re}\left(\Phi\left(v\ee^{\ii{\pi\over 4}}\right)\right)\right]\over\dd v^3}\Bigg|_{v={1\over\sqrt{|S|}}}
\Delta_3^3
=\left(|S|^2\over\sqrt{2}\right)\Delta_3^3\,.
\eeq
To ensure that this is $\ll 1$ as $S\to\infty$, we must have
\beq
\Delta_3\ll |S|^{-{2\over 3}}\,,
\eeq
which is consistent with \rf{Range}. Now using \rf{Expan_1} in the bound \rf{Bond_1}, and using that bound in the bound \rf{I_4.1_3}, we have
\beq
\mid T_{4.1a|}\mid <{1\over\sqrt{|S|}}\exp\left(-\sqrt{|S|^3\over 2}\Delta_3^2\right)\,.
\eeq
Combining this with \rf{I_4.1_1} we get
\beqn
\mid I_{4.1}\mid &<& \big|\ee^{\Phi(u_m)}\big|{1\over\sqrt{|S|}}\exp\left(-\sqrt{|S|^3\over 2}\Delta_3^2\right)\nn\\
&<&\big|\ee^{\Phi(u_m)}\big|{1\over\sqrt{|S|}}\exp\left(-{|S|^\alpha\over\sqrt{2}}\right)\,,
\label{Bond_2}
\eeqn
where $\alpha>0$. In the second ``$<"$ in \rf{Bond_2} we have used the range for $\Delta_3$ given in \rf{Range}.

Now we turn to $I_{4.2}$. Since we have shown above that $\Delta_3$ is small enough to expand $\Phi(u)$ to quadratic order in the regime 
\beq
{1\over\sqrt{|S|}}-\Delta_3<v<{1\over\sqrt{|S|}}-\Delta_3\,,
\eeq
the integral \rf{4.2} simply becomes a Gaussian integral: 
\begin{widetext}
\beqn
I_{4.2}&=&\ee^{\Phi(u_m)} \int_{{1\over\sqrt{|S|}}-\Delta_3}^{{1\over\sqrt{|S|}}+\Delta_3}\,  \exp\left[{1\over 2}{\dd^2\Phi\left(u\right)\over\dd u^2}\Bigg|_{u=u_m}
\left(u-u_m\right)^2\right] \, \dd v \nn\\
&=&\ee^{\Phi(u_m)} \int_{{1\over\sqrt{|S|}}-\Delta_3}^{{1\over\sqrt{|S|}}+\Delta_3}\,  \exp\left[-{\sqrt{2}\over 2}(1-\ii)|S|^{3\over 2}
\left(v-{1\over\sqrt{|S|}}\right)^2\right] \, \dd v 
\label{4.2.1}
\eeqn
\end{widetext}
Since $\Delta_3$ (see\rf{Range}) is much larger than the width ($\sim |S|^{-3/4}$) of the Gaussian integral \rf{4.2.1}, we can, with  exponential accuracy, extend the range of this integral to $\pm\infty$. This gives
\begin{widetext}
\beqn
I_{4.2}&=&\ee^{\Phi(u_m)} \int_{-\infty}^{\infty}\,  \exp\left[-{\sqrt{2}\over 2}(1-\ii)|S|^{3\over 2}
\left(v-{1\over\sqrt{|S|}}\right)^2\right] \, \dd v \nn\\
&=&{\sqrt{\pi}\over |S|^{3\over 4}}\exp\left[{\Phi(u_m)+\ii {\pi\over 8}}\right]
\nn\\ 
&=&{\sqrt{\pi}\over |S|^{5\over 4}}\exp\left[-\sqrt{2|S|}
+\ii \left(\sqrt{2|S|}+{3\pi\over 8}\right)\right]\,,
\label{4.2.2}
\eeqn
\end{widetext}
where in the last equality we have inserted $u_m=\ee^{\ii\pi\over 4}/\sqrt{|S|}$. Inserting \rf{4.2.2} into \rf{I_3<_3} and then \rf{I_3<_3} into \rf{Fh_large} recovers the result \rf{bigSfin} of section \rf{F}.

The argument that $I_{4.3}$ is negligible compared to $I_{4.2}$ closely parallels the argument given above for $I_{4.1}$, so we will leave it as an exercise for the reader.


\begin{thebibliography}{16}
\bibitem{fixed} Our approach here of freezing the magnitude of $\bv$ and only allowing its  orientation  to fluctuate differs from the approach of \cite{Toner_prl12,Chen_prl20,Chen_pre20}, who used a ``soft-spin" model in which the magnitude $|\bv|$ of the velocity field $\bv$ was allowed to fluctuate as well. However, in those papers,  and in reality, fluctuations of the magnitude of $\bv$ are ``massive"; that is,  decay at a finite rate, even in the long-wavelength limit. That is, such fluctuations  are not
``Goldstone modes", and, hence, are irrelevant in the long-distance, long-time limit. It therefore makes no difference whether we allow  them or not, and we find it more convenient  here to ban such fluctuations from the outset.
\bibitem{product rule} Terms in which the derivative acts on more than one velocity component - i.e., terms in which there is more than one velocity component to the right of the derivative operator- can be reduced by the product rule for derivatives to terms of the form \rf{1dgen}.
\end{thebibliography}
\end{document}